\documentclass[%
 reprint, 
 amsmath,amssymb,
 aps, physrev,
]{revtex4-2}

\usepackage{graphicx}
\usepackage{dcolumn}
\usepackage{bm}

\usepackage{braket}
\usepackage{bbold}
\usepackage{subcaption}
\usepackage{ulem}
\usepackage{xcolor}
\usepackage{ragged2e}
\usepackage{caption}

\DeclareCaptionJustification{justified}{\justifying}
\begin{document}

\preprint{APS/123-QED}

\title{\textbf{Optimizing Doppler laser cooling protocols for quantum sensing with 3D ion crystals in a Penning trap} 
}%

\author{John Zaris}
 \email{Contact author: john.zaris@colorado.edu}
\affiliation{%
 Department of Physics, University of Colorado, Boulder, Colorado 80309, USA
}%

\author{Wes Johnson}
\affiliation{%
 Department of Physics, University of Colorado, Boulder, Colorado 80309, USA
}%
\affiliation{%
 Sandia National Laboratories, Albuquerque, New Mexico, 87185, USA
}

\author{Athreya Shankar}
\affiliation{%
 Department of Physics, Indian Institute of Technology Madras, Chennai, India, 600036
}%
\affiliation{%
 Center for Quantum Information, Communication and Computing, Indian Institute of Technology Madras, Chennai 600036, India
}%

\author{John J. Bollinger}
\affiliation{%
 National Institute of Standards and Technology, Boulder, Colorado 80309, USA
}%

\author{Allison L. Carter}
\affiliation{%
 National Institute of Standards and Technology, Boulder, Colorado 80309, USA
}%

\author{Daniel H.E. Dubin}
\affiliation{%
 Department of Physics, University of California, San Diego, California 92093, USA
}

\author{Scott E. Parker}
\affiliation{%
 Department of Physics, University of Colorado, Boulder, Colorado 80309, USA
}
\affiliation{%
 Renewable and Sustainable Energy Institute, University of Colorado, Boulder, Boulder, Colorado 80309, USA
}

\date{\today}

\begin{abstract}
Large, 3D trapped ion crystals offer improved sensitivity in quantum sensing protocols, and are expected to be implemented as platforms in near-future experiments. However, numerical techniques used to study the laser cooling of such crystals are inefficient as the number of ions, $N$, in the crystal increases. Here we develop a powerful numerical framework to simulate laser cooling of up to $10^5$ ions stored in a Penning trap. We apply this framework to characterize and optimize the cooling of ellipsoidal 3D crystals. We document new pathways to enhanced cooling based on the addition of an axial component to the potential energy-dominated $\boldsymbol{E}\times\boldsymbol{B}$ modes. Furthermore, we observe greatly enhanced cooling of the perpendicular kinetic energy to below 1 mK in prolate ion crystals, enabling a simplified cooling beam setup for such crystals. We propose specific values of trap and laser beam parameters which lead to optimal cooling in a variety of examples. This work illustrates the feasibility of preparing large 3D crystals for high-sensitivity quantum science protocols, motivating their use in future experiments.

\end{abstract}

\maketitle


\section{\label{sec:introduction}Introduction}

Trapped ion crystals are key platforms in modern quantum science research, owing to the precise control which modern experiments exert over their internal and motional degrees of freedom.  In the last decade or so, much experimental work has been devoted to studying the use of two-dimensional (2D) ion crystals in Penning traps for quantum simulation and sensing applications. In particular, 2D crystals have been utilized to study quantum many-body physics \cite{britton2012,safavi-naini2018,cohn2018,shankar2022,qiao2022, pham2024}, growth of entanglement in interacting systems \cite{swingle2016,garttner2017}, creation of many-ion spin squeezed states \cite{bohnet2016,pezze2018}, and quantum sensing of motion and electric fields \cite{gilmore2017,affolter2020,gilmore2021, toscano2006}.  Experimental methods with single plane crystals in Penning traps continue to advance, recently allowing for site-resolved single-shot detection \cite{wolf2024} and single-site addressing of ions \cite{mcmahon2024}.  While the Penning trap has historically been the preferred apparatus for storing large 2D crystals, new experimental setups have successfully trapped 2D crystals with hundreds of ions in radiofrequency traps \cite{kiesenhofer2023, guo2024}, where single-site addressing is straightforward.

Currently, there is significant interest in extending experimental protocols to three-dimensional (3D) crystals with large numbers of ions.  The use of 3D crystals would allow for the simulation of new many-body Hamiltonians \cite{hawaldar2024}, development of novel metrological applications \cite{hausser2025,leibrandt2024}, and measurement of increasingly weak signals in quantum sensing experiments \cite{gilmore2021}. For example, due to the $1/\sqrt{N}$ sensitivity scaling of quantum sensing protocols, a 3D crystal with $N=10^6$ ions could lead to a 100-fold improvement over current measurements with 2D crystals of $N\approx 100$ ions. While the equilibrium structure and laser cooling of large 3D trapped ion crystals have been studied for decades \cite{itano1998,itano1988, drewsen1998, dubin1999}, modern methods of control and readout could now help to realize their potential in quantum science research.

To utilize their quantum properties, trapped ion crystals must first be laser-cooled to ultracold temperatures.  For instance, a traveling-wave potential can produce an optical dipole force (ODF) Hamiltonian which couples the internal state of the ions and their positions, as long as the ions are localized with positional fluctuations small compared to the ODF wavelength.  This is known as Lamb-Dicke confinement and is a requirement for many quantum protocols.  Laser cooling of 2D ion crystals in Penning traps is well-studied and has seen key improvements in recent years. Significant effort has been dedicated, in particular, to improving the cooling of the axial, or `drumhead' modes, which describe ion motion perpendicular to the plane of the crystal and which are typically utilized in quantum sensing and simulation experiments. Notably, near-ground state cooling of the drumhead mode branch has been achieved by implementing electromagnetically-induced transparency (EIT) cooling, which is expected to improve quantum sensing capabilities \cite{jordan2019, kiesenhofer2023}. On the other hand, the potential-energy dominated $\boldsymbol{E}\times\boldsymbol{B}$ modes have proven difficult to cool, leading to unwanted broadening of the drumhead modes \cite{shankar2020}. In fact, this is one of several problems in the study of ion crystal laser cooling for which numerical methods have proven invaluable. Novel techniques for improved $\boldsymbol{E}\times\boldsymbol{B}$ cooling have recently been proposed, and evaluated via molecular dynamics (MD) laser cooling simulations \cite{johnson2024, johnson2025}.   

The cooling dynamics of 3D crystals remain largely uncharacterized, both experimentally and theoretically.  Past experimental studies have observed crystallization in spheroidal crystals \cite{itano1988,itano1998,mortensen2006}, but cooling to the level of several mK has not been well-characterized. Furthermore, current numerical laser cooling simulations \cite{tang2019,poindron2023} are challenging with more than a few thousand ions, as the computation of Coulomb interactions becomes intractable due to their $N^2$ scaling. To enable the simulations of large 3D crystals, we recently implemented an efficient, compiled code which utilizes the fast multipole method (FMM) to accelerate the calculation of Coulomb interactions between ions \cite{zaris2024}.

In this paper, we numerically study the laser cooling of 3D ion crystals in a Penning trap using our MD code. In Section II, we review the relevant Penning trap physics and introduce the MD code, including our laser cooling model.  Furthermore, we describe the normal mode analysis typically used to study the excitation of ion crystals.  In Section III, we examine the cooling of low-frequency, potential energy dominated $\boldsymbol{E}\times\boldsymbol{B}$ modes.  In 2D crystals, these have proven difficult to cool as they experience limited coupling to the cooling lasers.  However, $\boldsymbol{E}\times\boldsymbol{B}$ modes in 3D crystals include both planar and axial components, and we show that this can lead to efficient potential energy cooling.  We then study the effect of several trap and laser beam parameters on the potential energy of 3D crystals.  In Section IV, we demonstrate enhanced cooling of the kinetic energy in prolate 3D crystals.  We discover a regime in which coupling between the axial and planar motion allows for efficient cooling with the use of a single laser beam directed parallel to the magnetic field.  Finally, in Section V, we describe numerical methods used to scale up our simulations to larger crystals and present cooling results for a $N=10^5$ ion crystal.

\section{Theoretical Framework}
\subsection{Molecular Dynamics Simulation}

In a Penning trap, ions are confined by a combination of static electric and magnetic fields.  By applying a voltage difference between the endcap and middle electrodes such that the endcaps are at a higher potential, a positively charged ion is subject to a confining potential in the axial $(\boldsymbol{\hat{z}})$ direction.  Near the center of the trap, this potential is approximately harmonic and parameterized by

\begin{equation}
\label{eq1}
    \phi_{trap}(\boldsymbol{x}) = \frac{1}{4}k_z(2z^2-x^2-y^2),
\end{equation}
where $\boldsymbol{x} = x\hat{\boldsymbol{x}}+y\hat{\boldsymbol{y}}+z\hat{\boldsymbol{z}}$ is the position of the ion and $k_z$ parameterizes the strength of the potential.  The axial trapping frequency is given by $\omega_z = \sqrt{qk_z/m}$, where $m$ is the ion's mass and $q$ is its charge.  This is the frequency of the axial center-of-mass (COM) mode of a trapped ion crystal.  The potential $\phi_{trap}$ is deconfining in the radial direction, so an axial magnetic field, $\boldsymbol{B} = B\hat{\boldsymbol{z}}$, is used to produce radial confinement. In the remainder or this work, we refer to motion in the plane perpendicular to $\boldsymbol{B}$ as `planar', `perpendicular', or `radial'.  

In order to control the global rotation frequency of an ion crystal, it is common to introduce a time-varying rotating wall potential given by

\begin{equation}
\label{eq2}
    \phi_{wall}(\boldsymbol{x},t) = \frac{1}{2}k_z\delta(x^2+y^2)\cos[2(\varphi+\omega_r t)],
\end{equation}
where $\delta$ parameterizes the strength of the rotating wall in relation to the trap potential, $\varphi$ is the azimuthal coordinate of the ion, and $\omega_r$ is the rotating wall frequency. For an $N$-ion crystal, the total potential energy is given by

\begin{align}
\label{eq3}
E_{pot} = \sum_{i=1}^N\Big[q_i\Big(&\phi_{trap}(\boldsymbol{x}_i)+\phi_{wall}(\boldsymbol{x}_i,t)\nonumber\\
    &+\frac{1}{8\pi\varepsilon_0}\sum_{j=1,j\neq i}^N\frac{q_j}{|\boldsymbol{x}_i-\boldsymbol{x}_j|}\Big)\Big].
\end{align}

In order to study the laser cooling of 3D ion crystals in a Penning trap, we use a previously developed MD simulation framework which has been carefully benchmarked and optimized to efficiently simulate large crystals \cite{zaris2024}. Our MD simulation evolves ions according to the fields listed above.  It utilizes the fast multipole method (FMM), via the FMM3D library \cite{greengard1987,greengard1997,cheng1999}, to accelerate the calculation of Coulomb interactions.  Briefly, the FMM forms groups of nearby ions at different levels of refinement and calculates their multipole expansions to approximate the electric field at the locations of more distant ions.  This avoids the explicit calculation of the Coulomb interaction between each pair of ions.  For large ion number $N$, the number of FMM calculations required to find the Coulomb force on all ions scales with $N$, much better than the $N^2$ scaling of the direct Coulomb calculation. Previous benchmarking studies have characterized differences in ion crystal dynamics when the FMM is used in place of the direct Coulomb calculation \cite{zaris2024}.  While the trajectories of individual ions are not reproduced, bulk properties of the crystal, such as its energy, agree to high precision. As in previous simulations \cite{zaris2024}, here we set the requested precision, $\epsilon$, to $10^{-7}$ in the FMM3D library.

\begin{figure}
    \centering
    \includegraphics[scale=0.35]{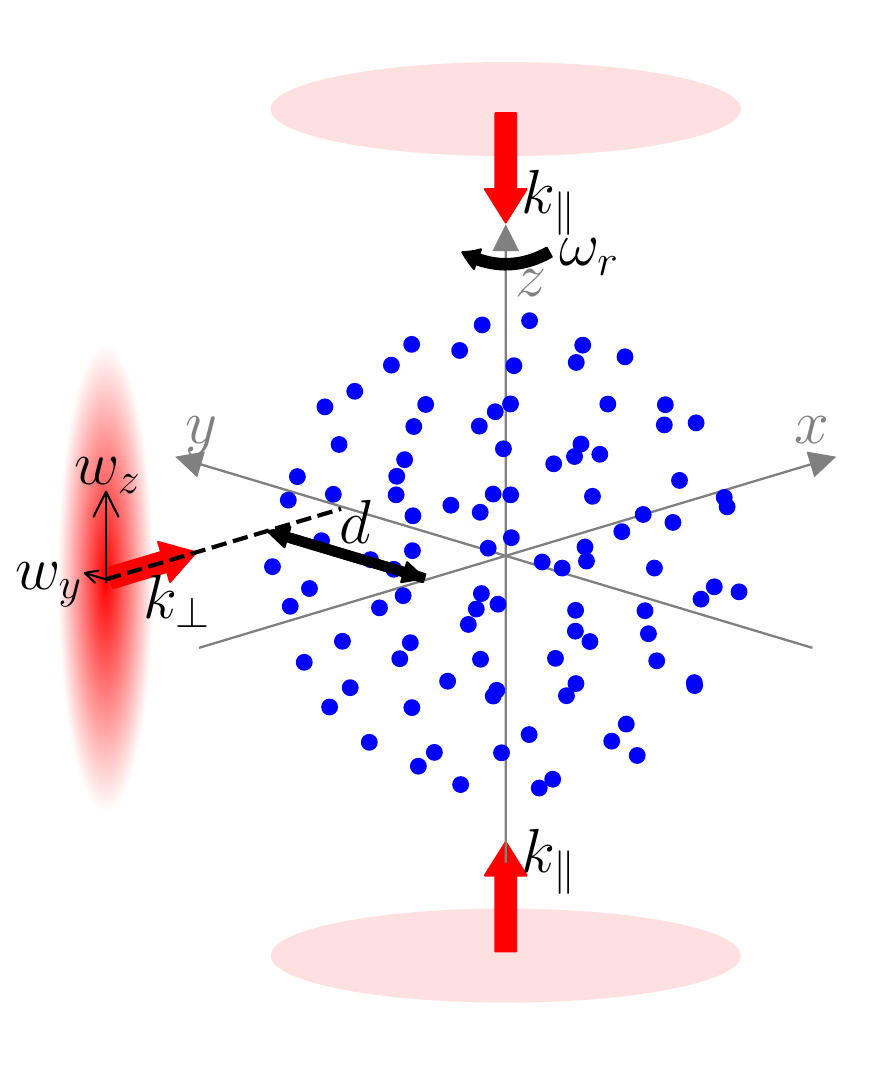}
    ~ 
    \caption{The setup to cool the crystal relies on two axial and one perpendicular cooling beam.  The axial beams have a low intensity which we approximate as uniform across the crystal in our simulations. They are detuned by $\Delta_{\parallel_1}=\Delta_{\parallel_2} = -\gamma_0/2$ from resonance with the $2s\;^2S_{1/2}\rightarrow 2p\;^2P_{3/2}$ transition.  The perpendicular beam is offset from the trap center by a distance $d$ and has detuning $\Delta_{\perp}$. Unlike in the case of a 2D crystal, here, the perpendicular beam waist in both directions, $w_y$ and $w_z$, affect the cooling rate. Figure is reproduced from \cite{zaris2024}.}
    \label{fig1}
\end{figure}

In the NIST Penning trap experiment modeled in this work, $^9Be^+$ ions are Doppler cooled using laser beams tuned near the $2s\;^2S_{1/2}\rightarrow 2p\;^2P_{3/2}$ cycling transition. An axial beam, whose intensity is low (to prevent recoil heating in the radial directions) and approximated to be uniform across the crystal, cools the ion motion parallel to $\boldsymbol{B}$. We note that the NIST experiment uses only a single axial beam.  However, we use two counterpropagating axial beams, denoted by $\parallel_1$ and $\parallel_2$, in our simulations to prevent heating of the axial COM mode when the beams are turned off \footnote{When a single axial beam is used in our simulations, it exerts a force which displaces the crystal axially.  When the axial beam is turned off (to, for instance, evolve the crystal in the absence of Doppler cooling beams), the axial COM mode becomes excited.  The use of two beams prevents the displacement of the crystal and the resulting COM mode excitation. In the NIST experiment, the axial beam is turned off by reducing its intensity adiabatically, which prevents heating}. A perpendicular beam, with a higher peak intensity and smaller waist, and which is offset from the trap center by distance $d$, is used to cool the perpendicular motion. The beam's intensity is assumed to be Gaussian and parameterized as $I_{\perp}(\boldsymbol{x}) = I_{\perp,0}\exp\big(\frac{-2(y-d)^2}{w_y^2}\big)\exp\big(\frac{-2z^2}{w_z^2}\big)$, where $I_{\perp,0}$ is the central intensity and $w_y$ ($w_z$) is the beam waist in the $\boldsymbol{\hat{y}}$ ($\boldsymbol{\hat{z}}$) direction.  The Doppler cooling setup is illustrated in Fig.~\ref{fig1}. It is useful to define the saturation parameter of each beam  $l\in\{\parallel_1,\parallel_2,\perp\}$ in terms of its intensity $I_l(\boldsymbol{x})$ as $S_l(\boldsymbol{x}) = I_l(\boldsymbol{x})\sigma_0/\hbar\omega_0\gamma_0$, where $\sigma_0$ is the scattering cross section on resonance, $\omega_0$ is the cooling transition frequency, and $\gamma_0 = 2\pi\times 18$ MHz is the natural linewidth of the cooling transition.  Note that the saturation parameters of the axial beams are uniform in space.  Then, the theoretical rate at which an ion with position $\boldsymbol{x}$ and velocity $\boldsymbol{v}$ scatters photons from laser beam $l$ is approximately given by \cite{itano1988}

\begin{equation}
\label{eq4}
 \gamma_l(\boldsymbol{x},\boldsymbol{v}) =  S_l(\boldsymbol{x})\gamma_0\frac{(\gamma_0/2)^2}{(\gamma_0/2)^2(1+2S_l(\boldsymbol{x}))+(\Delta_l-\boldsymbol{k}_l\cdot\boldsymbol{v})^2},
\end{equation}
where $\Delta_l$ is the detuning of the laser beam from the transition and $\boldsymbol{k}_l$ is the beam's wavevector. In Eq.~(\ref{eq4}), we ignore the shift in $\Delta_l$ due to the recoil term, which is small. Throughout this work, we assume that the saturation parameter at maximum laser intensity for the axial and perpendicular beams are given by $S_{0,\parallel_1}=S_{0,\parallel_2} = 5\times10^{-3}$ and $S_{0,\perp} = 0.5$, respectively. 

The laser cooling model implemented in our code treats the scattering of photons by a given ion as a Poisson process \cite{tang2019}. Therefore, the number of photons from laser beam $l$ scattered by ion $i$ during a given timestep of duration $\Delta t$ is computed by randomly sampling from a Poisson distribution with mean value $\gamma_l(\boldsymbol{x}_i,\boldsymbol{v}_i)\Delta t$.  Each photon is absorbed along the direction of the beam's $k$-vector, while the direction of the emitted photons is randomized and the ions' velocities are modified accordingly. This algorithm makes use of certain key simplifications.  First, we approximate the time between photon absorption and emission as infinitesimal, and apply only a single momentum kick to an ion per timestep.  The validity of this assumption rests on the condition $\omega_c/\gamma_0 << 1$, where $\omega_c = qB/m$ is the cyclotron frequency of an ion (for the NIST parameter values used here, $\omega_c/\gamma_0\sim0.4$).  Additionally, when computing the total change in momentum of an ion, we independently calculate the momentum kick due to each laser beam.  This method ignores changes in the saturation parameter for ions illuminated by multiple beams as well as interference between the counterpropagating axial beams, which would produce a standing wave in the axial direction.  As mentioned, the use of two axial beams is a numerical tool and the NIST experiment uses only a single axial beam, so no standing wave exists, in practice. Finally, we ignore correlations between scattering events, which is allowable since the average number of photons scattered by a given ion during a single timestep is $\ll 1$.

Our code advances the ion motion using a cyclotronic integrator, which is a second-order, symplectic method.  We use a timestep of $\Delta t=1$ ns in this work, which is sufficiently small to satisfy the scattering condition and also to resolve the highest frequency ion motion. More details on this algorithm can be found in \cite{tang2019,zaris2024}.  In the following sections, we study how 3D crystal Doppler laser cooling results depend on various laser beam parameters. While the axial beams are typically very effective at cooling the axial motion to near the Doppler limit ($T_{Doppler} = \hbar\gamma_0/2k_B$), the cooling of the perpendicular motion is highly sensitive to the perpendicular beam's parameters, including its waist, detuning, and offset from the origin.  

\subsection{Ion crystal normal modes}

\begin{figure}
    \centering
    \begin{subfigure}{0.4\textwidth}
        \subcaption{}
        \centering
        \includegraphics[scale=0.26]{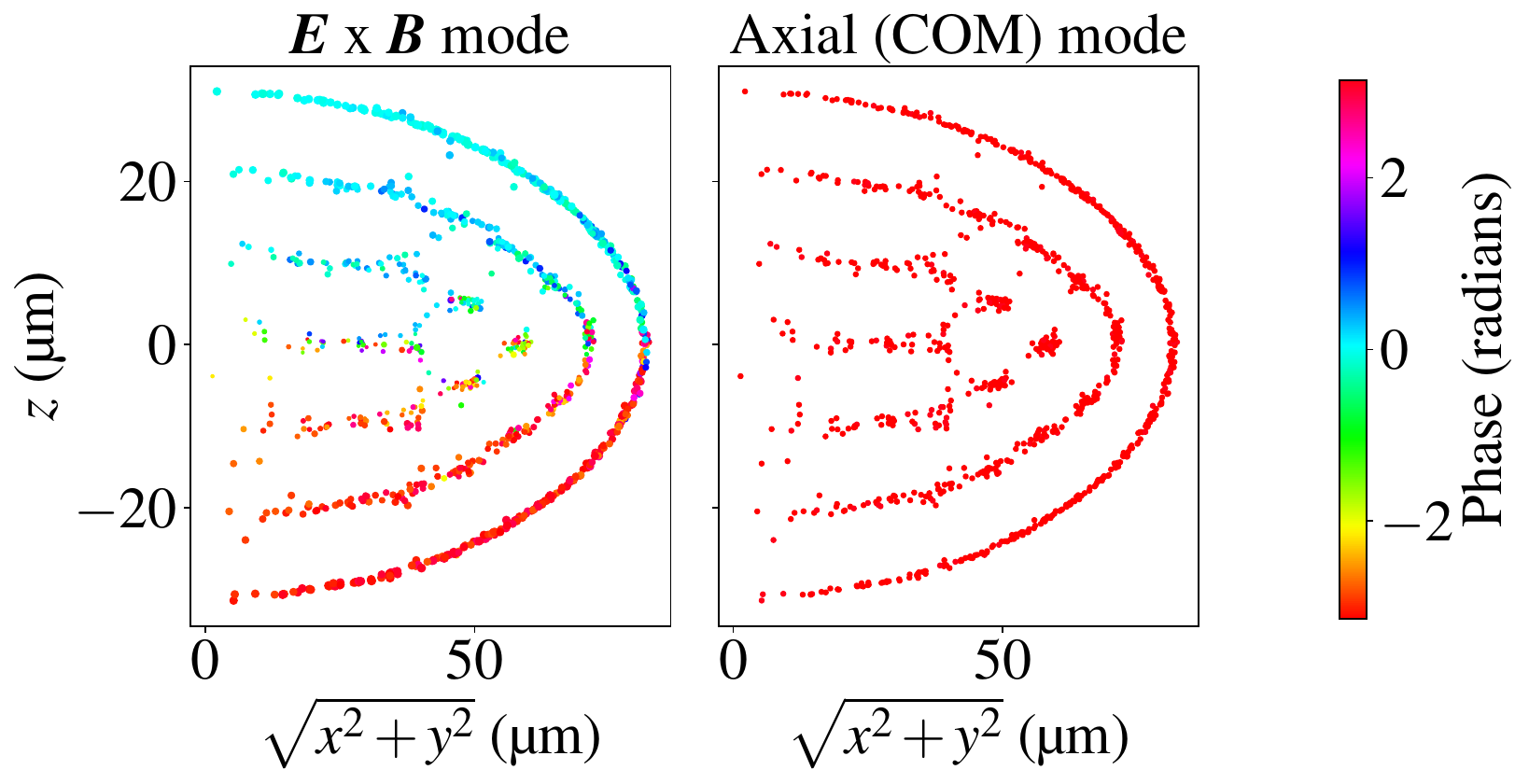}
        \label{fig2a}
    \end{subfigure}
    ~ 
    \bigskip
    \centering
    \hspace{0cm} 
    \begin{subfigure}{0.4\textwidth}
        \subcaption{}
        \centering
        \includegraphics[scale=0.37]{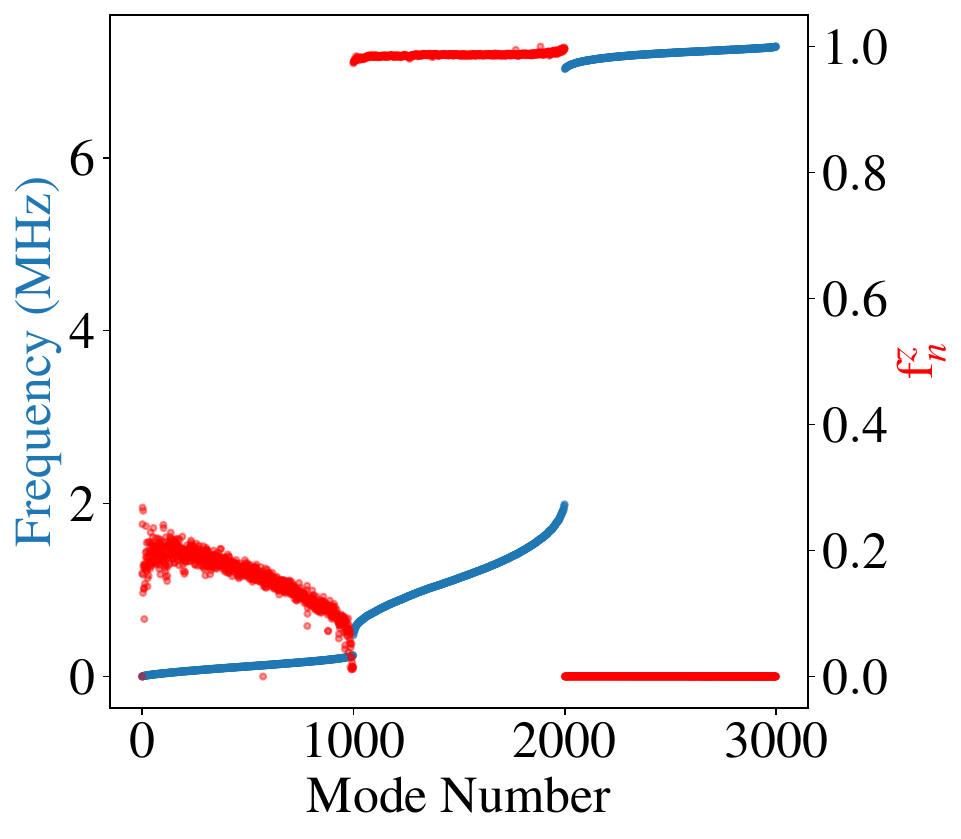}
        \label{fig2b}%
    \end{subfigure}
    \caption{(a) Two examples of 3D crystal normal modes are illustrated. The same equilibria configuration for a $N=1000$ ion crystal is shown in each panel.  The shell structure characteristic of ultracold mesoscopic crystals is evident. The relative phase of each ion in the mode is represented by its color and the relative amplitude of the ion's motion is represented by its size.  The left-hand plot shows an $\boldsymbol{E}\times\boldsymbol{B}$ mode in which the ions above $z=0$ have a nearly uniform phase and ions below $z=0$ have a different, but also nearly uniform phase (mode 2).  For this plot, the phase is calculated using $\phi_p = arg\{u_n^x\}$, where $u_n^x$ is the $x$-component of the positional part of eigenmode $n$. The right-hand panel illustrates the axial center of mass mode (mode 1884), in which all ions move in phase at frequency $\omega_z$. Here, the phase is $\phi_p = arg\{u_n^z\}$. (b) The normal mode frequencies for the crystal shown in (a) are plotted in blue.  The mode branches occupy distinct frequency bands and are separated by frequency gaps, as is seen in 2D crystals. In order of increasing frequency, they are the $\boldsymbol{E}\times\boldsymbol{B}$, axial, and cyclotron branches. In more prolate 3D crystals, the gap between branches can vanish. The axial component of each mode, $f_n^z$, is shown in red.  It is computed using Eq.~\ref{eq8}.  The $\boldsymbol{E}\times\boldsymbol{B}$ modes gain a large axial component, even in 3D crystals with moderate aspect ratios.  The axial modes have a small, but noticeable, planar component in this case.  Note that the axial center-of-mass mode, for which $f^z=1$, is no longer the highest frequency mode, as is the case in 2D crystals.}
    \label{fig2}
\end{figure}

Ion motion in an ultracold crystal is described by the crystal's normal modes.  This treatment is valid as long as the ion crystal remains near its equilibrium configuration.  The method of computing normal modes of various trapped ion configurations is described in many papers \cite{wang2013,dubin2020, jain2020,shankar2020, hawaldar2024}. We will provide a brief description of the analysis for a general 3D crystal here.  We introduce the state vector,

\begin{equation}
\label{eq5}
 \ket{q} = \begin{pmatrix}
\ket{\delta x}\\ 
\ket{v} \\ 
\end{pmatrix},
\end{equation}
which has length $6N$ and contains the positions and velocities of the ions in the crystal.  We follow \cite{shankar2020} and use bra-ket notation in this discussion. As will be explained here, the eigenmodes of ion crystals in a Penning trap are described using complex vectors. The use of bra-ket notation will then prove convenient when working in the normal mode basis.  After expanding the system Lagrangian about equilibrium and computing the Euler-Lagrange equations of motion, one obtains

\begin{equation}
\label{eq6}
\frac{d\ket{q}}{dt} = \begin{pmatrix}
    \mathbb{0}_{3N}&\mathbb{1}_{3N}\\-\mathbb{K}/m&-2\mathbb{W}/m 
    \end{pmatrix}\ket{q}\equiv \mathbb{D}\ket{q}.
\end{equation}
Each of the submatrices above has size $3N\times 3N$.  The matrix $\mathbb{K}$ is known as the stiffness matrix and is real and symmetric.  The effect of the Lorentz force is encapsulated in the real, antisymmetric matrix $\mathbb{W}$.  The exact forms of $\mathbb{K}$ and $\mathbb{W}$ are provided in \cite{zaris2024}. Diagonalizing this system produces $3N$ normal modes $\ket{u_n}$ with frequencies $\omega_n$ such that the time evolution of the state vector can be expressed as

\begin{equation}
\label{eq7}
\ket{q} = \sum_{n=1}^{3N}\big[A_ne^{-i\omega_nt}\ket{u_n}+A_n^*e^{i\omega_nt}\ket{u_n^*}\big],
\end{equation}
where $A_n$ is the complex amplitude of mode $n$ and $\ket{u_n^*}$ is the complex conjugate of $\ket{u_n}$. Each $6N$ dimensional eigenmode includes $3N$ position amplitudes and $3N$ velocity amplitudes, which desrcibe the motion of every ion in each spatial direction.  To refer to various components of the eigenvectors, it is often convenient to write $\ket{u_n} = (\ket{u_n^r}, \ket{u_n^v})^T$, where $\ket{u_n^r} = (\ket{u_n^x}, \ket{u_n^y},\ket{u_n^z})^T$ and $\ket{u_n^v}= (\ket{u_n^{v_x}}, \ket{u_n^{v_y}},\ket{u_n^{v_z}})^T$ are the position and velocity components, respectively.

Here we briefly review the general differences in the mode spectra between 2D and 3D crystals. In an $N$-ion 2D crystal, $N$ normal modes describe the purely axial motion of the ions while $2N$ modes describe purely planar motion.  The planar mode branches are further divided into low frequency $\boldsymbol{E}\times\boldsymbol{B}$ modes and high frequency cyclotron modes. For most trapping parameters, the three mode branches are well separated in frequency space, with axial mode frequencies located between $\boldsymbol{E}\times\boldsymbol{B}$ mode and cyclotron mode frequencies. In 3D crystals, the normal modes are no longer purely planar or axial, due to the mixing between degrees of freedom in $\mathbb{K}$.  However, the modes still tend to form three distinct branches.  We will refer to the branches in 3D crystals using the same names as before, but it is useful to keep in mind that the axial modes now have a planar component and the $\boldsymbol{E}\times\boldsymbol{B}$ and cyclotron modes have an axial component. As we will see in the next section, oblate 3D crystals generally have mode branches which are well-separated in frequency space and their modes exhibit minimal planar-axial coupling.  On the other hand, prolate 3D crystals have smaller frequency gaps between branches and significant planar-axial coupling. 

We illustrate the normal modes of a small oblate 3D ion crystal in Fig.~\ref{fig2}.  In Fig.~\ref{fig2a} we show two crystal eigenvectors in which sections of the crystal move in phase.  For instance, the axial COM mode is characterized by the in-phase motion of all ions. Notably, this mode is still purely axial, even in 3D crystals.    In Fig.~\ref{fig2b}, we plot the mode frequencies of the same crystal. For this set of trap parameters, the three mode branches are well-separated. Also in Fig.~\ref{fig2b}, we plot the axial fraction of each mode, computed according to  

\begin{equation}
\label{eq8}
f^z_n = \frac{\braket{u_n^z|u_n^z}}{\braket{u_n^r|u_n^r}}.
\end{equation}
The $\boldsymbol{E}$ x $\boldsymbol{B}$ modes have a significant axial component, typical of many 3D crystals.  On the other hand, in this regime, the axial modes describe nearly pure axial motion and the cyclotron modes remain almost entirely planar.

\section{\label{sec:pe_cooling}Cooling across the 2D-3D transition}

We begin by investigating laser cooling across the transition from a 2D crystal to a 3D crystal.  This transition occurs when the ratio of the confining potential in the radial and axial directions reaches a critical value.  This ratio, denoted by $\beta$, is explicitly given by

\begin{equation}
\label{eq9}
\beta =\frac{\omega_r\omega_c-\omega_r^2-\frac{1}{2}\omega_z^2}{\omega_z^2} = \frac{\omega_r(\omega_c-\omega_r)}{\omega_z^2}-\frac{1}{2}.
\end{equation}
The value of $\beta$ corresponding to the 2D-3D transition depends on the number of ions, $N$, in the crystal \cite{dubin1993}. In the NIST Penning trap experiment, $\beta$ is typically modulated by varying $\omega_r$. In recent work with 2D trapped ion crystals, it has proven difficult to cool the low-frequency $\boldsymbol{E}\times\boldsymbol{B}$ modes to less than a few milliKelvin. This is because Doppler laser cooling directly removes kinetic energy, but these modes are dominated by potential energy.  In this section, we investigate how the mode spectrum of 3D crystals affects the potential energy temperatures achievable using Doppler laser cooling.  We also characterize the kinetic energy cooling of these crystals and study the effects of several trap parameters on cooling results.

\subsection{Simulation setup}

Our laser cooling simulations proceed in three main steps.  First, the equilibrium configuration of the ion crystal is found.  Then, the crystal positions and velocities are perturbed such that the potential and kinetic energies are initialized at temperature $T_i$.  Finally, the crystal is evolved in time via our MD code, which includes laser cooling. 

\begin{figure*}
    \centering
    \begin{subfigure}{1.0\textwidth}
        \subcaption{}
        \centering
        \includegraphics[scale=0.5]{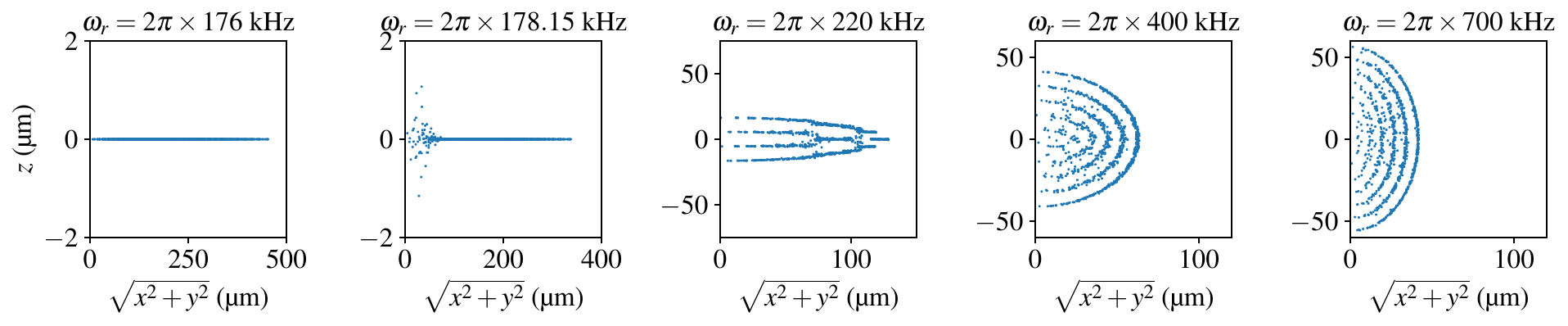}
        \label{fig3a}%
    \end{subfigure}
    ~ 
    \bigskip
    \centering
    \begin{subfigure}{0.35\textwidth}
        \subcaption{}
        \centering
        \includegraphics[scale=0.3]{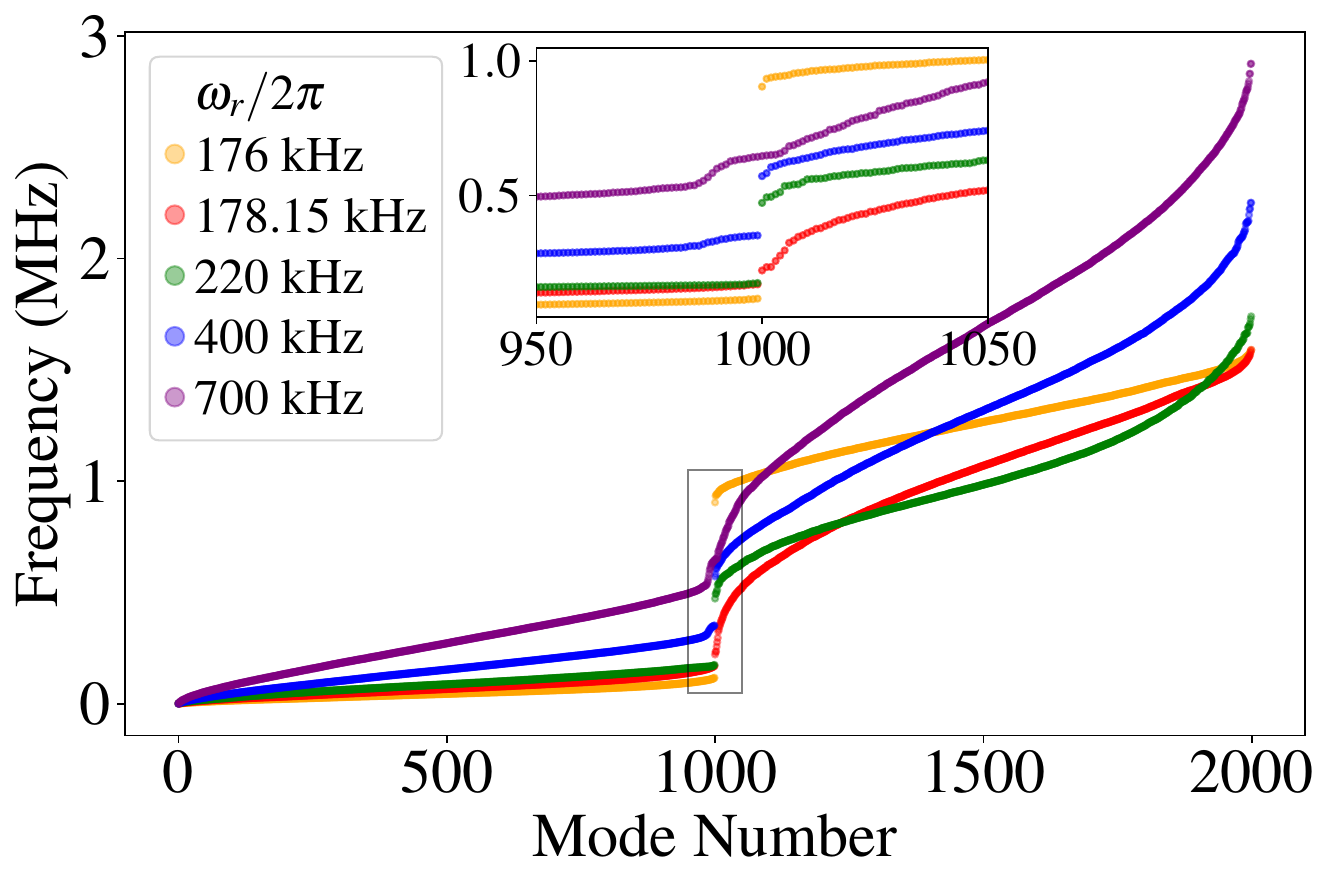}
        \label{fig3b}%
    \end{subfigure}
    ~ 
    \hspace{0cm} 
    \begin{subfigure}{0.25\textwidth}
        \subcaption{}
        \centering
        \includegraphics[scale=0.3]{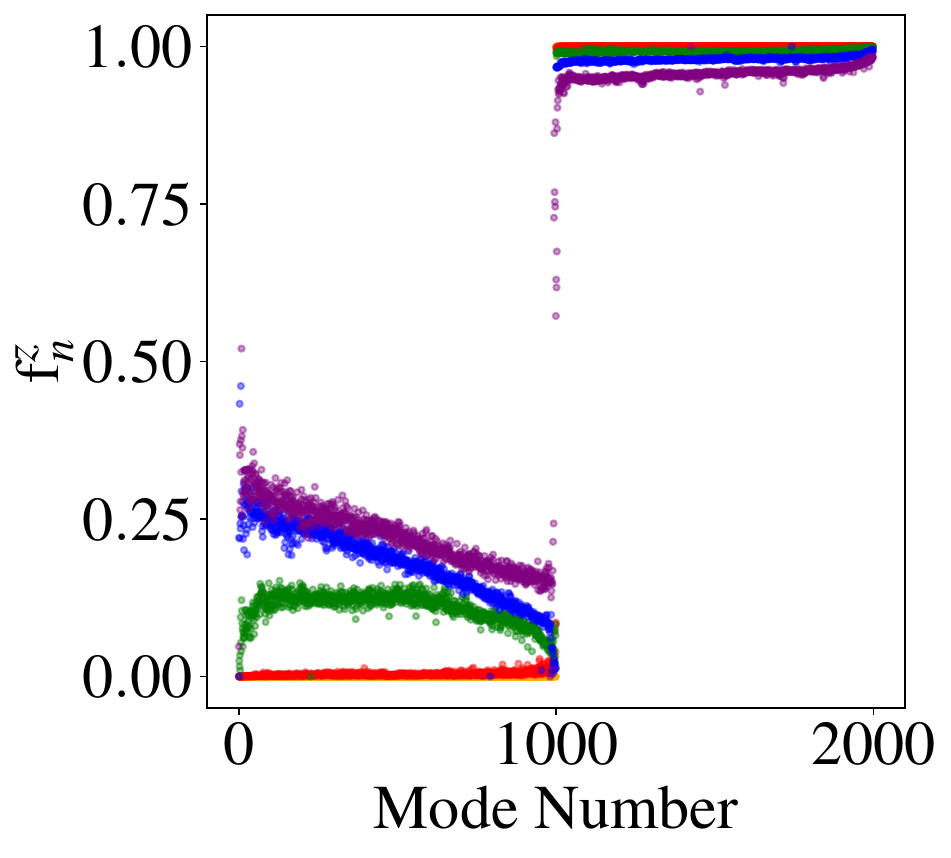}
        \label{fig3c}%
    \end{subfigure}
    ~
    \hspace{0cm} 
    \begin{subfigure}{0.25\textwidth}
        \subcaption{}
        \centering
        \includegraphics[scale=0.3]{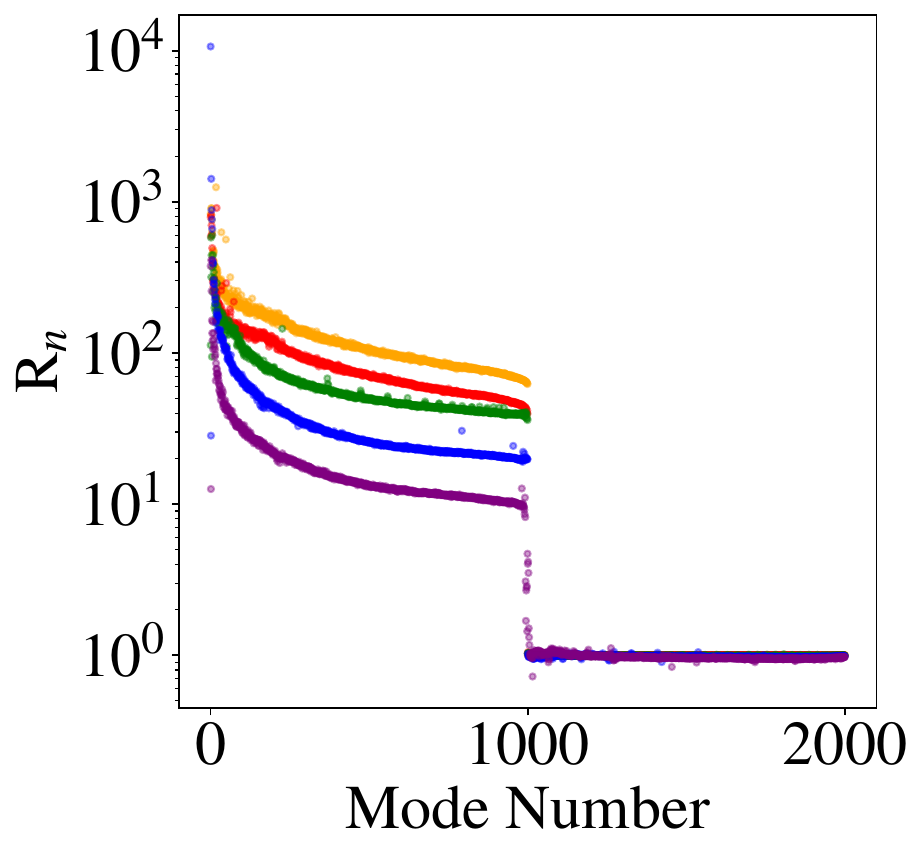}
        \label{fig3d}%
    \end{subfigure}
    \caption{(a) The shape of $N=1000$ ion crystals are plotted for different values of $\omega_r$.  The equilibrium configurations plotted correspond to the following trap parameters: $B_z =4.4588$ T, $\omega_z = 2\pi\times1.59$ MHz, $\delta=0.0104$.  The transition from a 2D crystal to a 3D crystal occurs near $\omega_r = 2\pi\times 178.15$ kHz.  The equilibrium crystal corresponding to this frequency is nearly 2D, except the most central ions have popped out of the plane and have axial coordinates of up to $\sim 1$ $\mu$m.  As $\omega_r$ is increased further, the axial extent of the crystal increases and the shell structure becomes evident. (b) The $\boldsymbol{E}\times\boldsymbol{B}$ (mode numbers 1-1000) and axial (mode numbers 1001-2000) normal mode frequencies are plotted for the ion crystals in (a).  Most 2D crystals have $\boldsymbol{E}\times\boldsymbol{B}$, axial, and cyclotron modes which occupy distinct frequency bands (orange curve).  However, near the 2D-3D transition, the highest frequency $\boldsymbol{E}\times\boldsymbol{B}$ and lowest frequency axial modes approach one another (red curve).  The frequency gap reappears immediately after the transition, but again vanishes as $\omega_r$ continues to increase.  Therefore, resonant mode coupling in large $\beta$ 3D crystals may result in enhanced Doppler laser cooling.  (c)  The axial component of the $\boldsymbol{E}\times\boldsymbol{B}$ and axial modes is shown for each of the crystals.  As $\omega_r$ increases, $\boldsymbol{E}\times\boldsymbol{B}$ modes gain larger axial components and axial modes gain a larger planar component. The $\boldsymbol{E}\times\boldsymbol{B}$ modes of 3D crystals may, therefore, be cooled by the axial beams in the NIST setup. (d) The ratio of potential energy to kinetic energy for each of the $\boldsymbol{E}\times\boldsymbol{B}$ and axial modes is plotted.  For crystals with larger $\omega_r$, the $\boldsymbol{E}\times\boldsymbol{B}$ modes have a larger kinetic energy component.  Since Doppler laser cooling directly reduces the kinetic energy, this may result in enhanced cooling of large $\beta$ crystals.}
    \label{fig3}
\end{figure*}

In order to study laser cooling of 3D crystals, we simulate traps with various values of $\omega_r$.  Our first set of simulations investigates five $N=1000$ ion crystals with $\omega_r/2\pi = 176,\;178.15,\;220,\;400,$ and $700$ kHz ($\beta = 0.017,\;0.023,\;0.14,\;0.64,$ and $1.4$), respectively.  The remaining trap parameters are as follows: $B=4.4588$ T, $\omega_z = 2\pi\times1.59$ MHz, and $\delta = 0.0104$. The equilibria of these crystals are found using the BFGS method provided by SciPy and are displayed in Fig.~\ref{fig3a}. We briefly note that slightly lower energy local minimum configurations may be found using a variety of methods, including numerical damping and the modified basin hopping algorithm in \cite{hawaldar2024}.  We will discuss energy minimization techniques in more detail in Section V. The crystal equilibrium corresponding to $\omega_r=2\pi\times176$ kHz is found to be planar. At $178.15$ kHz, the outer ions are all located at $z=0$, but the central ions have popped out of the plane and have axial coordinates on the order of $1$ micron.  Further increasing the rotation frequency to $220$ kHz, the equilibrium consists of several planes of ions.  The number of planes decreases with the distance from the trap center since the axial extent of the crystal is largest at $r=0$.   As $\omega_r$ continues to increase, the axial extent of the crystal also grows.  At $\omega_r=2\pi\times400$ and $2\pi\times700$ kHz, the equilibria consist of concentric ellipsoidal shells of ions, which are less defined near the trap center. 

The frequency spectra of these crystals are displayed in Fig.~\ref{fig3b}.  In the majority of trap parameter space, 2D crystals exhibit large frequency gaps, typically on the order of $1$ MHz, between the highest frequency $\boldsymbol{E}\times\boldsymbol{B}$ mode and the lowest frequency axial mode. The gap disappears near $\omega_r = 2\pi\times178.15 \text{ kHz}$ because the 2D to 3D transition that occurs at this rotation frequency is a buckling transition of second order: the planar lattice  becomes unstable due to an axial mode that approaches zero frequency as $\omega_r$ increases toward the transition value \cite{dubin1993}. As the frequency gap vanishes, resonant mode coupling between branches is expected to produce enhanced cooling of the $\boldsymbol{E}\times\boldsymbol{B}$ modes \cite{johnson2024}. For $\omega_r$ above the 2D to 3D transition value, the gap opens up again because the stable 3D equilibrium is characterized by axial modes which have a finite (that is, minimum) frequency. However, in these crystals, individual modes experience mixing between the axial and planar degrees of freedom, as shown in Fig.~\ref{fig3c}.  Note that the $\boldsymbol{E}\times\boldsymbol{B}$ modes in the 3D crystals have large values of $f^z$, or significant axial components. 

When $\omega_r$ is well above the transition value, the simulations observe that the frequency gap closes again. This can be understood by comparing the maximum frequency of $\boldsymbol{E}\times\boldsymbol{B}$ modes with the minimum frequency of axial modes in a plasma spheroid. When the plasma is a spheroid,  the maximum frequency $\omega_{E max} $ of  $\boldsymbol{E}$ x $\boldsymbol{B}$ modes is roughly
\begin{equation}
\label{eq10}
\omega_{E max} \approx \frac{\omega_p^2}{2 \Omega_v},
\end{equation}
 where $\Omega_v \equiv \omega_c - 2 \omega_r$ is the vortex frequency, the effective cyclotron frequency in the rotating frame of the plasma, and $\omega_p$ is the ion plasma frequency, given in terms of the plasma rotation frequency by
\begin{equation}
\label{eq11}
\omega_p^2 = 2 \omega_r (\omega_c - \omega_r). 
\end{equation}
Eq.~(\ref{eq10})  follows by considering the large wavenumber and large vortex frequency limit of the dispersion relation \cite{dubin1991} for plasma waves in a magnetized spheroid. For an ion crystal it is only a rough estimate because correlation effects affect the restoring forces.

\begin{figure}
    \centering
    \includegraphics[scale=0.5]{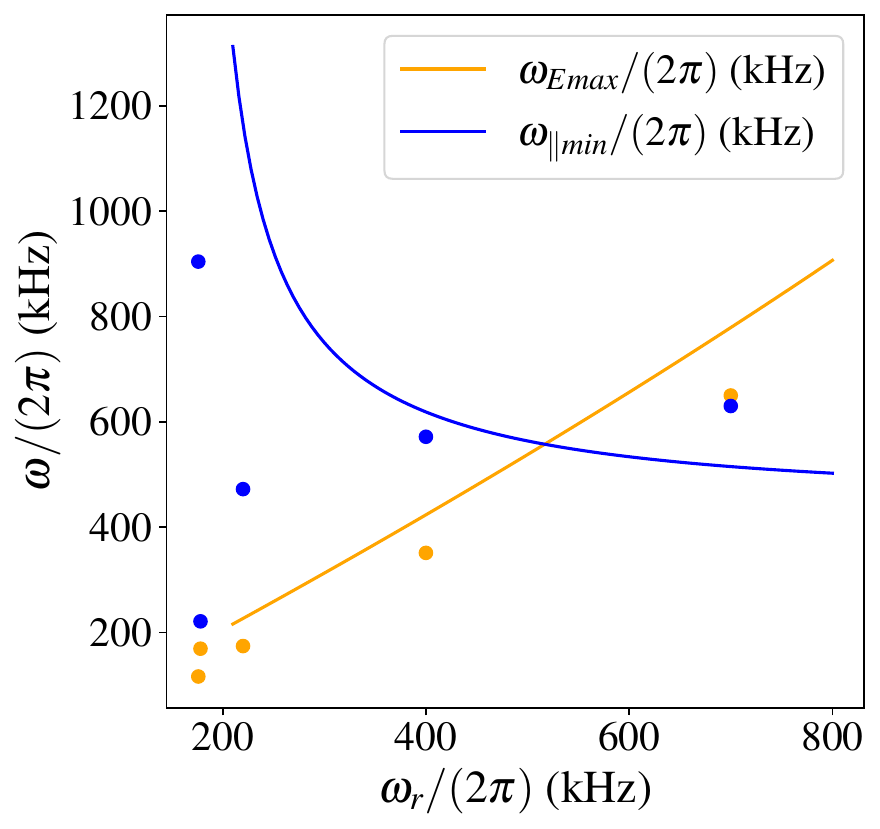}
    ~ 
    \caption{Theory estimates for the maximum $\boldsymbol{E}\times\boldsymbol{B}$ mode frequency and the minimum axial frequency versus plasma rotation frequency are shown in orange and blue curves, respectively. These plots are obtained using Eq.~(\ref{eq10}) and Eq.~(\ref{eq13}). The theoretical treatment predicts that $\omega_{\parallel min}>\omega_{E max}$ for $\omega_r \lesssim$ 500 kHz.  For larger values of $\omega_r$, $\omega_{\parallel min}$ becomes smaller than $\omega_{E max}$.  This behavior is observed in the mode frequencies of the crystals from Fig.~\ref{fig3}, which we plot here using orange and blue dots. Note that the difference $\omega_{\parallel min}-\omega_{E max}$ nearly vanishes in the simulation data from $\omega_r=178.15$ kHz.  This occurs near the 2D to 3D transition.  The theory plots do not account for this behavior, but this effect is well-understood (see text).}
    \label{fig4}
\end{figure}

The  minimum frequency $\omega_{|| min}$ of the axial modes in a plasma spheroid can also be roughly estimated, from the dispersion relation for magnetized plasma waves in a uniform plasma \cite{trivelpiece1959},
\begin{equation}
\label{eq12}
\omega^2 = \omega_p^2 \frac{\tilde{k}_z^2}{\tilde{k}_\bot^2 + \tilde{k}_z^2},
\end{equation}
where $\tilde{k}_z$ and $\tilde{k}_\bot$ are the axial and perpendicular wave numbers respectively (with respect to the magnetic field direction). The wave numbers are quantized by the finite size of the plasma, with values that may be estimated as $\tilde{k}_z \approx m_z \pi/(2Z)$ and $\tilde{k}_\bot \approx m_\bot \pi/R$ where $R$ is the radius of the plasma, $Z$ is its axial half-length, and $m_z$ and $m_{\perp}$ are positive integers.  This approximate theory breaks down in the 2D regime where $Z$ is zero, but provides order of magnitude estimates for normal mode frequencies when $Z>a$, where $a$ is the inter-ion spacing (Wigner-Seitz radius). Also,  the integers $m_z$ and $m_\bot$  cannot be arbitrarily large, since the theory is only sensible for inverse wave numbers less than of order  $a$. The lowest frequency axial modes then have the smallest possible $\tilde{k}_z$, of order $1/Z$, and the largest possible $\tilde{k}_\bot$, of order $1/a$, yielding a minimum axial frequency of roughly
\begin{equation}
\label{eq13}
\omega_{|| min} \approx C \omega_p \frac{a}{Z}
\end{equation}
where $C$ is a constant of order unity. The value of $C$ can be estimated by comparing Eq.~(\ref{eq13}) to simulation results, which gives $C\approx 2$. Eqs.~(\ref{eq10}) and (\ref{eq13}) are plotted in Fig.~\ref{fig4} versus the plasma rotation frequency and compared to the simulation results for the crystals shown in Fig.~\ref{fig3}.  For the simulated crystals with $176 \text{ kHz}\leq \omega_r/2\pi \leq 400$ kHz, the $\boldsymbol{E}\times\boldsymbol{B}$ modes are the $N$ lowest frequency modes while the axial modes are the $N$ middle frequency modes.  Therefore, when the modes are listed in order of increasing frequency $\omega_{E max}$ and $\omega_{\parallel min}$ are the frequencies of modes $N$ and $N+1$, respectively.  However, for the $\omega_r=700$ kHz crystal, the lowest frequency $N$ modes no longer correspond exactly to the $\boldsymbol{E}\times\boldsymbol{B}$ modes, which we define as the $N$ modes with the largest values of $R_n$.  In this case, we find that $\omega_{E max}$ and $\omega_{\parallel min}$ are the frequencies of mode numbers $1003$ and $995$, respectively.  For a plasma of $N=1000$ ions, Fig.~\ref{fig4} shows how the frequency gap disappears as the maximum $\boldsymbol{E}\times\boldsymbol{B}$ frequency increases and the minimum axial frequency decreases.

Having determined the crystal equilibria, we next initialize the crystals at $T_i=10$ mK.  To initialize the kinetic energy, the ions are randomly assigned initial velocities, $v$, according to the Maxwell-Boltzmann distribution,

\begin{equation}
\label{eq14}
    f(v) = \bigg(\frac{m}{2\pi k_BT_i}\bigg)^{3/2}\exp\bigg(-\frac{mv^2}{2k_BT_i}\bigg),
\end{equation}
where $k_B$ is the Boltzmann constant. Since the $\boldsymbol{E}\times\boldsymbol{B}$ modes are dominated by potential energy, they are not excited by initializing the ion velocities. Instead, the potential energy is independently initialized using the Metropolis-Hastings algorithm to perturb the ion positions \cite{shankar2020}. Starting from equilibrium, each ion is perturbed by independently displacing its position along each Cartesian axis by a random value uniformly distributed in $[-\delta x/2,\delta x/2]$, with $\delta x=0.5\;\mu$m. The energy before and after the change in the ion's position are denoted $E_{0}$ and $E_{new}$, respectively, and the list of ion coordinates are $\{\boldsymbol{x}^n_0\}$ and $\{\boldsymbol{x}^n_{new}\}$. The MH algorithm accepts this change in the ion's position if $E_{new} < E_0$, or if rand$(0,1) < \exp[-(E_{new}-E_{0})/k_BT_i]$.  In this case, we set $\{\boldsymbol{x}^n_0\} = \{\boldsymbol{x}^n_{new}\}$ and $E_{0}=E_{new}$. Otherwise, the displacement is rejected. Then, the next ion in the crystal is randomly displaced and this process is repeated for each ion, constituting one full scan of the MH algorithm.  We complete on the order of 1000 scans in order to initialize the potential energy of the crystal at $T_i$.

After the crystal is initialized at $T_i$, it is evolved in time using our MD code, allowing us to study the laser cooling of an ion crystal for a given set of laser and trap parameters.

\subsection{Potential energy cooling}

Since $\boldsymbol{E}\times\boldsymbol{B}$ modes contain mostly potential energy, they have proven difficult to cool in the case of 2D crystals \cite{johnson2024}.  We now study the reduction in potential energy, or potential energy cooling, experienced by the 3D crystals in Fig.~\ref{fig3} during Doppler laser cooling.  Prior to discussing laser cooling simulations, we identify several factors which may lead to efficient potential energy cooling in 3D crystals, thereby motivating this investigation. 

First, as seen in Fig.~\ref{fig3c}, the $\boldsymbol{E}\times\boldsymbol{B}$ modes of 3D crystals have large axial components, $f^z$. This coupling between perpendicular and axial degrees of freedom in 3D crystal modes, which we will refer to as `$f^z$-coupling', may result in efficient $\boldsymbol{E}\times\boldsymbol{B}$ mode cooling.  Because axial motion is efficiently cooled by the axial laser beams, the planar component of the $\boldsymbol{E}\times\boldsymbol{B}$ modes in 3D crystals may experience significant direct cooling.  However, this reasoning is further complicated by the consideration that the $\boldsymbol{E}\times\boldsymbol{B}$ modes are dominated by potential energy.  This fact is illustrated in Fig.~\ref{fig3d}, where we plot the ratio, $R_n$, of potential energy to kinetic energy in each normal mode, given by

\begin{equation}
\label{eq15}
R_n = \frac{\braket{u_n^{r}|\mathbb{K}|u_n^r}}{m\braket{u_n^{v}|u_n^v}}.
\end{equation}
Since Doppler laser cooling primarily cools kinetic energy, the efficiency with which the axial beams can cool the axial component of the $\boldsymbol{E}\times\boldsymbol{B}$ modes may be limited. 

Crystals with larger values of $\beta$ may also see enhanced potential energy cooling due to `resonant mode coupling' between the $\boldsymbol{E}\times\boldsymbol{B}$ and axial modes. In general, when two modes have nearly the same frequency, their energies tend to equilibrate. In the case of 2D crystals, it has been shown that when the frequency gap between the $\boldsymbol{E}\times\boldsymbol{B}$ and axial branches closes, the $\boldsymbol{E}\times\boldsymbol{B}$ modes are sympathetically cooled as energy is removed from the axial modes \cite{johnson2024}. Additionally, $\boldsymbol{E}\times\boldsymbol{B}$ modes rapidly equilibrate with one another \cite{tang2019}.  Therefore, only a small number of $\boldsymbol{E}\times\boldsymbol{B}$ modes must be resonantly coupled to the axial branch to produce cooling across the entire $\boldsymbol{E}\times\boldsymbol{B}$ bandwidth.  As illustrated in Fig.~\ref{fig3b}, the frequency gap between mode branches vanishes in 3D crystals as $\omega_r$ is increased.  At $\omega_r=2\pi\times700$ kHz, the gap is no longer visible. We briefly mention that resonant mode coupling may also occur via three-mode interactions \cite{marquet2003,johnson2025_2}. In this case, three modes are coupled when the sum of two mode frequencies is approximately equal to a third mode frequency.  If two $\boldsymbol{E}\times\boldsymbol{B}$ mode frequencies sum to an axial mode frequency, then energy can be transferred from the $\boldsymbol{E}\times\boldsymbol{B}$ branch to the axial branch.

MD laser cooling simulations inherently include $f^z$-coupling and resonant mode coupling and, therefore, can elucidate the effects of trap and laser beam parameters on the attainable ion crystal temperatures. Our first set of cooling simulations compares the potential energy cooling of the crystals shown in Fig.~\ref{fig3}.  The values of the trap parameters used in these simulations are listed in the previous section.  The laser parameters are as follows: $w_y$ = 20$\sqrt{2}$ $\mu$m, $w_z$ = 100$\sqrt{2}$ $\mu$m, $\Delta_{\parallel} = -\gamma_0/2$, $S_{\parallel} = 0.005$, $S_{\perp} = 0.5$.  For each crystal, we have run simulations with a variety of values for $\Delta_{\perp}$ and $d$, which are relatively easy to tune experimentally.  The exact values used in the parameter scans are different for each crystal, as we use the theory developed in \cite{torrisi2016} to choose ranges which result in optimal kinetic energy cooling. We expect to achieve the best potential energy cooling in a similar parameter space \cite{zaris2024}. The final temperatures for the full parameter scans (after 16 ms of cooling) are provided in Appendix~\ref{app:a}.

Fig.~\ref{fig5} illustrates the decrease in potential energy over time during laser cooling. For clarity, we only plot the cooling results for the $(\Delta_{\perp},d)$ pair which results in the best potential energy cooling for each crystal.  The exact values are listed in Table \ref{table1}. However, we note that the potential energy is quite similar for a variety of $(\Delta_{\perp},d)$ pairs near the optimal values, as seen in Appendix~\ref{app:a}. The temperature corresponding to the potential energy of an ion crystal $\{\boldsymbol{x}_i(t)\}_{i=1}^N$, at time $t$, is calculated using

\begin{equation}
\label{eq16}
    T^{PE}(t)= \frac{2}{3}\frac{\sum_{i=1}^N(E_{pot}(\boldsymbol{x}_i(t))-E_{pot}(\boldsymbol{x}_i^0))}{Nk_b},
\end{equation}
where $\boldsymbol{x}_i^0$ is the equilibrium ion configuration. Here, we subtract the equilibrium potential energy, $E_{pot}(\boldsymbol{x}_i^0)$, from the total potential energy before dividing by the factor $Nk_b$ to convert to units of temperature.  The factor of $2/3$ accounts for the fact that the potential energy is shared (not necessarily equally) between three spatial degrees of freedom. We note that, in some cases with highly effective laser cooling, the potential energy of the ion crystal after simulated cooling is lower than the equilibrium potential energy found in our initialization process using the SciPy library.  This behavior is a result of the complicated energy landscape of 3D crystals, which includes many local energy minimum states.  In order to ensure that our potential energy values are positive, we perform another SciPy minimization following the laser cooling simulations.  The configuration generated by this final energy minimization is then what we define as $\{\boldsymbol{x}_i^0\}$.

\begin{table}[b]
\caption{\label{table1}%
The values of perpendicular laser beam parameters $\Delta_{\perp}$ and $d$ used in the simulations shown in Fig.~\ref{fig5} are listed here.  As explained in the text, these values result in optimal potential energy cooling for the respective crystal.
}
\begin{ruledtabular}
\begin{tabular}{llll}
\textrm{$\omega_r/2\pi$ (kHz)}&
\textrm{$\delta$}&
\textrm{$\Delta_{\perp}/2\pi$ (MHz)}&
\textrm{$d$ ($\mu$m)}\\
\colrule
176 & 0.0104 & -50 & 8.2\\
178.15 & 0.0104 & -150 & 5.8\\
220 & 0.0104 & -75 & 1.0\\
220 & 0.104 & -75 & 13.0\\
400 & 0.0104 & -20 & 20.2\\
400 & 0.104 & -55 & 20.2\\
700 & 0.0104 & -350 & 10.6\\
700 & 0.104 & -150 & 25.0\\
\end{tabular}
\end{ruledtabular}
\end{table}

\begin{figure}
    \centering
    \includegraphics[scale=0.5]{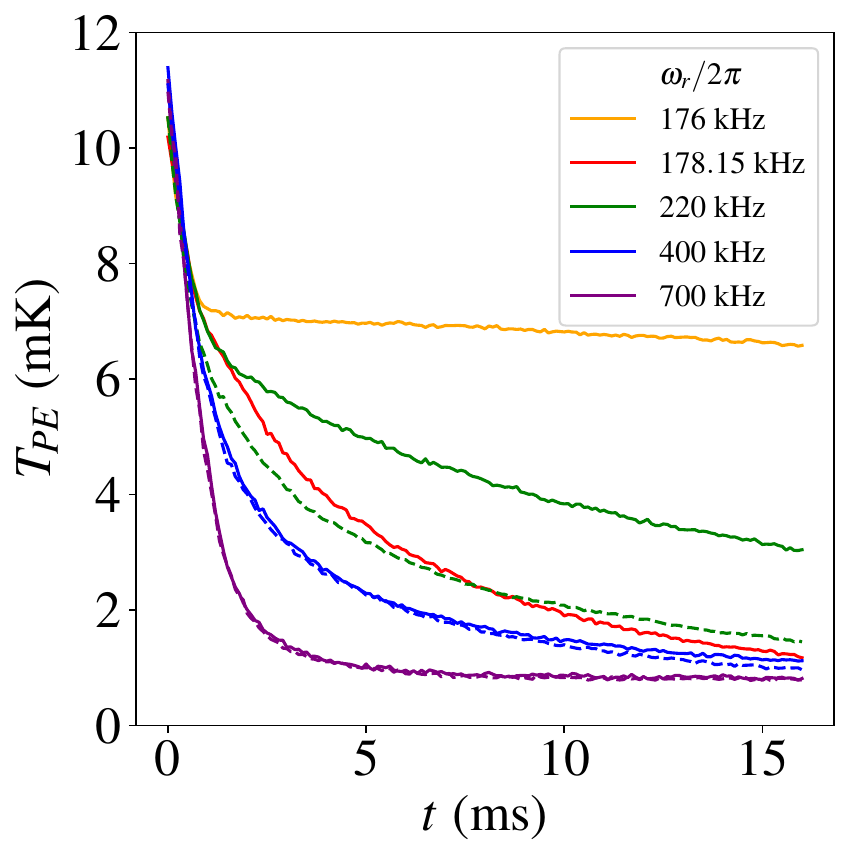}
    ~ 
    \caption{The potential energy cooling is illustrated for crystals with different values of $\omega_r$ and $\delta$. The kinetic and potential energies are initialized at 10 mK and then laser cooling is simulated for 16 ms. \textbf{Solid lines ($\delta=0.0104$):} The potential energy of 2D crystals is generally difficult to cool, as shown by the yellow curve.  By tuning the trap parameters such that the crystal is near the 2D-3D transition, the potential energy can be cooled efficiently (red curve).  After the transition, an oblate 3D crystal is produced and, again, the potential energy cools relatively slowly (green curve).  By further increasing $\beta$, a combination of the increased axial component of the $\boldsymbol{E}\times\boldsymbol{B}$ modes and a reduced gap between mode branches leads to enhanced cooling (blue and purple curves).  \textbf{Dashed lines ($\delta=0.104$):} Increasing the strength of the rotating wall potential, $\delta$, can reduce the potential energy of 3D crystals. This is seen most dramatically in the case of the $\omega_r=2\pi\times220$ kHz crystal.}
    \label{fig5}
\end{figure}

We run simulations using two different values of $\delta$ ($\delta=0.0104$ and $\delta = 0.104$), and begin by discussing the results of the small $\delta$ case. For the crystal with the smallest $\beta$ (or $\omega_r$), the potential energy quickly cools to $\sim 7$ mK, a result which we attribute to the rapid cooling of the axial modes. After this initial period, the potential energy decreases at a prohibitively slow rate, illustrating the slow cooling of the $\boldsymbol{E}\times\boldsymbol{B}$ modes.  Increasing $\beta$ to near the 2D-3D transition, significantly enhanced potential energy cooling is observed. This regime has been shown to be uniquely well-suited to laser cooling of 2D crystals \cite{johnson2024}. Even though there is a frequency difference of approximately $50$ kHz between the $\boldsymbol{E}\times\boldsymbol{B}$ and axial branches in this crystal, the enhanced cooling is likely due to resonant mode coupling, as there is minimal $f^z$-coupling in the $\boldsymbol{E}\times\boldsymbol{B}$ branch.

Further increasing the $\omega_r$ to $2\pi\times220$ kHz, the crystal becomes ellipsoidal and gains an appreciable axial extent. The associated cooling curve suggests that, when $\delta$ is small, low $\beta$ 3D crystals suffer similar cooling obstacles as seen in 2D crystals.  In particular, the potential energy quickly cools to $\sim 6$ mK within a few milliseconds, but then cools at a significantly slower rate.  Still, the cooling is notably more effective than for the $\omega_r=2\pi\times176$ kHz 2D crystal, likely due to $f^z$-coupling, as the $\boldsymbol{E}\times\boldsymbol{B}$ modes in this crystal have substantial $f^z$ values.  At $\omega_r =2\pi\times400$ kHz, the crystal's potential energy can be quickly cooled to $\sim$ 1 mK, due to stronger $f^z$-coupling.  Resonant mode coupling between the $\boldsymbol{E}\times\boldsymbol{B}$ and axial mode branches seems less likely to cause this rapid cooling, as there still exists a significant gap between branches at this value of $\omega_r$.  However, three-mode coupling may play a role, and additional work is needed to definitively evaluate the relative contribution of these effects.  For instance, cooling simulations using linearized versions of the forces would eliminate mode coupling and could, therefore, be compared to our full dynamics results to help to quantify the effects of resonant coupling.  Finally, at $\omega_r=2\pi\times700$ kHz, the potential energy is cooled even faster.  In this case, resonant mode coupling may play a significant role since there is no longer a frequency gap between branches.

\subsection{Laser torque considerations}
\begin{figure}
    \centering
    \begin{subfigure}{0.4\textwidth}
        \subcaption{}
        \centering
        \includegraphics[scale=0.26]{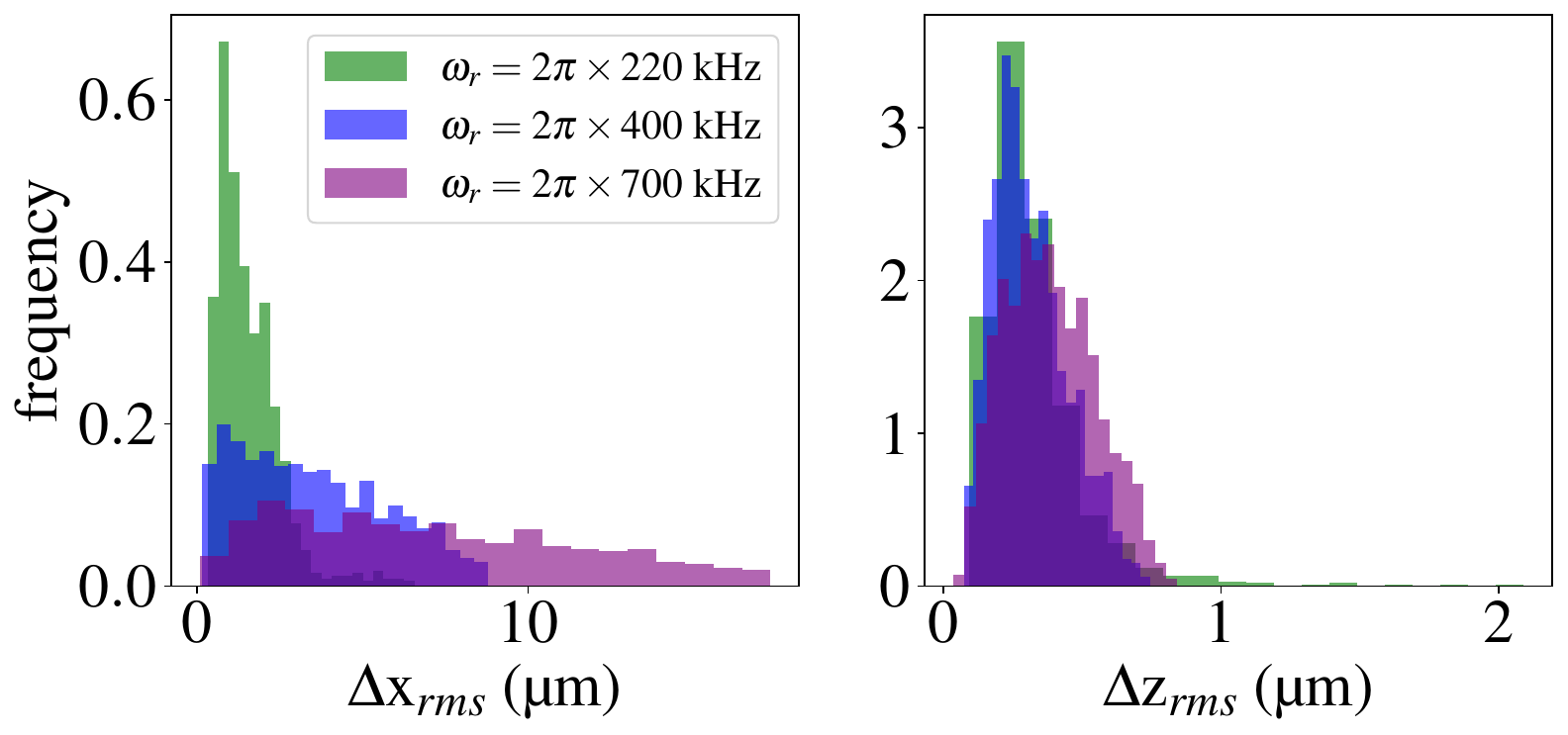}
        \label{fig6a}
    \end{subfigure}
    ~ 
    \bigskip
    \centering
    \hspace{0cm} 
    \begin{subfigure}{0.4\textwidth}
        \subcaption{}
        \centering
        \includegraphics[scale=0.26]{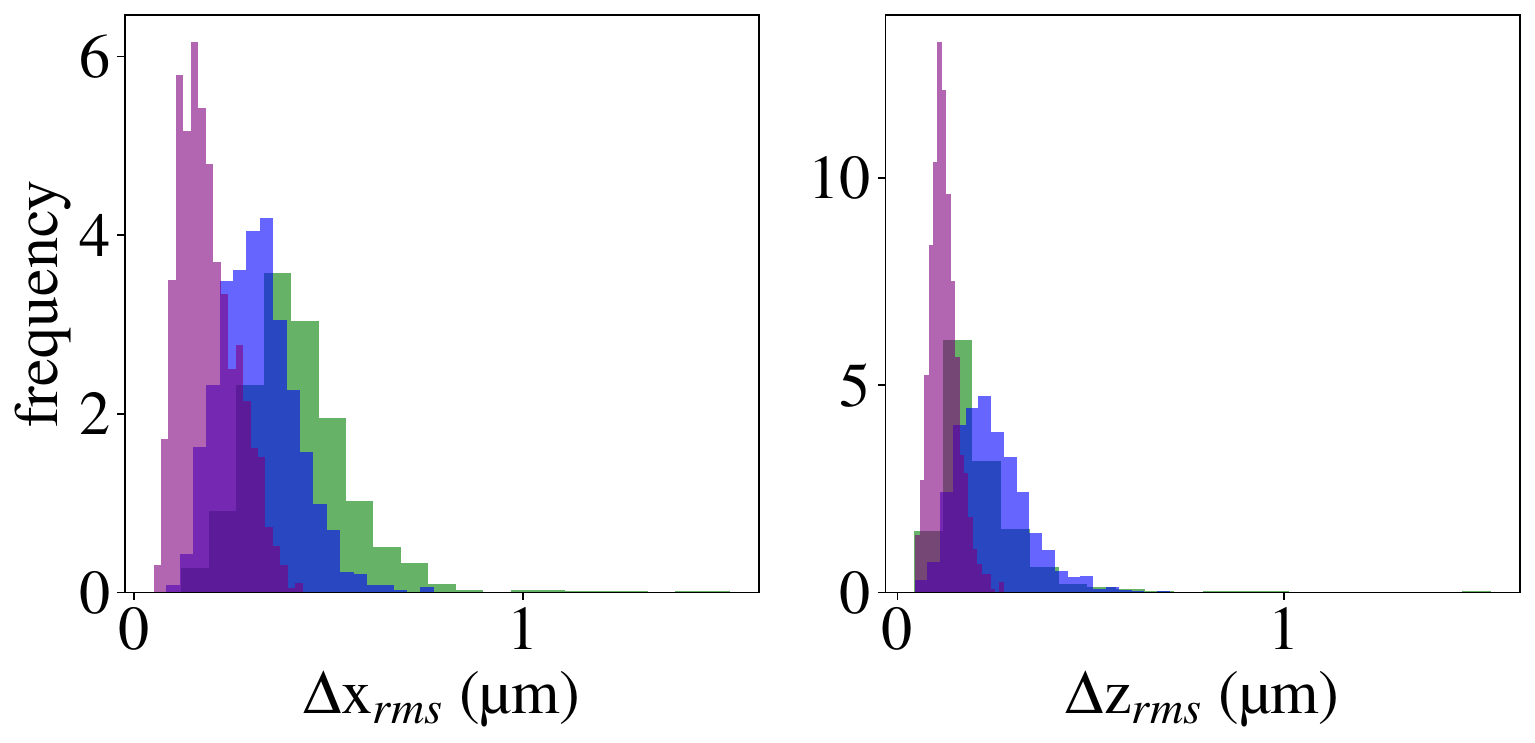}
        \label{fig6b}%
    \end{subfigure}
    \caption{The rms ion displacements during the last $1$ ms of laser cooling for the three 3D crystals from Fig.~\ref{fig5}.  The upper plots show the displacements in the small $\delta$ ($\delta=0.014$) case, while the lower plots illustrate the large $\delta$ ($\delta=0.104$) case.  Increasing the rotating wall strength significantly improves ion confinement in the perpendicular direction (here, we show the rms displacement in the $\hat{\boldsymbol{x}}$ direction).  It also improves axial confinement, to a lesser extent. The $\omega_r=2\pi\times700$ kHz crystal exhibits the worst confinement in the small $\delta$ case, but the best confinement in the large $\delta$ case.  This demonstrates that a strong rotating wall is essential when cooling large $\beta$ crystals, which have smaller extent in the perpendicular directions.}
    \label{fig6}
\end{figure}

Ideally, trapped ion crystals rotate in phase with the time-varying rotating wall potential.  When this occurs, the ion motion in the rotating frame is minimized, which is key to the implementation of quantum sensing protocols.  However, there exist physical effects which can cause the crystal to rotate relative to rotating wall. First, since ions absorb perpendicular beam photons from a single direction, but emit them in a random direction, on average the beam exerts a torque on the crystal. A sufficiently large value of $\delta$ results in a rotating wall which is strong enough to provide a restoring torque, such that the global rotation frequency of the crystal remains very close to $\omega_r$ during laser cooling.  If $\delta$ is not sufficiently large, then the laser torque can cause the ion crystal to slip in the azimuthal direction. This phenomenon, known as slip-stick dynamics, has been observed experimentally in Penning trap ion crystals \cite{mitchell2001}.  Over long time scales, discrete slips of the crystal lead to the global rotation frequency to deviate from $\omega_r$. Secondly, a crystal's rotation frequency can vary, even in the absence of cooling laser beams, if it's rocking mode is excited.  The rocking mode consists of an azimuthal oscillation of the ion crystal's orientation in the rotating frame. In this section, we first study the effect of the rotating wall strength, $\delta$, on a crystal's potential energy.  Next, we analyze the effect of $\delta$ on the confinement of ions in the rotating frame.

To evaluate the effect of $\delta$ on the potential energy, we simulate laser cooling in a trap with $\delta=0.104$ for the cases of $\omega_r/2\pi = 220,400,$ and $700$ kHz.  All other parameters are held fixed, although we again test a variety of $\Delta_{\perp}$ and $d$ values, and plot only the cooling results for the parameter pair which has the best potential energy cooling.  These cooling results are illustrated as dotted lines in Fig.~\ref{fig5} and show that increasing $\delta$ can improve potential energy cooling. The reason for this improvement is not completely understood.  However, we note that increasing $\delta$ does decrease the mode gap in the $\omega_r=2\pi\times 220$ kHz crystal from $\sim300$ kHz to $\sim200$ kHz. While this gap is still relatively large, this decrease may produce three-mode interactions.  In the large $\delta$ crystal, the frequency of the first axial mode is only $\sim 16$ kHz greater than double the frequency of the last $\boldsymbol{E}\times\boldsymbol{B}$ mode (compared to $\sim124$ kHz in the small $\delta$ crystal).

To discern how well the rotating wall is able to stabilize the crystal's rotation frequency, it is useful to calculate the root-mean-square (rms) displacement of the ions in the frame which rotates at frequency $\omega_r$.  If the rms displacement over a duration $\tau_{rms} > 1/\omega_r$ is small compared to the inter-ion spacing, then the crystal exhibits little to no slipping out of the rotating frame. The perpendicular and axial rms displacements of ion $i$ in the rotating frame are found using 

\begin{equation}
\label{eq17}
    \Delta \eta_{rms}^i = \sqrt{\langle |\eta_i-\langle \eta_i\rangle|^2\rangle},
\end{equation}
where $\eta=x$ or $z$, respectively, and $\langle\rangle$ represents the time average over the last millisecond of cooling. Note that while we compute the perpendicular confinement by looking at motion along the $\hat{\boldsymbol{x}}$ direction, one could also choose the $\hat{\boldsymbol{y}}$ direction, or even both perpendicular directions.

In Fig.~\ref{fig6},  we plot histograms of the rms displacement of the ions in the $\omega_r/2\pi \in\{220,400,700\}$ kHz crystals with $\delta\in\{0.0104,0.104\}$.  For the small $\delta$ case, the perpendicular displacements of the ions are relatively large, especially for larger values of $\omega_r$, suggesting that the rotating wall potential is not strong enough to enforce the global rotation frequency of the crystal.  The perpendicular displacements are much smaller in the large $\delta$ case, meaning that the ions remain nearly fixed in the rotating frame and the crystal rotates globally at a frequency very close to $\omega_r$. The improvement in ion confinement is most dramatic for the crystals with larger $\omega_r$. This behavior is expected; the crystals with larger $\omega_r$ have smaller extents in the perpendicular direction, so the torque exerted by the rotating wall is smaller. Therefore, a larger value of $\delta$ is required to enforce global rotation at $\omega_r$ in such crystals.  The axial displacement of ions is also smaller when using a large $\delta$, but this effect is less dramatic because the laser torque causes perpendicular motion.

\subsection{Kinetic energy cooling}
\begin{figure}
    \centering
    \includegraphics[scale=0.5]{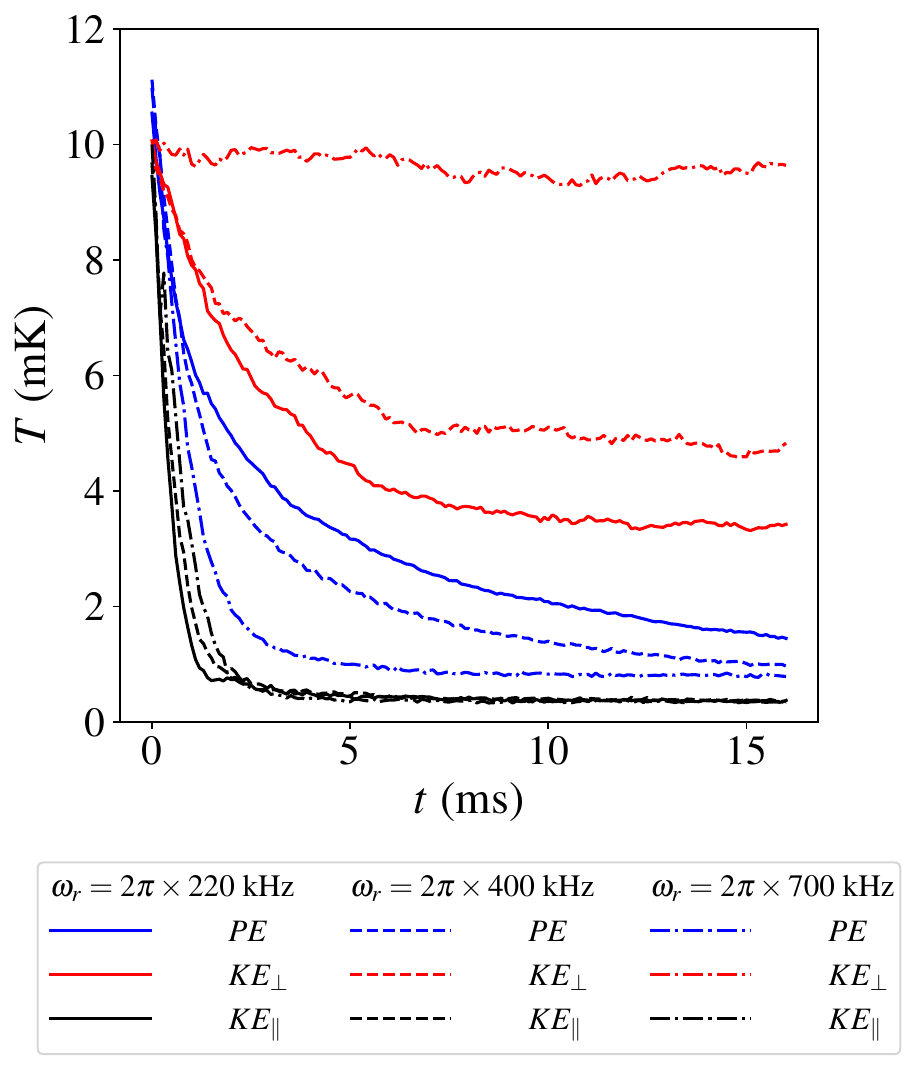}
    ~ 
    \caption{The kinetic and potential energy cooling of the large $\delta$ crystals from Fig.~\ref{fig5}. The energies are initialized at 10 mK before laser cooling is simulated. While the potential energy cooling improves as $\omega_r$ increases, the perpendicular kinetic energy cooling worsens.  This behavior agrees with the kinetic energy cooling theory developed in \cite{torrisi2016}. The axial kinetic energy is well-cooled to $<1$ mK in all cases.}
    \label{fig7}
\end{figure}

While we have focused, so far, on the cooling of potential energy in 3D crystals, it is equally important to ensure that the kinetic energy is sufficiently well-cooled. For instance, in 2D crystals, excited axial modes have been shown to add noise to quantum sensing measurements \cite{shankar2020}.

In Fig.~\ref{fig7} we plot the kinetic energy (divided into perpendicular and axial components) as the $\delta=0.104$ crystals from Fig.~\ref{fig5} are laser cooled.  The potential energies are plotted again, for completeness.  In all cases, the axial kinetic energy, $KE_{\parallel}$, is cooled to $<1$ mK within a few milliseconds.  However, the perpendicular kinetic energy, $KE_{\perp}$ is not as well-cooled.  The $\omega_r=2\pi\times220$ kHz crystal reaches the lowest $KE_{\perp}$, which is still relatively high, at $\sim 4$ mK.  While increasing $\omega_r$ produces better $PE$ cooling, it also leads to worse $KE_{\perp}$ cooling.  The higher $KE_{\perp}$ is presumably due to elevated cyclotron motion, and is consistent with theory that accounts for the dispersion of Doppler shifts across the waist of the perpendicular laser cooling beam \cite{torrisi2016}. Nevertheless, in order to prepare experimentally useful crystals, it will be important to further reduce $KE_{\perp}$ while maintaining small potential energies.

\begin{figure}
    \centering

    \begin{subfigure}{0.2\textwidth}
        \centering
        \hspace*{0cm}  
        \includegraphics[scale=0.22]{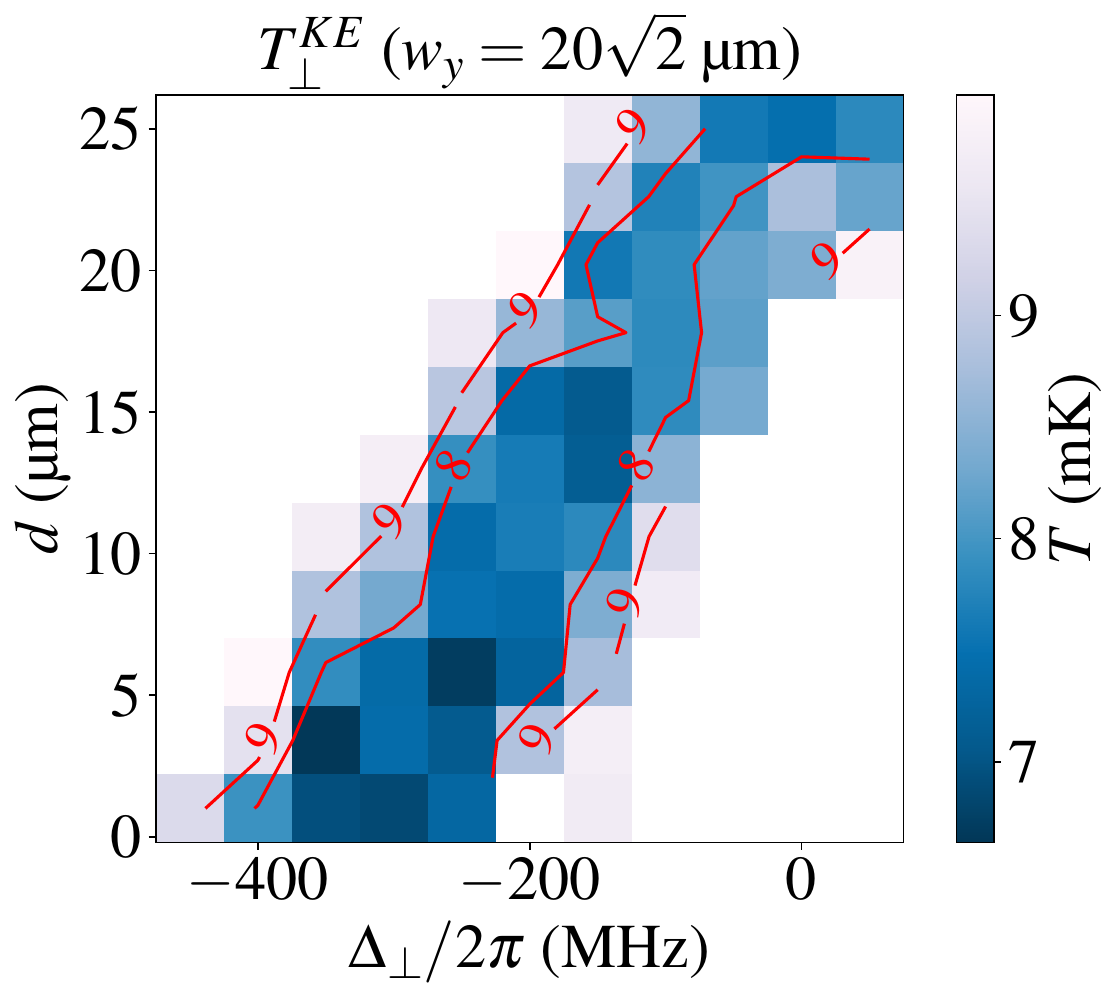}
        \label{fig8a}
    \end{subfigure}%
    \hfill
    \begin{subfigure}{0.2\textwidth}
        \centering
        \hspace*{-0.7cm}  
        \includegraphics[scale=0.22]{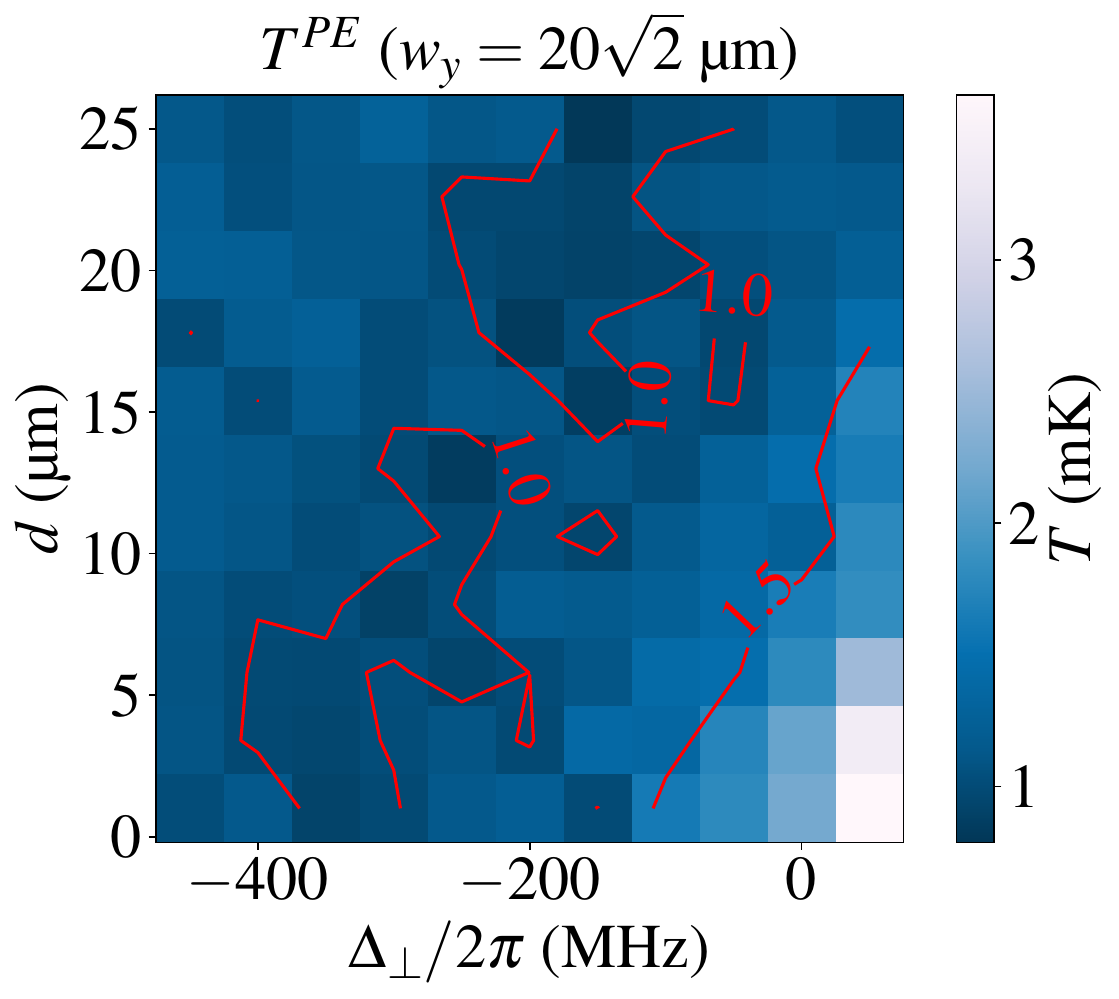}
        \label{fig8b}
    \end{subfigure}

    \vspace{0.1em}

    \begin{subfigure}{0.2\textwidth}
        \centering
        \hspace*{0.15cm}  
        \includegraphics[scale=0.22]{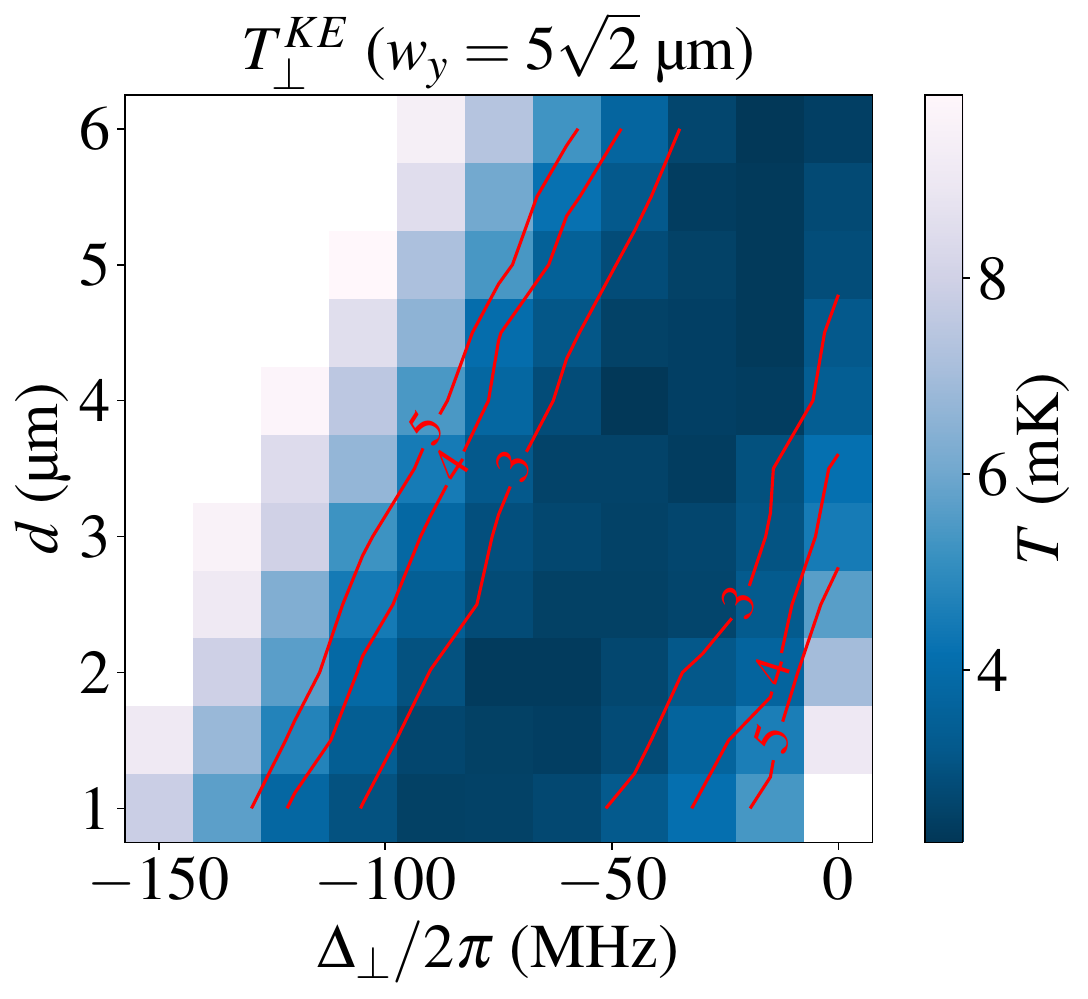}
        \label{fig8c}
    \end{subfigure}%
    \hfill
    \begin{subfigure}{0.2\textwidth}
        \centering
        \hspace*{-0.6cm}  
        \includegraphics[scale=0.22]{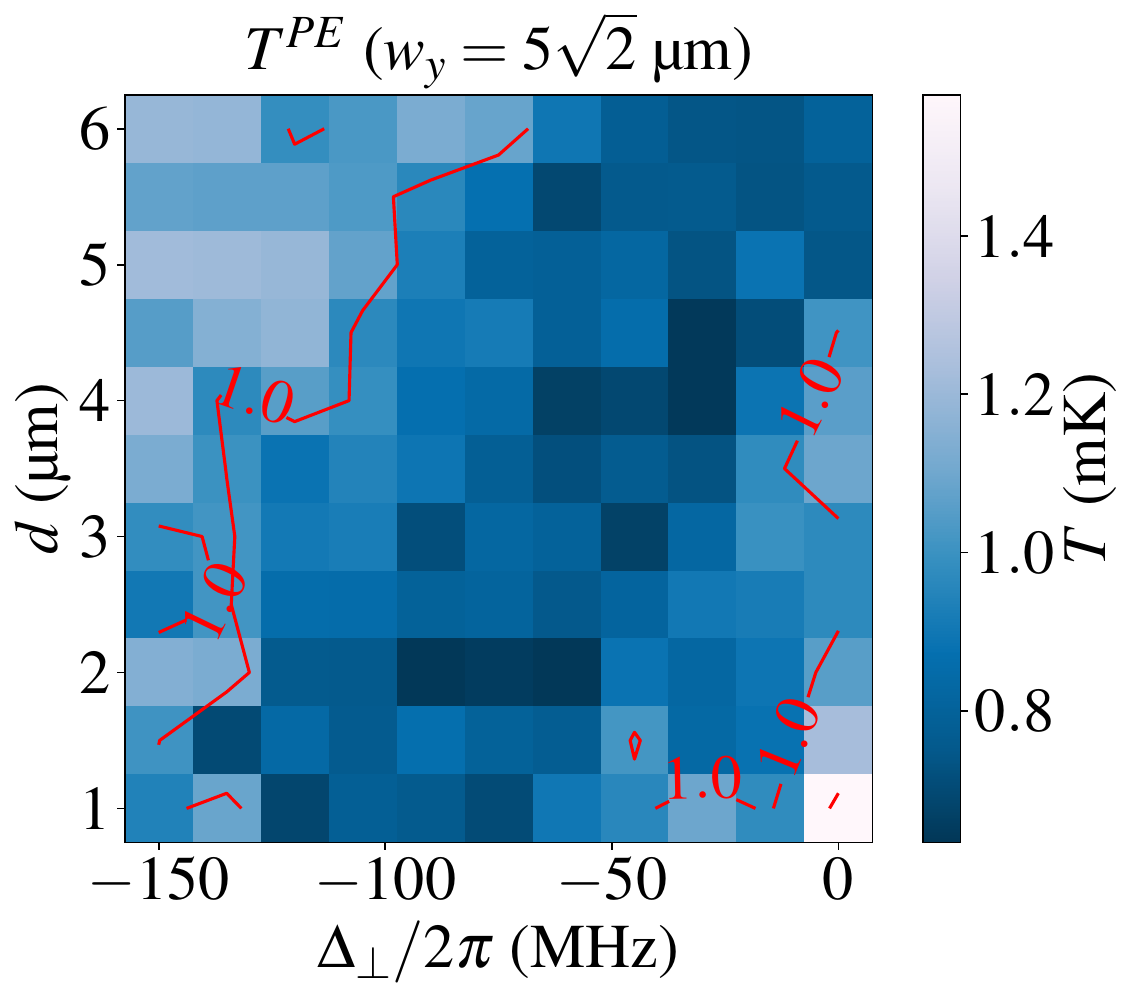}
        \label{fig8d}
    \end{subfigure}

    \caption{The perpendicular kinetic energy and total potential energy of the large $\delta$, $\omega_r=2\pi\times700$ kHz is shown.  Each plot shows the final temperature after 16 ms of laser cooling as a function of perpendicular beam detuning, $\Delta_{\perp}$ and offset, $d$.  The upper and lower plots are obtained using a perpendicular beam waist of $w_y = 20\sqrt{2}\;\mu$m and $w_y = 5\sqrt{2}\;\mu$m, respectively.  Note that the simulated ($\Delta_{\perp},d$) parameter space differs. The kinetic energy cooling dramatically improves by decreasing $w_y$, which suggests an experimental method to improve the cooling of larger $\beta$ 3D crystals.  The potential energy cooling also shows a slight improvement. White boxes represent simulations in which the temperature does not equilibrate after 16 ms.}
    \label{fig8}
\end{figure}

In order to achieve this aim, we propose to reduce the perpendicular beam waist, $w_y$, from 20$\sqrt{2}$ $\mu$m to 5$\sqrt{2}$ $\mu$m, while holding the peak laser intensity constant. Previous simulations showed that narrower waists often result in reduced potential and kinetic energies \cite{torrisi2016, zaris2024}. Since it is more challenging experimentally to vary the beam waist, simulations can help determine whether there is a sufficient benefit to warrant such a change to the setup. We repeat the large $\delta$ simulations for $\omega_r/2\pi =700$ kHz, this time setting $w_y=5\sqrt{2}\; \mu$m.  Changing the beam waist also changes the optimal $(\Delta_{\perp}, d)$ parameter space for kinetic energy cooling, and this is reflected in our simulations. In Fig.~\ref{fig8}, we compare the kinetic and potential energies for crystals with $w_y=20\sqrt{2} \;\mu$m and $w_y=5\sqrt{2} \;\mu$m.  Here, we show the final temperatures after 16 ms of laser cooling for each $(\Delta_{\perp},d)$ parameter pair. By reducing $w_y$, we find that $KE_{\perp}$ is dramatically reduced over a large parameter space.  Temperatures below 3 mK appear achievable, and even lower temperatures can likely be reached by further reducing $w_y$.

\section{\label{sec:cyc_cooling}Enhanced Cooling in Prolate Crystals}

In the previous section, we studied crystals in which $f^z$-coupling led to enhanced potential energy cooling.  By further increasing $\beta$, $f^z$-coupling can be introduced in the high frequency cyclotron mode branch.  We characterize the $KE_{\perp}$ cooling which is achievable in this regime, and find that the kinetic energy is cooled below the theoretical limits presented in \cite{torrisi2016, zaris2024}. Furthermore, we discover a highly coupled parameter space in which efficient cooling can be achieved using only axial beams.  In this case, the use of a perpendicular beam, whose parameters require careful optimization, is unnecessary. 

\subsection{$KE_{\perp}$ cooling beyond the uncoupled approximation}

Earlier, we found that the perpendicular kinetic energy was not cooled as well for crystals with larger $\beta$ (see Fig.~\ref{fig7}). This is, in fact, expected, and these kinetic energy cooling results are in agreement with the predictions of the theoretical treatment developed by Torrisi et al. in \cite{torrisi2016}. This model considers the dispersion of Doppler shifts sampled by the perpendicular cooling beam due to the ion crystal rotation and the finite waist of the laser beam.  It predicts the post-cooling kinetic energy of 2D ion crystals given the values of key laser parameters, but has also been shown to accurately predict the temperatures of many 3D crystals \cite{zaris2024}.  Since the theoretical model does not account for $f^z$-coupling or resonant mode coupling, it is not guaranteed to produce correct cooling predictions in all 3D crystal regimes.  However, since the cyclotron modes, which provide the dominant contribution to the perpendicular kinetic energy, remain almost entirely planar in a large parameter space, it is not surprising that the theory predictions hold for many 3D crystals. Nevertheless, as the crystal becomes increasingly prolate, effects of coupling appear. 

\begin{figure}
    \centering

    \begin{subfigure}{0.2\textwidth}
        \centering
        \includegraphics[scale=0.22]{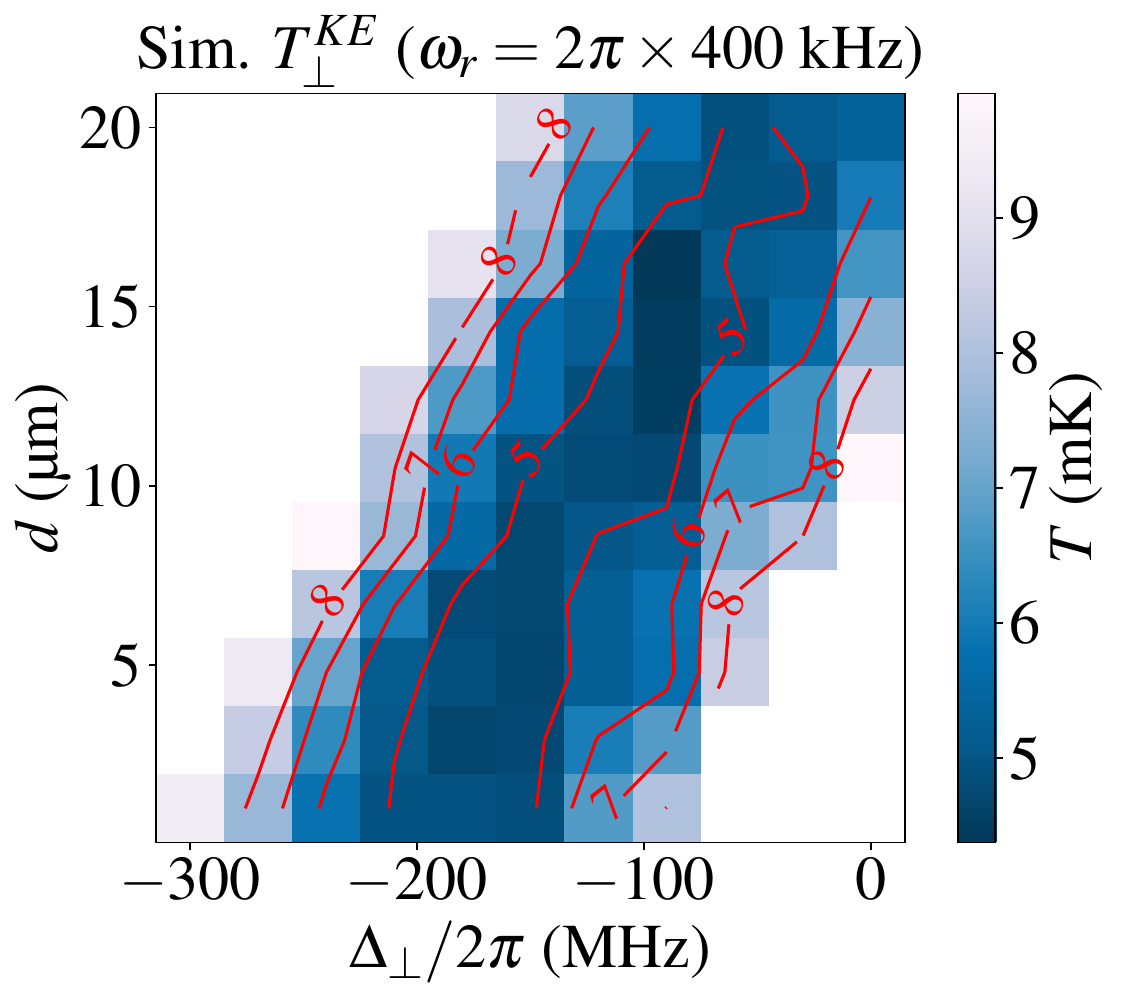}
        \label{fig9a}
    \end{subfigure}%
    \hfill
    \begin{subfigure}{0.2\textwidth}
        \centering
        \hspace*{-0.7cm}
        \includegraphics[scale=0.22]{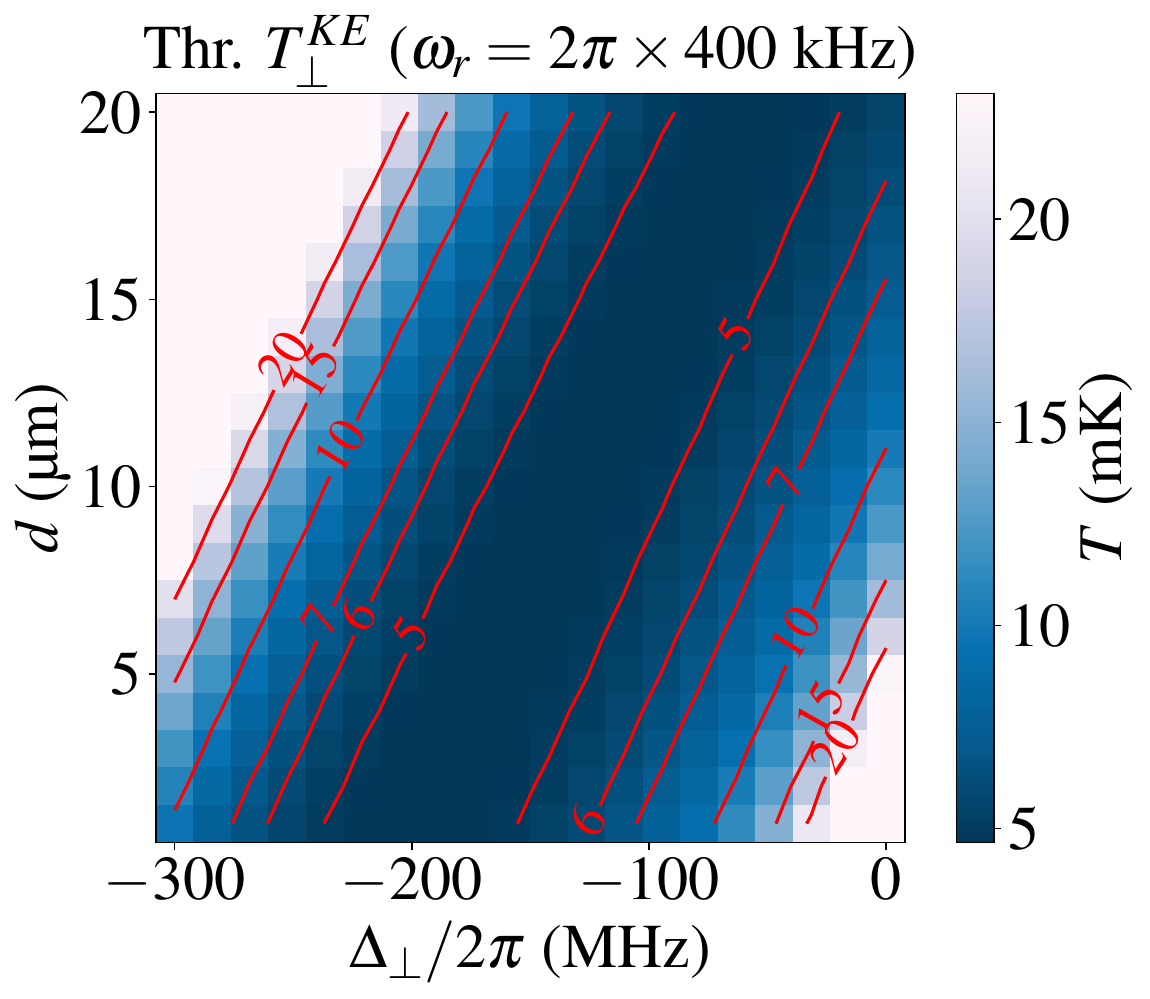}
        \label{fig9b}
    \end{subfigure}

    \vspace{0.1em}

    \begin{subfigure}{0.2\textwidth}
        \centering
        \includegraphics[scale=0.22]{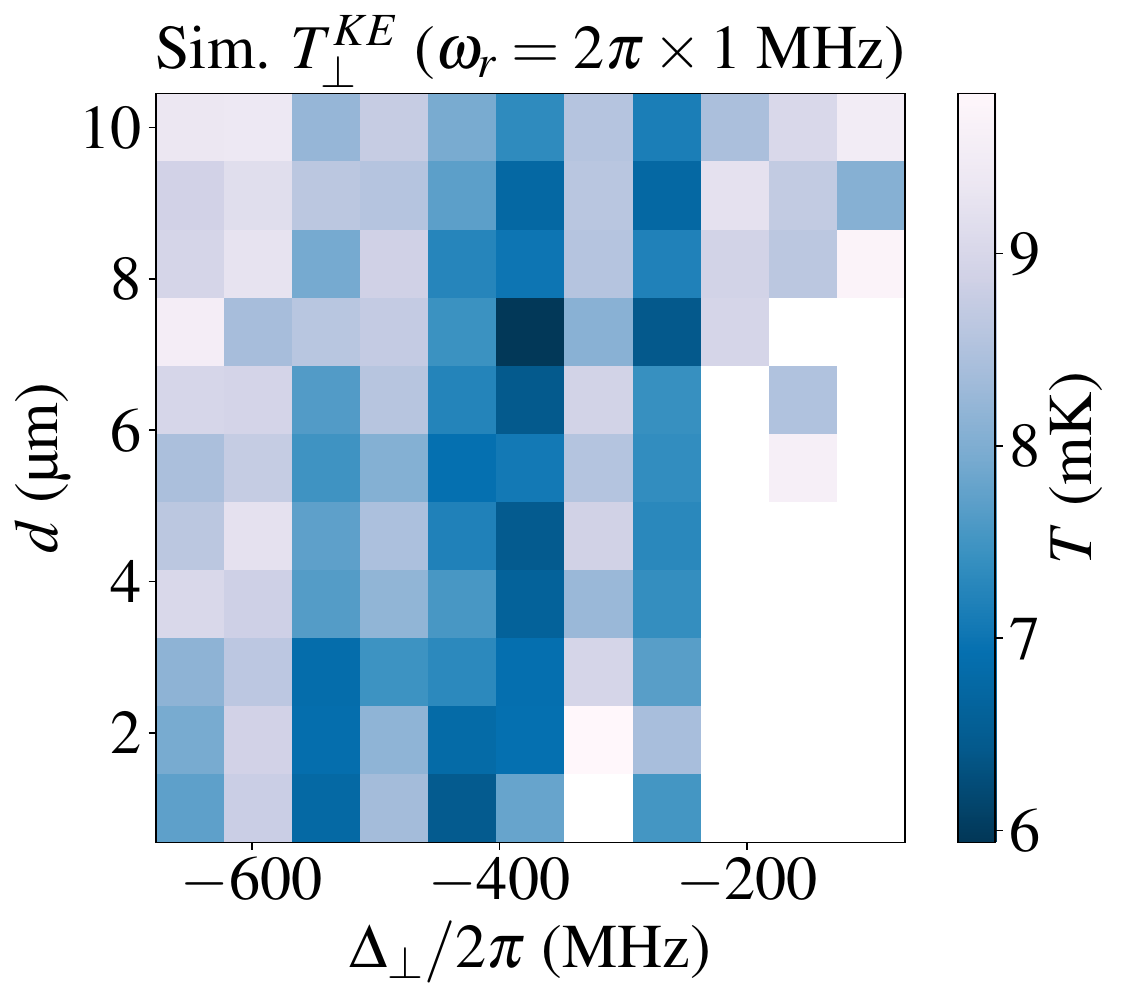}
        \label{fig9c}
    \end{subfigure}%
    \hfill
    \begin{subfigure}{0.2\textwidth}
        \centering
        \hspace*{-0.7cm}
        \includegraphics[scale=0.22]{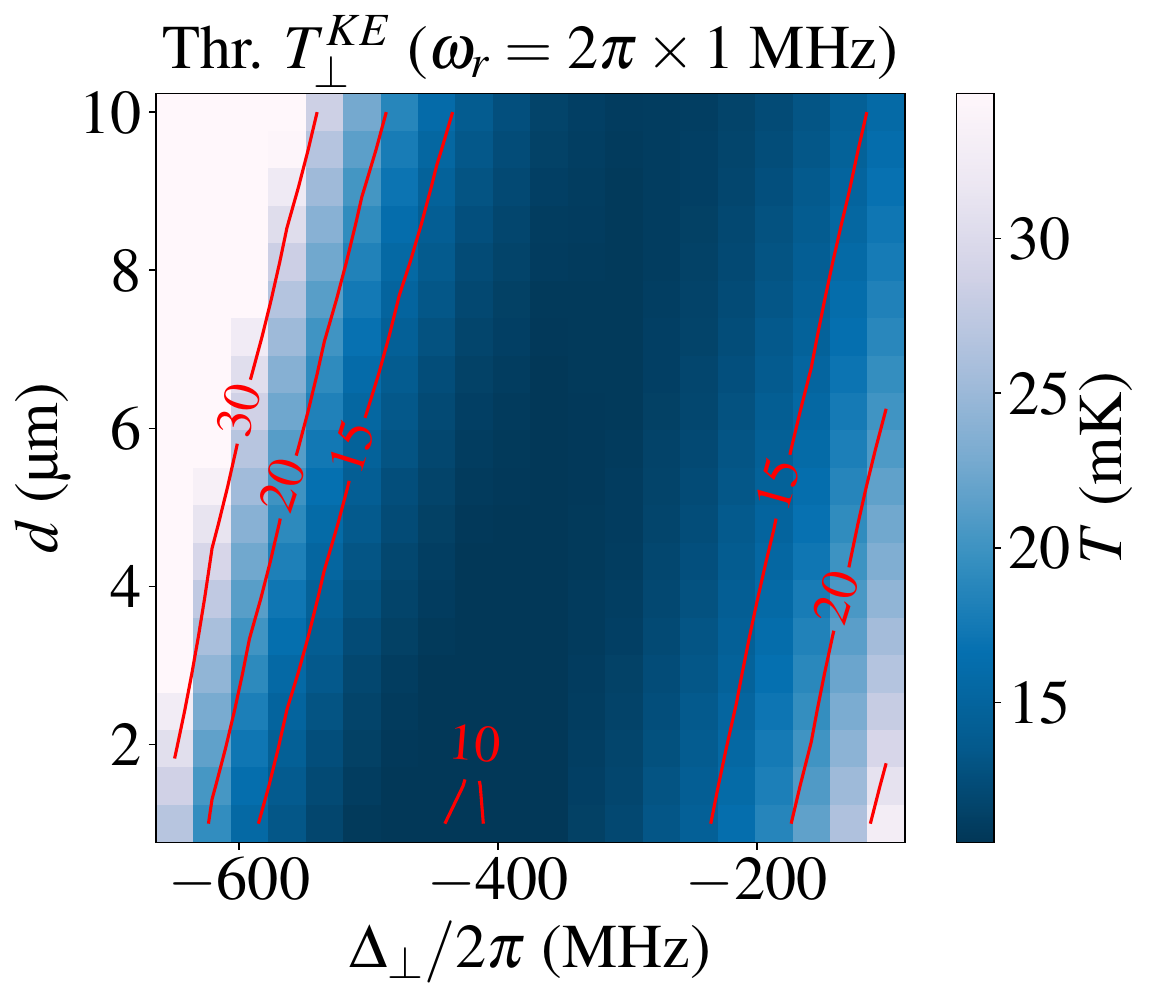}
        \label{fig9d}
    \end{subfigure}

    \vspace{0.1em}

    \begin{subfigure}{0.2\textwidth}
        \centering
        \includegraphics[scale=0.22]{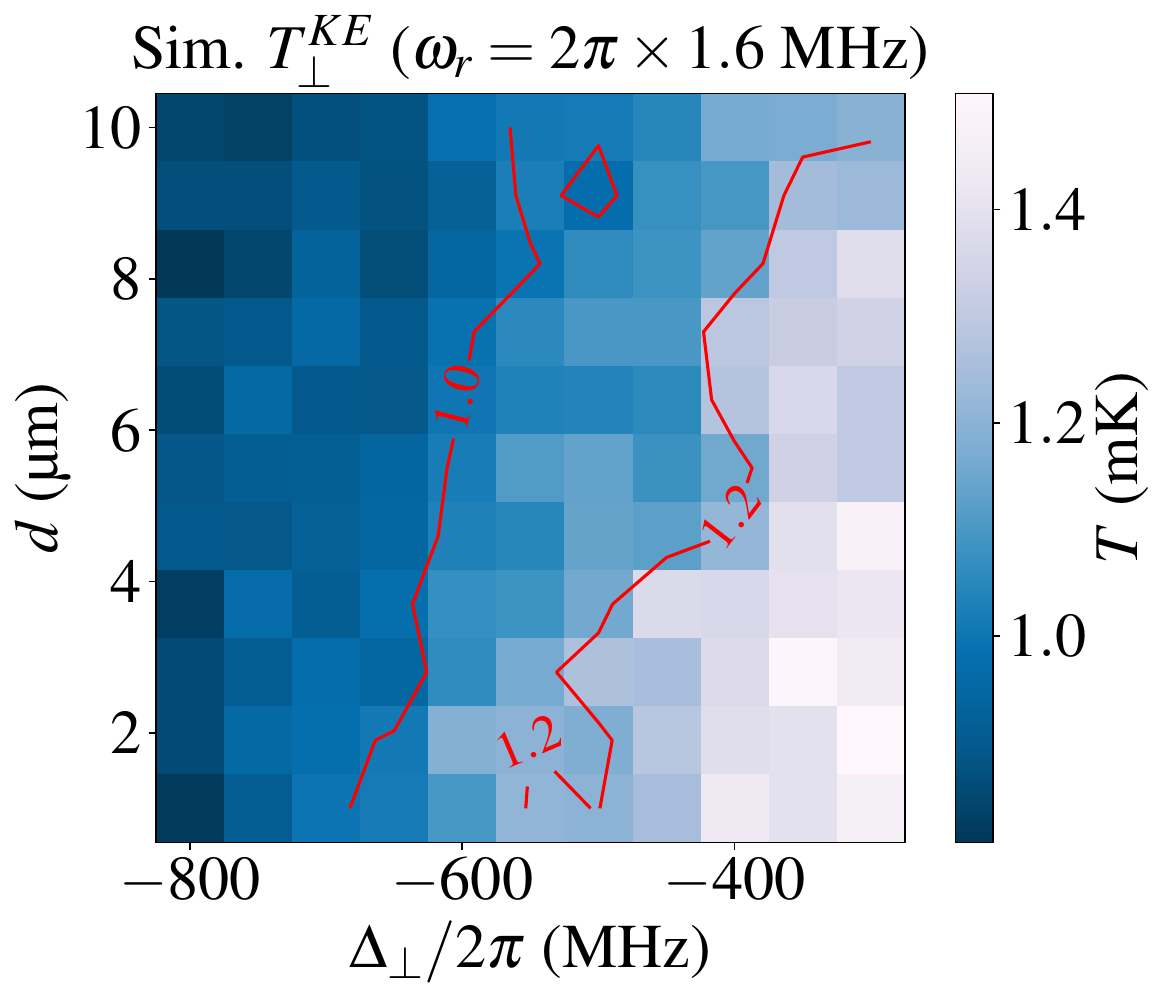}
        \label{fig9e}
    \end{subfigure}%
    \hfill
    \begin{subfigure}{0.2\textwidth}
        \centering
        \hspace*{-0.7cm}
        \includegraphics[scale=0.22]{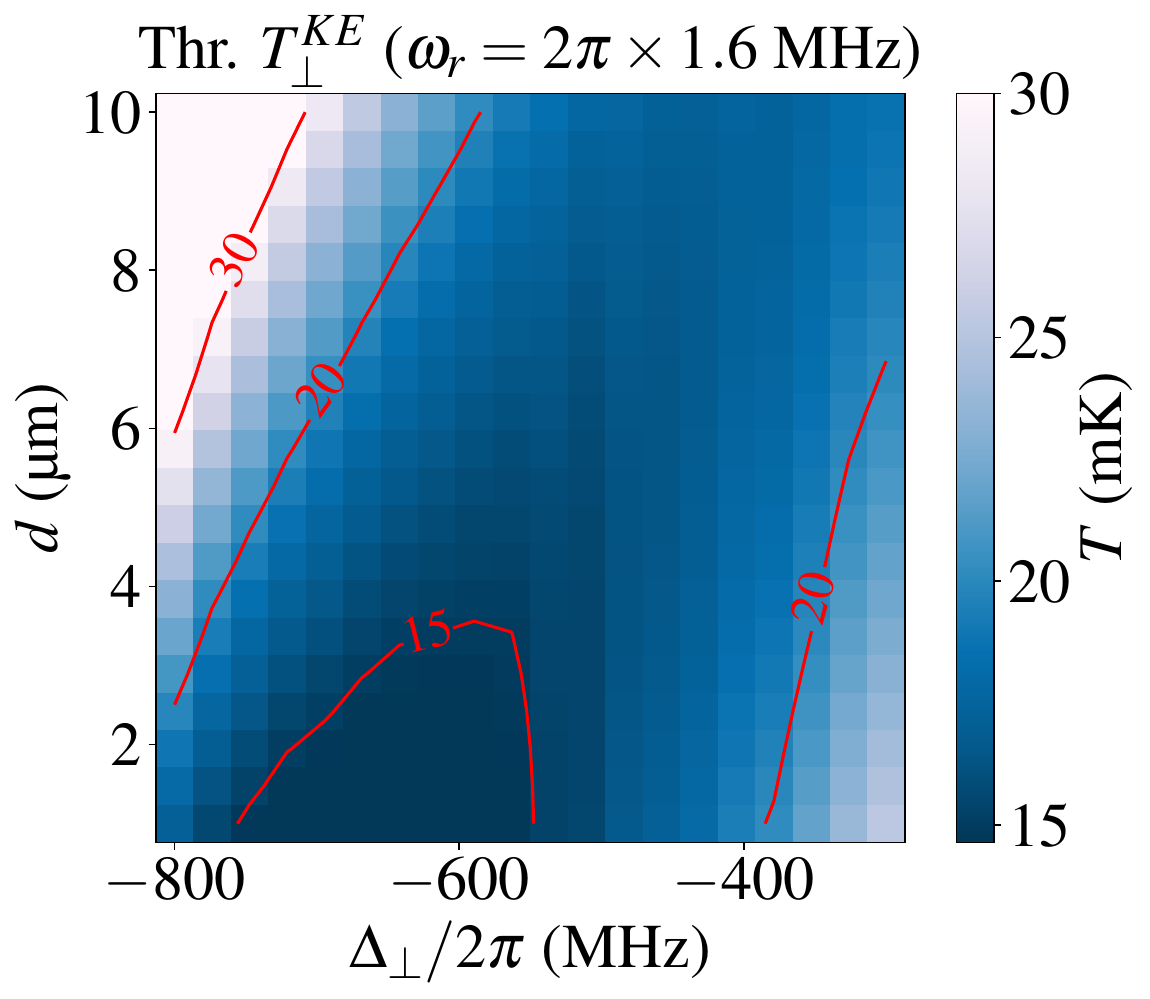}
        \label{fig9f}
    \end{subfigure}

    \caption{The simulated perpendicular kinetic energy after laser cooling is compared to theoretical predictions for three $N=1000$ ion crystals with different values of $\omega_r$.  Our simulations (left-hand figures) use 10 ms of laser cooling, and the theoretical predictions (right-hand figures) are obtained using the model from \cite{torrisi2016}, which has been extended to 3D crystals in \cite{zaris2024}.  The rows of figures correspond to $\omega_r/2\pi = 400$ kHz, 1 MHz, and 1.6 MHz, respectively.  The theoretical model does not account for cooling of the cyclotron motion by the axial laser beam, and is therefore not guaranteed to produce reliable results for 3D crystals.  Still, we see that the simulated and predicted temperatures agree quite well for lower $\beta$ crystals, such as the $\omega_r=2\pi\times400$ kHz one. As $\omega_r$ increases, the theoretical treatment predicts increased perpendicular kinetic energies.  However, at $\omega_r=2\pi\times1$ MHz, the simulated kinetic energies are significantly smaller than these predictions.  We attribute this result to the effect of increased coupling in this regime.  This behavior is amplified in the $\omega_r=2\pi\times1.6$ MHz case, where we see cooling to below $1$ mK throughout a large parameter space. White boxes in the simulation plots represent runs in which the temperature does not equilibrate after 10 ms, while white boxes in the theoretical plots correspond to high temperatures which are excluded to enhance the color gradient in the regions of interest.}
    \label{fig9}
\end{figure}

In order to understand the limits of the Torrisi theory, we simulate the cooling of $N=1000$ ion crystals with $\omega_r/2\pi\in\{400,1000,1600\}$ kHz.  Here, we use $\delta=0.104$ and $w_y = 20\sqrt{2}\;\mu$m.  All other parameters have the same values as in the last section, except for $\Delta_{\perp}$ and $d$, whose ranges are once again chosen to minimize the predicted perpendicular kinetic energy.  The $KE_{\perp}$ values after 10 ms of laser cooling are shown in Fig.~\ref{fig9}.  Also shown are the corresponding theoretical temperatures obtained using the Torrisi model.  This model does not account for planar anisotropy (i.e. nonzero $\delta$).  Therefore, to obtain the theoretical temperatures, we define the ion crystal's planar extent as the mean of its maximum and minimum planar extents.  This approximation is not expected to significantly affect the results presented in this section.  

We find excellent agreement between the simulated and theoretical results for $\omega_r=2\pi\times400$ kHz.  At this rotation frequency, there is minimal coupling between planar and axial motion.  As $\omega_r$ is increased, the Torrisi theory predicts that cooling of $KE_{\perp}$ becomes less effective when the beam width is held constant, as seen in Fig.~\ref{fig9}.   However, at higher rotation frequencies, coupling effects ignored in the theoretical model lead to improved cooling in our simulations. This is seen at $\omega_r= 2\pi\times1000$ kHz and $\omega_r=2\pi\times1600$ kHz, where the simulated kinetic energy is significantly lower than the theoretical predictions. Coupling effects become stronger as the rotation frequency increases, and we find that the simulated $KE_{\perp}$ cooling is most effective at $\omega_r=2\pi\times1600$ kHz.  In the absence of coupling, this crystal is predicted to exhibit the weakest cooling of $KE_{\perp}$. As such, MD simulations are essential to obtain accurate cooling results in the highly coupled regime.  Finally, we note that the results for $\omega_r=2\pi\times1600$ kHz suggest that prolate ion crystals can quickly be cooled to temperatures of approximately 1 mK using a large range of laser parameters.  This reduced sensitivity to laser beam parameters should allow excellent cooling to be achieved easily in experimental setups.

\subsection{Cooling without perpendicular beam}

\begin{figure}
    \centering
    \begin{subfigure}{0.5\textwidth}
        \subcaption{}
        \centering
        \includegraphics[scale=0.45]{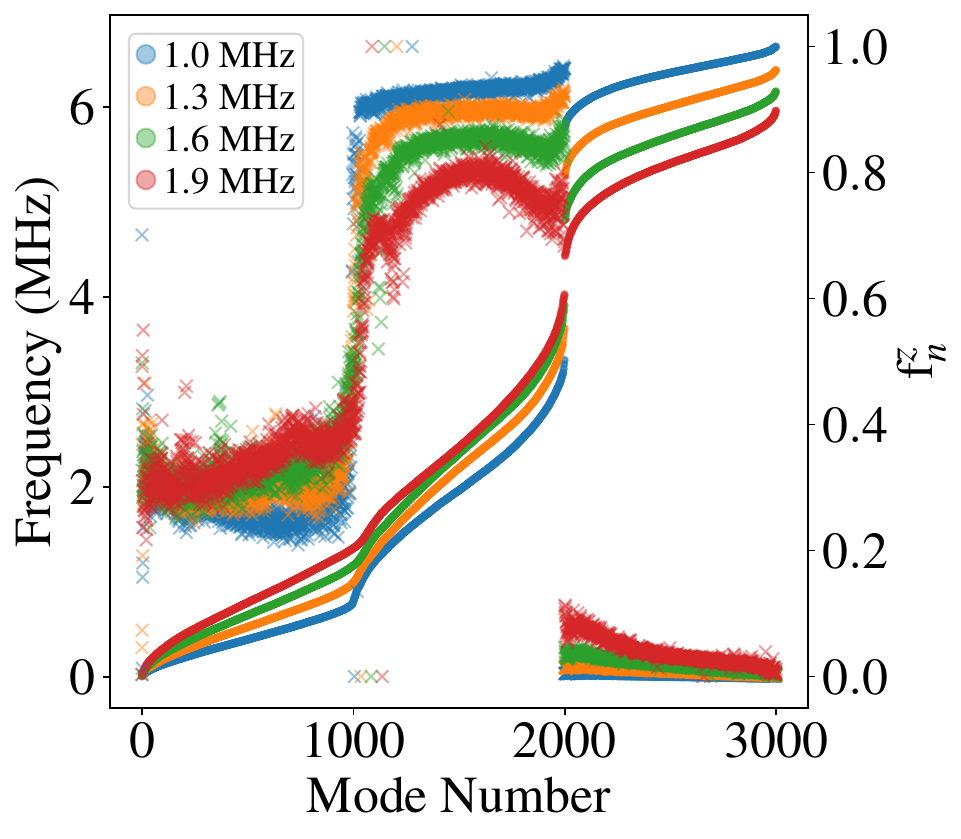}
        \label{fig10a}
    \end{subfigure}
    ~ 
    \bigskip
    \centering
    \hspace{0cm} 
    \begin{subfigure}{0.5\textwidth}
        \subcaption{}
        \centering
        \includegraphics[scale=0.5]{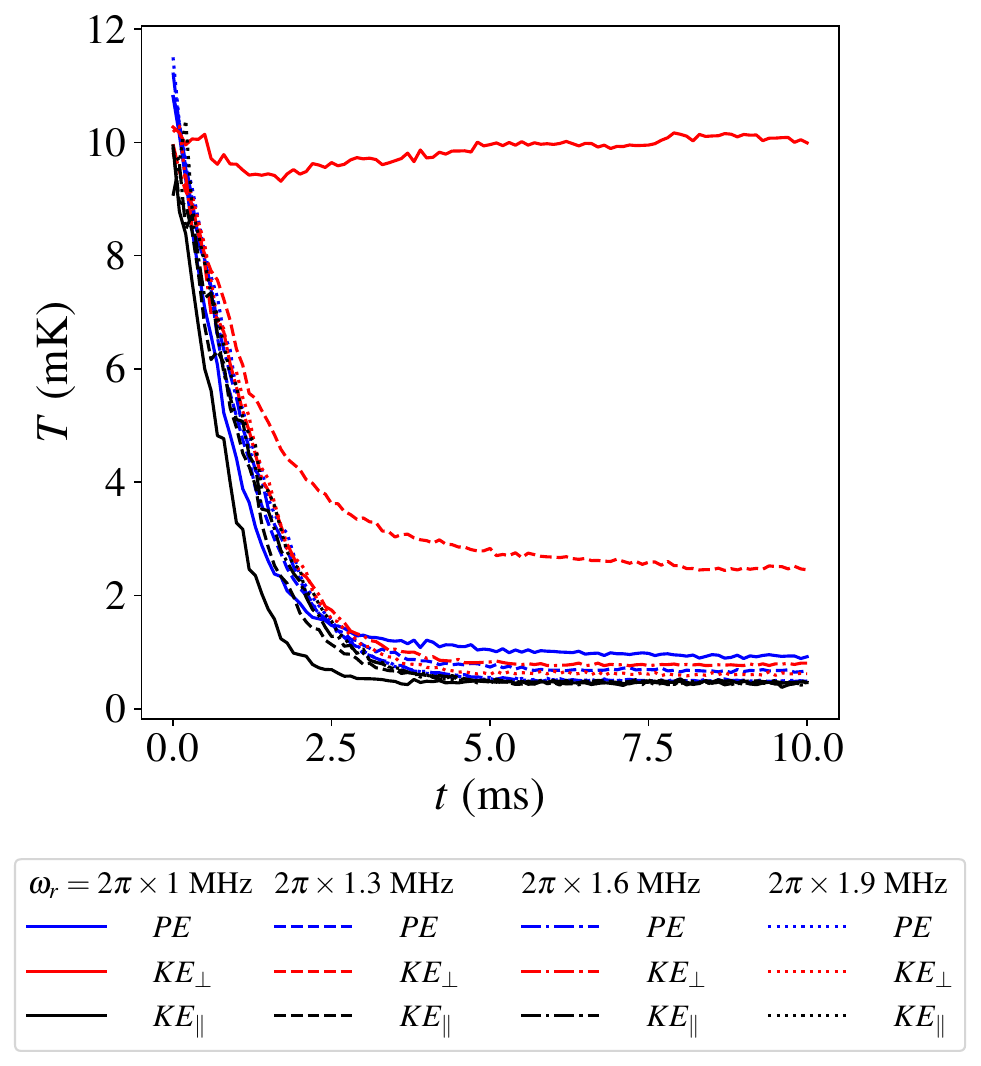}
        \label{fig10b}%
    \end{subfigure}
    ~ 
    \caption{(a) Mode spectra and mode $f^z$ values for several high $\beta$ crystals with different values of $\omega_r/2\pi$.  The cyclotron modes gain a significant axial component. (b) Cooling curves for these crystals using only axial beams. In all cases shown here, potential and axial kinetic energies are reduced to temperatures of $<$ 1 mK within several milliseconds.  For the larger values of $\omega_r$, the perpendicular kinetic energy is also cooled to $<$ 1 mK.}
    \label{fig10}
\end{figure}

As $\beta$ is further increased, the axial component of the cyclotron modes continues to grow.  In Fig.~\ref{fig10a}, we plot $f^z$, as well as the mode spectra, for several $N=1000$ ion crystals with $\omega_r/2\pi$ between $1$ MHz and $1.9$ MHz (using $\delta = 0.173$). The remaining trap and axial beam parameters are the same as in the previous sections. Within this parameter space, the perpendicular cooling beam becomes unnecessary to achieve milliKelvin temperatures.  In Fig.~\ref{fig10b}, we plot the cooling of the kinetic and potential energy of these high $\beta$ crystals when only the axial cooling beams are applied.  At $\omega_r=2\pi\times 1$ MHz, the axial kinetic and potential energies are rapidly cooled to $\lesssim 1$ mK, but the perpendicular kinetic energy is not cooled below its initial temperature of $10$ mK. However, as $\omega_r$ increases, the perpendicular kinetic energy is cooled much more effectively by the axial beams.  At $\omega_r/2\pi = 1.6$ MHz and $1.9$ MHz, $KE_{\perp}$ reaches $<1$ mK within a few milliseconds of cooling.  

While the perpendicular beam is primarily responsible for cooling the cyclotron motion in lower $\beta$ crystals, the axial beams can effectively cool these modes in large $\beta$ crystals, often to even lower temperatures.  Therefore, the use of such crystals could eliminate the need to optimize the perpendicular beam parameters and experimentally configure the perpendicular beam.  As such, prolate crystals may be attractive platforms for future quantum science experiments.

\section{\label{sec:large_ion_number}Scaling up to larger ion number}

In order to develop transformative quantum sensing applications with 3D crystals, it will be desirable to trap larger ion crystals of $\sim 10^5-10^6$ ions.  With this experimental goal in mind, our MD code has been designed for numerical efficiency when simulating large $N$ crystals. In this section, we carry out a preliminary study of laser cooling of a $N=10^5$ ion crystal.  

Certain methods from the preceding sections must be modified when studying such large crystals. First, it is prohibitively inefficient to use line-search based energy minimization techniques like BFGS to find equilibrium configurations for large $N$.  For this reason, we have implemented a dynamical minimization technique, which utilizes numerical damping, to relax a cloud of ions with randomized initial positions to a local equilibrium configuration. This method relies on evolving the ions according to the Penning trap equations of motion while removing a small fraction of each ion's velocity in the rotating frame at each timestep.  In order to accelerate the damping process, we modify our MD code to remove the high frequency cyclotron motion which limits the speed of numerical damping.  First, we evolve the ions according to the equations of motion in the rotating frame, rather than those in the lab frame.  The coordinate transformation to the rotating frame is given by

\begin{equation}
\label{eq18}
    \begin{pmatrix} x_r\\y_r\end{pmatrix} = \begin{pmatrix} \cos(\omega_r t) & -\sin(\omega_r t)\\\sin(\omega_r t) & \cos(\omega_r t)\end{pmatrix}\begin{pmatrix} x\\y\end{pmatrix}.
\end{equation}
In both the lab frame and rotating frame, the magnetic field produces terms in the system's Hamiltonian which are proportional to mixed position-velocity products like $\dot{x}y_r$.  Our MD cooling simulation employs a cyclotronic integrator \cite{patacchini2009} to account for the resulting motion.  However, in our damping algorithm, we ignore these terms since the ion velocities vanish at equilibrium. This is equivalent to setting the effective magnetic field in the rotating frame equal to zero \cite{wang2013} and evolving the ions in a purely electrostatic potential.  The potential energy is explicitly given by

\begin{align}
\label{eq19}
\phi_{r}(\{\boldsymbol{x}_n\}) = \sum_{j}\Big[&\frac{1}{2}m\omega_z^2z_j^2\nonumber\\&-\frac{1}{2}m\big(\omega_r^2-\omega_c\omega_r+\frac{1}{2}\omega_z^2\big)(x_j^2+y_j^2)\nonumber\\&+\frac{1}{2}qk_z\delta(x_j^2-y_j^2) +\frac{kq^2}{2}\sum_{k\neq j}\frac{1}{r_{jk}}\Big],
\end{align}
where we have dropped the `r' subscript on the ion coordinates for simplicity. The previously described simplification removes cyclotron motion, thereby changing the ion dynamics. However, it does not affect the potential energy landscape of the system. At each timestep ($\Delta t=1$ ns), we reduce the momentum of each ion, in the rotating frame, by a factor of $10^{-4}$. In general, slower momentum reduction causes the crystal to relax into lower energy local equilibrium states, but also requires longer damping simulations.  We note briefly that other dynamical energy minimization techniques have been developed for large ion crystals, including the simulation of guiding center motion \cite{dubin1988} and the implementation of Nose-Hoover thermostats \cite{totsuji2002}.

Initializing the crystal's potential energy using a Metropolis-Hastings algorithm becomes extremely slow for large $N$. Therefore, we implement an initialization routine based on Langevin dynamics in the rotating frame.  Specifically, starting from the equilibrium configuration, we evolve the system according to the Langevin equation

\begin{equation}
\label{eq20}
    m\frac{d^2\boldsymbol{x}_i}{dt^2} = \boldsymbol{F}_i - \gamma\dot{\boldsymbol{x}}_i + \sqrt{2\gamma k_BT}\boldsymbol{\xi}_i(t).
\end{equation}
Here, $T$ is set to the desired initial temperature, $\boldsymbol{F}_i = -q\nabla\phi_r(\boldsymbol{x}_i)$ is the force on ion $i$ (in the rotating frame), and $\gamma$ is the damping factor.  The Gaussian noise term, $\boldsymbol{\xi}_i(t) = \xi_{i,x}(t)\hat{\boldsymbol{x}} +\xi_{i,y}(t)\hat{\boldsymbol{y}} +\xi_{i,z}(t)\hat{\boldsymbol{z}}$, satisfies the correlation function $\langle \xi_{i,\alpha}(t)\xi_{i,\beta}(t')\rangle = \delta_{\alpha,\beta}\delta(t-t')$.  The damping algorithm described earlier is equivalent to Eq.~(\ref{eq20}) with $T=0$. We use a velocity Verlet algorithm to integrate Eq.~(\ref{eq20}) in time. Since the crystal is initialized in a harmonic potential (Eq.~(\ref{eq19})), the kinetic and potential energies are excited equally. By comparing the potential energies of an $N=10^5$ ion crystal obtained using the Langevin thermostat and the corresponding equilibrium crystal found using numerical damping, we verify that the potential energy is initialized at approximately 10 mK.  However, the exact potential energy difference between the crystals is slightly less than 10 mK.  This is likely because the damping algorithm is unable to find the global energy-minimizing ion configuration for large crystals, leading to an `equilibrium' potential energy which is small, but nonzero.  For this reason, we define the potential energy of the Langevin-initialized crystal as $T=10$ mK when plotting cooling results.

\begin{figure}
    \centering

    \begin{subfigure}{0.2\textwidth}
        \subcaption{}
        \centering
        \includegraphics[scale=0.22]{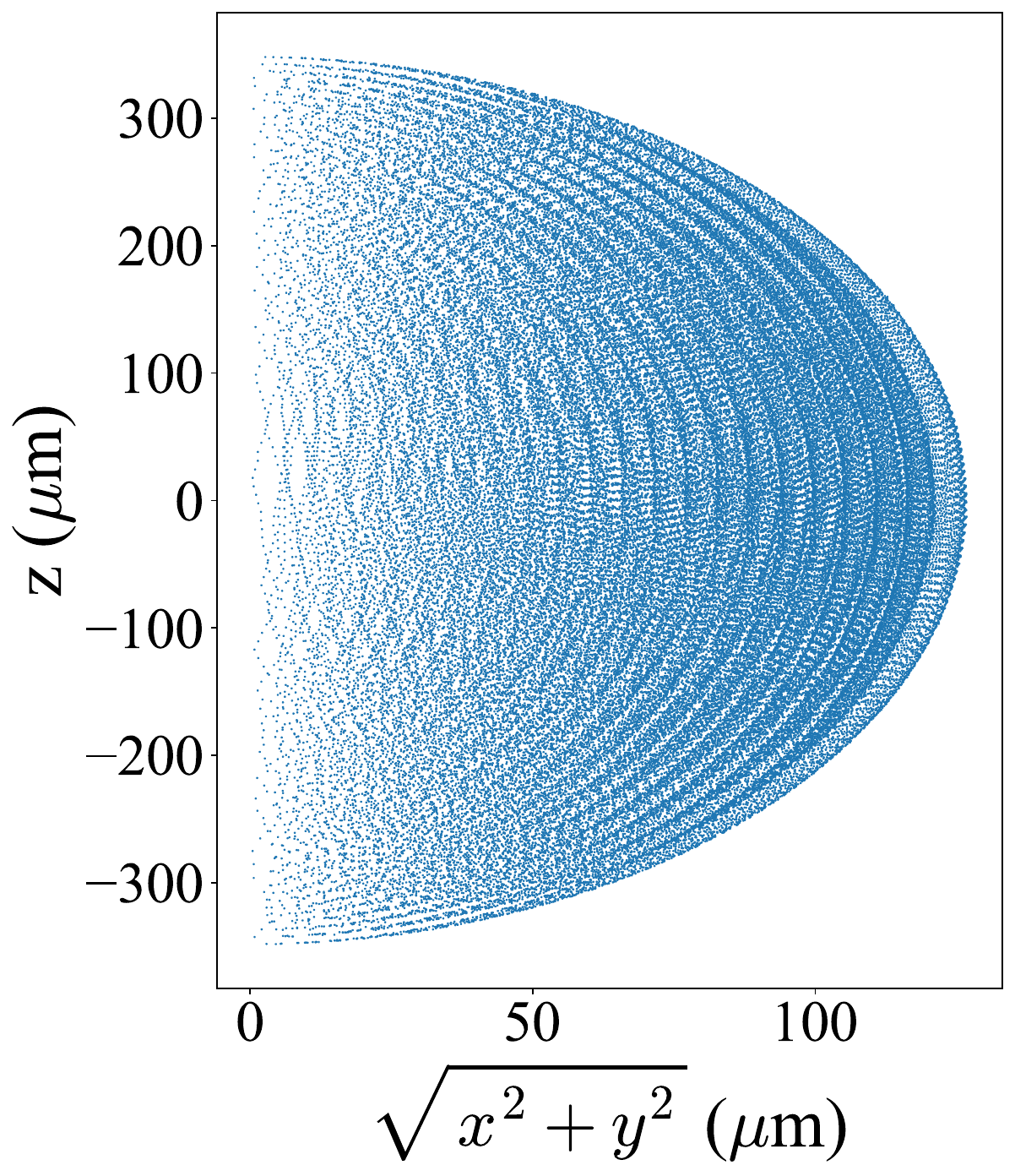}
        \label{fig11a}
    \end{subfigure}%
    \hfill
    \begin{subfigure}{0.2\textwidth}
        \subcaption{}
        \centering
        \hspace*{-0.2cm}
        \includegraphics[scale=0.28]{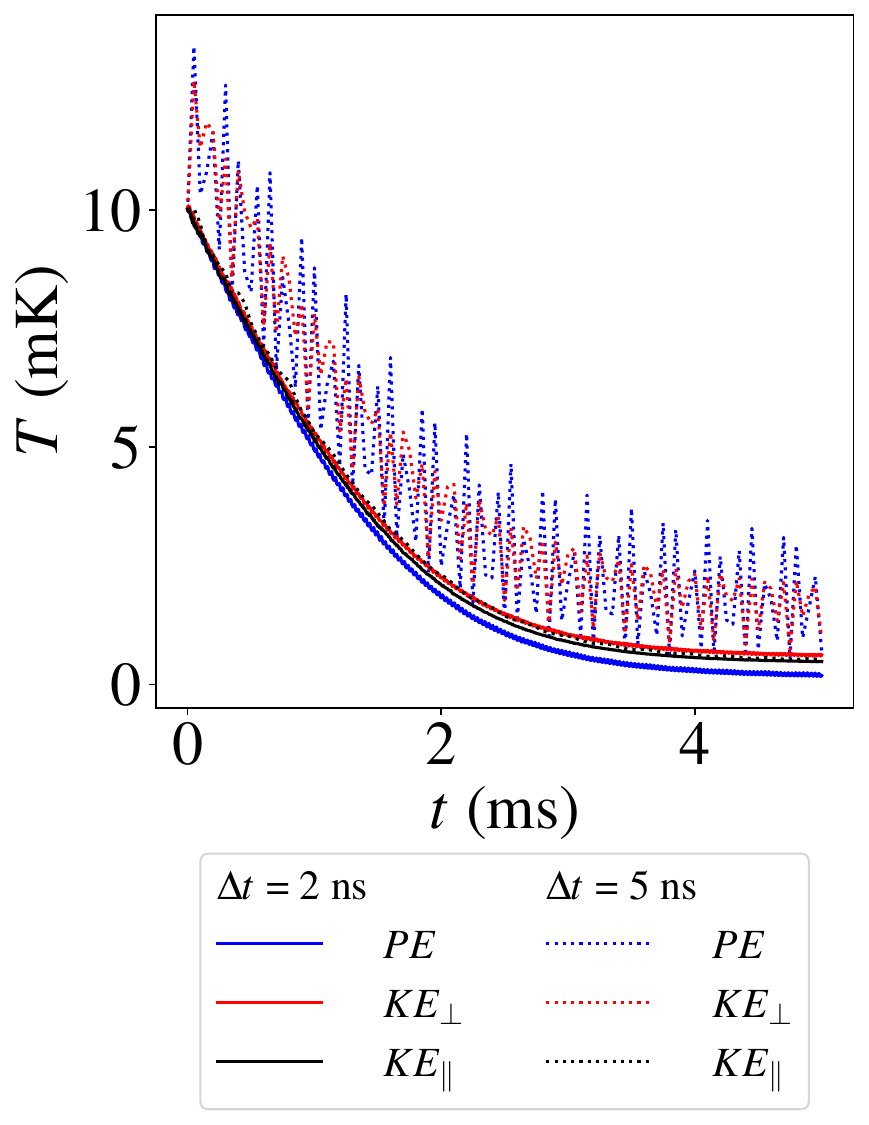}
        \label{fig11b}
    \end{subfigure}
    ~ 
    \caption{(a) Local equilibrium ion configuration of an $N=10^5$ crystal with $B=4.4588$ T, $\omega_z = 2\pi\times1.59$ MHz, $\omega_r = 2\pi\times1.9$ MHz, and $\delta= 0.173$  (b) Cooling of the crystal's energy using only axial beams.  Using $\Delta t=$ 2 ns, all energies are cooled to below 1 mK within 5 ms, similar to the results seen in Fig.~\ref{fig10}.  The potential energy configuration corresponding to 0 mK is found using numerical damping, as described in the text.  Here, laser cooling over 5 ms causes the potential energy to be cooled below this level.  We do not carry out a reminimization here, as was done for previous simulations.}
    \label{fig11}
\end{figure}

Following this scheme, we prepare a $N=10^5$ ion crystal with the following trap parameters: $B=4.4588$ T, $\omega_z = 2\pi\times1.59$ MHz, $\omega_r = 2\pi\times1.9$ MHz, and $\delta= 0.173$.  The equilibrium ion configuration is shown in Fig.~\ref{fig11a}. The concentric shell structure which characterizes ellipsoidal crystals \cite{dubin1999} is evident, although the shells appear smeared out due to the large value of $\delta$ which creates planar anisotropy.  We initialize the crystal at 10 mK using $10^5$ timesteps of the Langevin thermostat and then simulate Doppler cooling. In Fig.~\ref{fig11b}, we plot the resulting cooling curves.  Since the simulation time increases approximately linearly with ion number, it is helpful to consider increasing the simulation timestep.  Increasing the timestep to 2 ns, in which case $\omega_c\Delta t = 0.095$, leads to smooth cooling curves, as shown in Fig.~\ref{fig11b}. However, further increasing the timestep to $\Delta t=5$ ns ($\omega_c\Delta t = 0.24$) leads to numerical instabilities and results in noisier cooling curves. Based on the $\Delta t = 2$ ns simulation, we find that all energies are cooled to $<1$ mK within 5 ms.  Note that the potential energy reaches negative values because we have not reminimized the crystal after cooling.

\section{Conclusion}

In this work, we have developed computational tools to efficiently simulate dynamics of 3D crystals with up to $10^5$ ions (see section V).  Using MD simulations, we have characterized Doppler laser cooling of 3D crystals, providing results that will guide experimental efforts to prepare large ion crystals at low temperatures. By studying the normal mode structure of 3D crystals, we have shown that cooling generally improves when modes gain an axial component.  In particular, $PE$ ($KE_{\perp}$) cooling improves when the $\boldsymbol{E}\times\boldsymbol{B}$ (cyclotron) modes have larger values of $f^z$.  Cooling can be further enhanced by tuning parameters like the rotating wall strength and the perpendicular beam waist.  Interestingly, we have discovered that, in the high $\beta$ regime, it should be possible to quickly cool the crystal energies to below 1 mK without the need for a perpendicular cooling beam. This suggests a way to achieve enhanced cooling in 3D crystals while simultaneously simplifying the experimental setup. 

We have produced enhanced cooling by increasing the rotating wall frequency, $\omega_r$.  However, given the expression for $\beta$ (Eq.~(\ref{eq9})), it should also be possible to engineer the mode spectrum by varying the axial trapping frequency, $\omega_z$. We note that as the rotation frequency increases to near the Brillouin limit ($\omega_r=\omega_c/2$), cooling may be degraded due to anomalous heating effects.  Previous experiments with nonneutral plasmas have observed increased heating at large rotating wall frequencies due to slipping of the plasma relative to the field \cite{huang1997, anderegg1998}. This slipping is likely due to small trap asymmetries and is, therefore, not captured in our simulations. This effect may provide a practical limit on the range of feasible $\omega_r$ values. 

Future studies may aim to further investigate the mechanism responsible for improved cooling in 3D crystals.  In particular, by utilizing a linearized version of the MD code, in which mode coupling vanishes, one could determine the relative effects of (1) the narrowing frequency gap between mode branches and (2) the increasing values of $f^z$ in `planar' modes. In future work, we will also aim to simulate electromagnetically-induced transparency (EIT) cooling of trapped ion crystals, which results in near-ground state cooling of the axial modes \cite{jordan2019, kiesenhofer2023}. In particular, we will investigate if EIT cooling can result in sub-Doppler cooling of $\boldsymbol{E}\times\boldsymbol{B}$ and cyclotron modes in 3D crystals.

\begin{acknowledgments}

The authors thank Dietrich Leibfried, Diep Nguyen, and Yuan Shi for reading the manuscript and providing insightful feedback.  We also appreciate helpful discussions with Bryce Bullock and Jennifer Lilieholm.  J.Z, W.J, and S.E.P. were supported by U.S. Department of Energy grant DE-SC0020393.  A.S. acknowledges support by the Department of Science and Technology, Govt. of India through the INSPIRE Faculty Award (DST/INSPIRE/04/2023/001486), by the Anusandhan National Research Foundation (ANRF), Govt. of India through the Prime Minister’s Early Career Research Grant (PMECRG) (ANRF/ECRG/2024/001160) and by IIT Madras through the New Faculty Initiation Grant (NFIG). A.L.C. and J.J.B acknowledge support from AFOSR grant FA9550-25-1-0080 and from the DOE NQIS Research Center, Quantum System Accelerator.
\end{acknowledgments}

\appendix

\section{Doppler cooling parameter scans}
\label{app:a}

\begin{figure}
    \centering

    \begin{subfigure}{0.20\textwidth}
        \centering
        \includegraphics[scale=0.23]{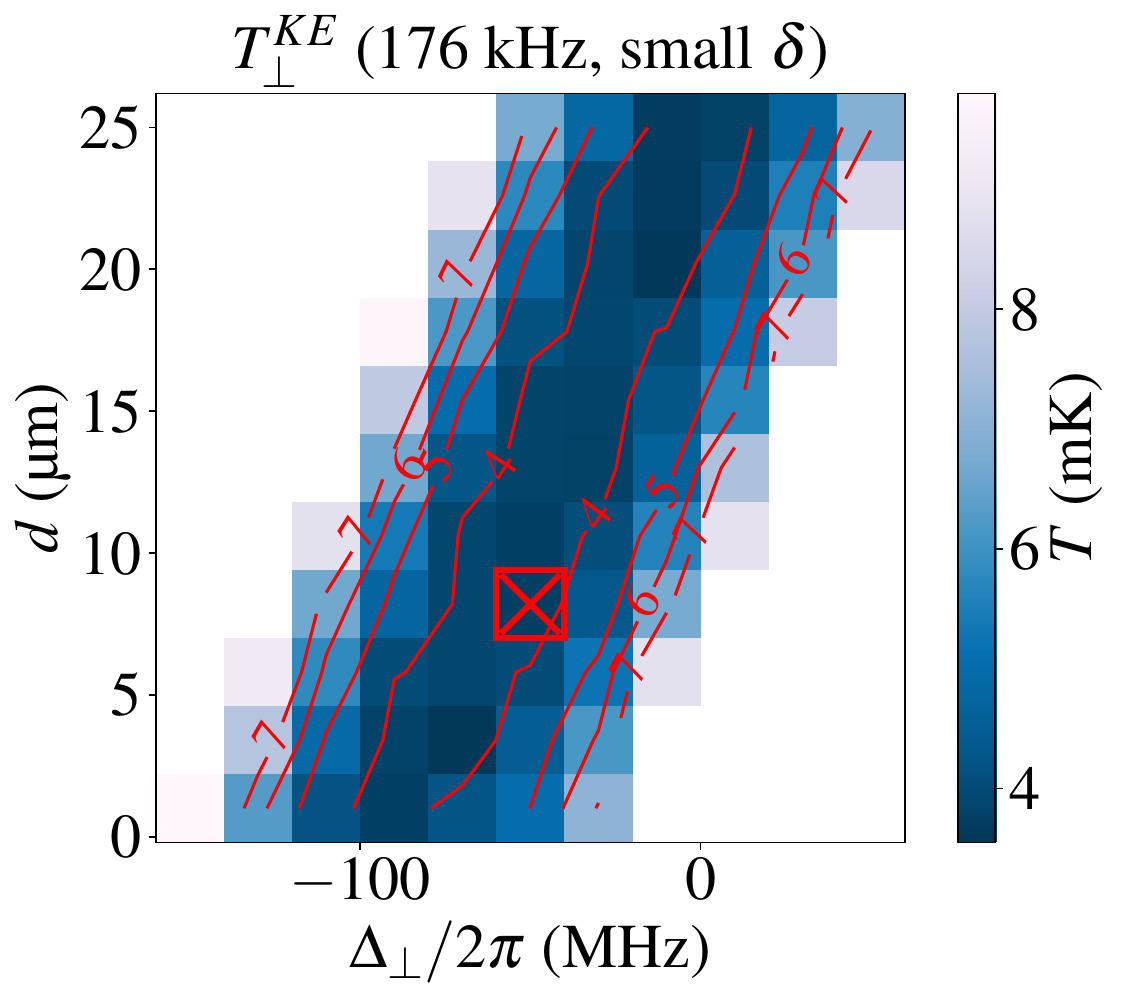}
    \end{subfigure}%
    \hfill
    \begin{subfigure}{0.2\textwidth}
        \centering
        \hspace*{-0.7cm}
        \includegraphics[scale=0.23]{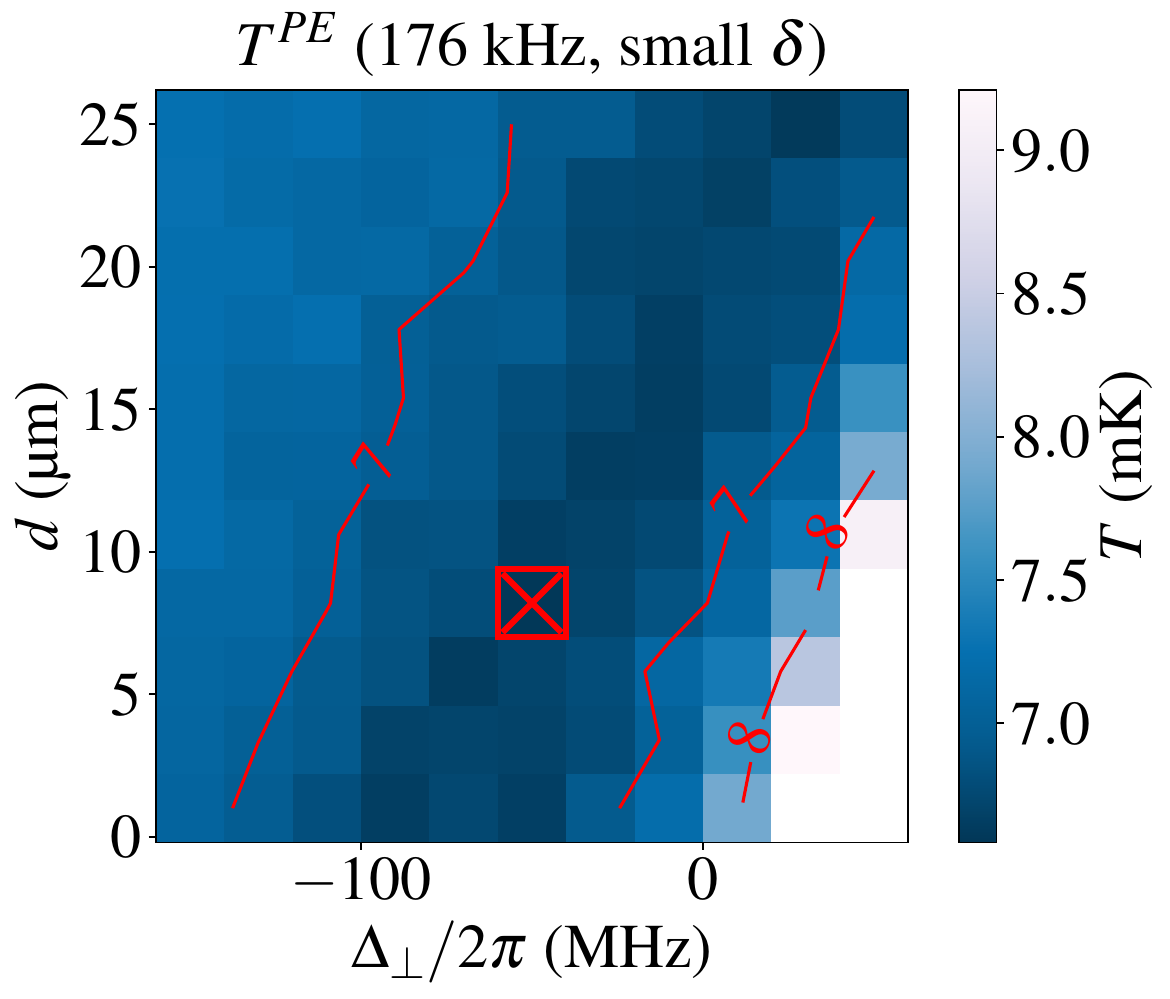}
    \end{subfigure}

    \vspace{0.1em}

    \begin{subfigure}{0.2\textwidth}
        \centering
        \includegraphics[scale=0.23]{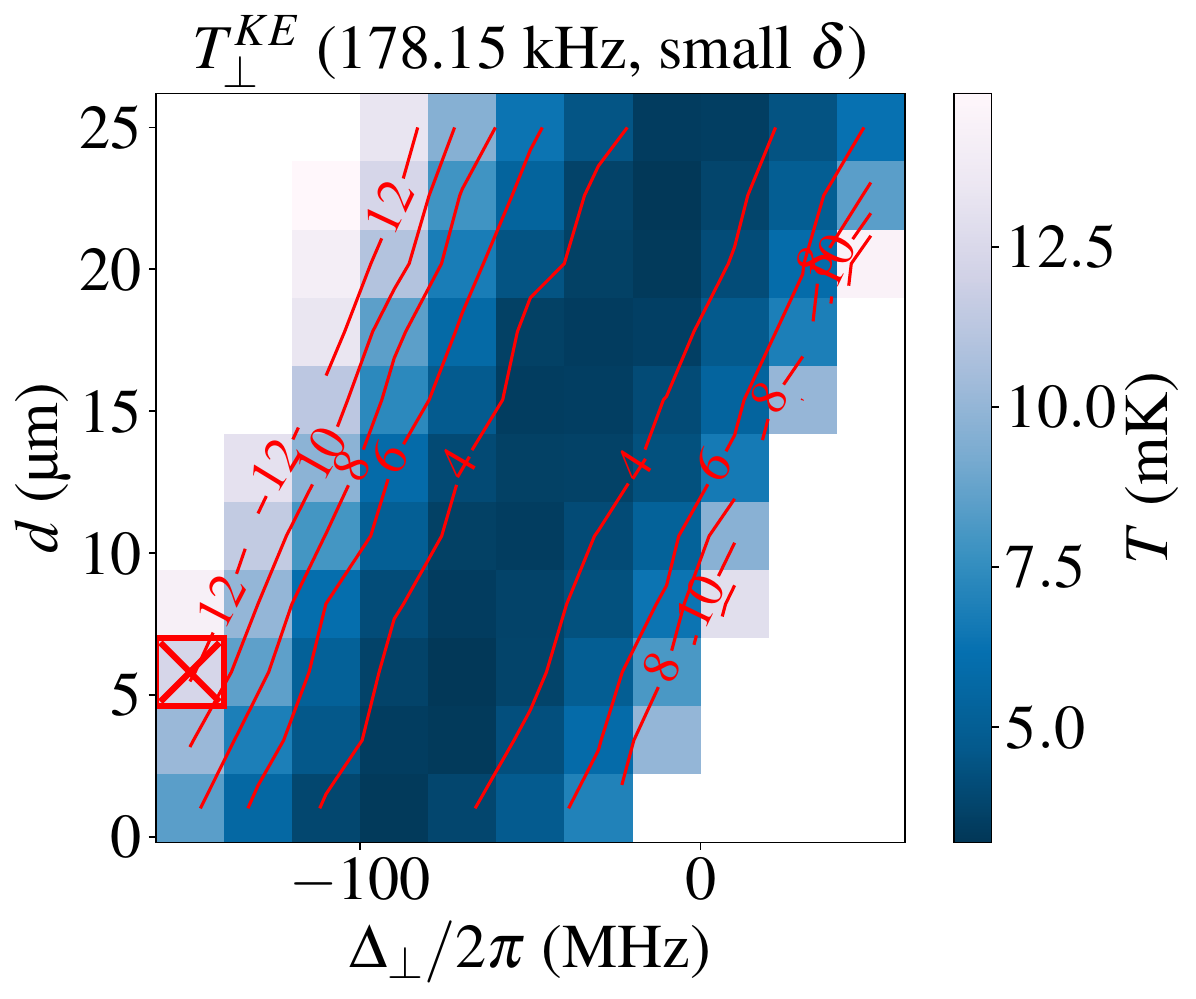}
    \end{subfigure}%
    \hfill
    \begin{subfigure}{0.2\textwidth}
        \centering
        \hspace*{-0.7cm}
        \includegraphics[scale=0.23]{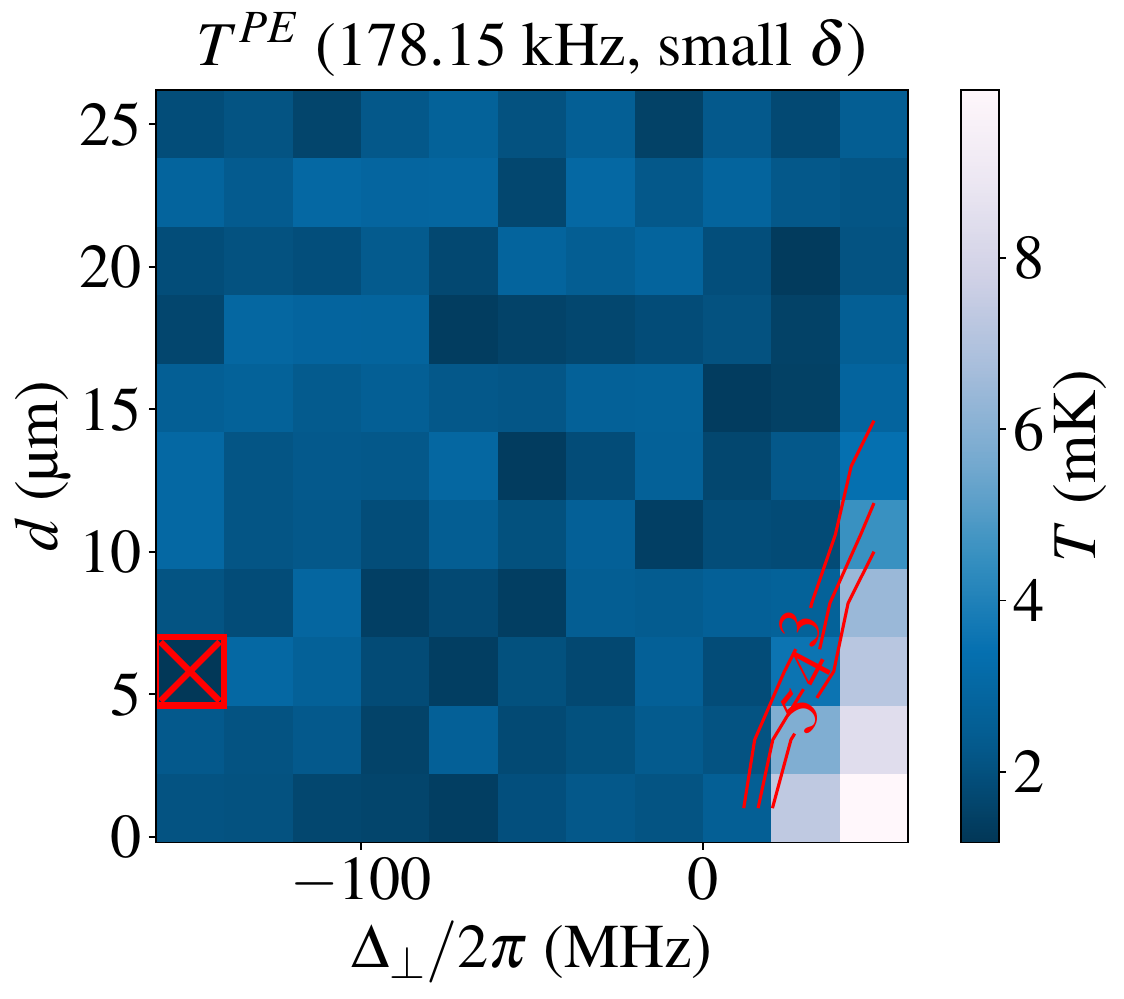}
    \end{subfigure}

    \vspace{0.1em}

    \begin{subfigure}{0.2\textwidth}
        \centering
        \includegraphics[scale=0.23]{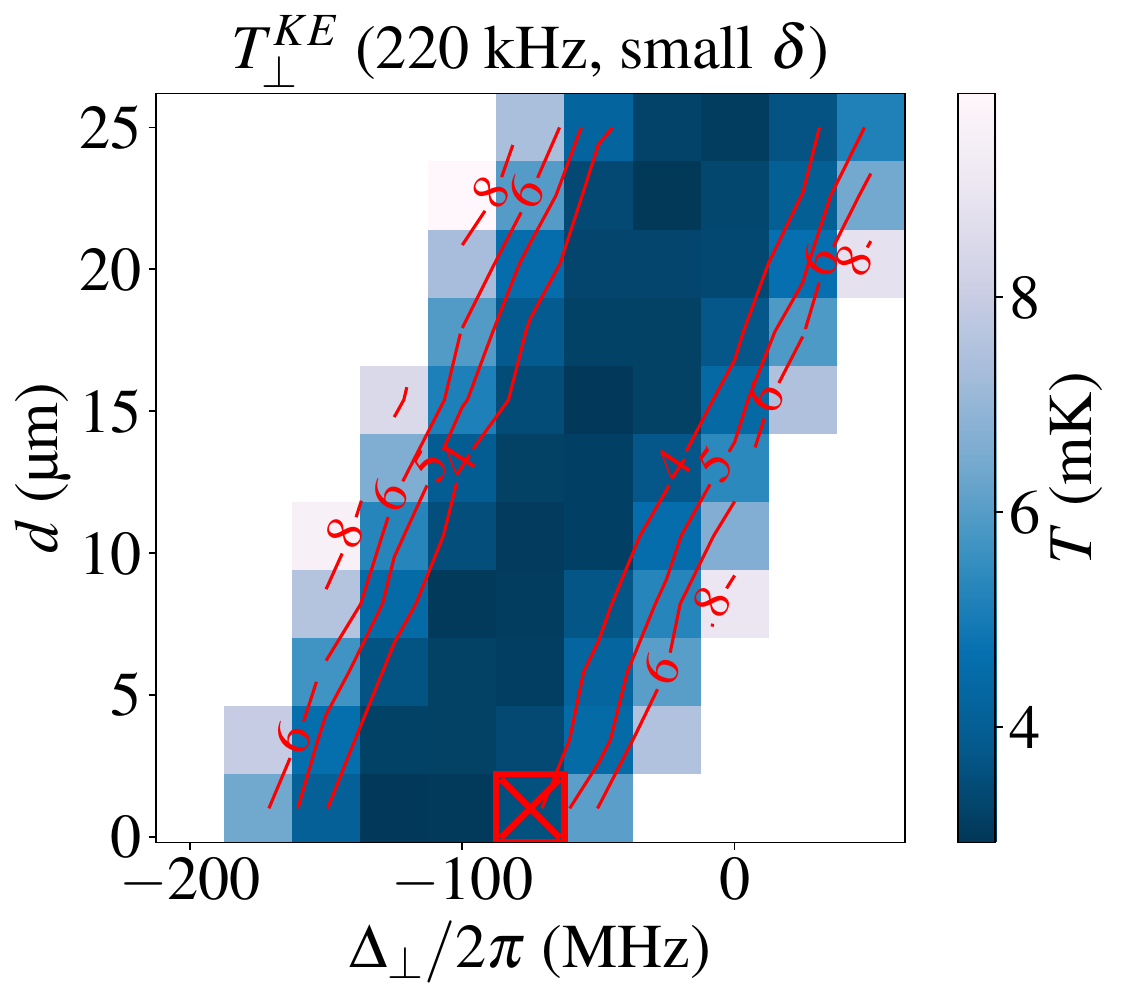}
    \end{subfigure}%
    \hfill
    \begin{subfigure}{0.2\textwidth}
        \centering
        \hspace*{-0.7cm}
        \includegraphics[scale=0.23]{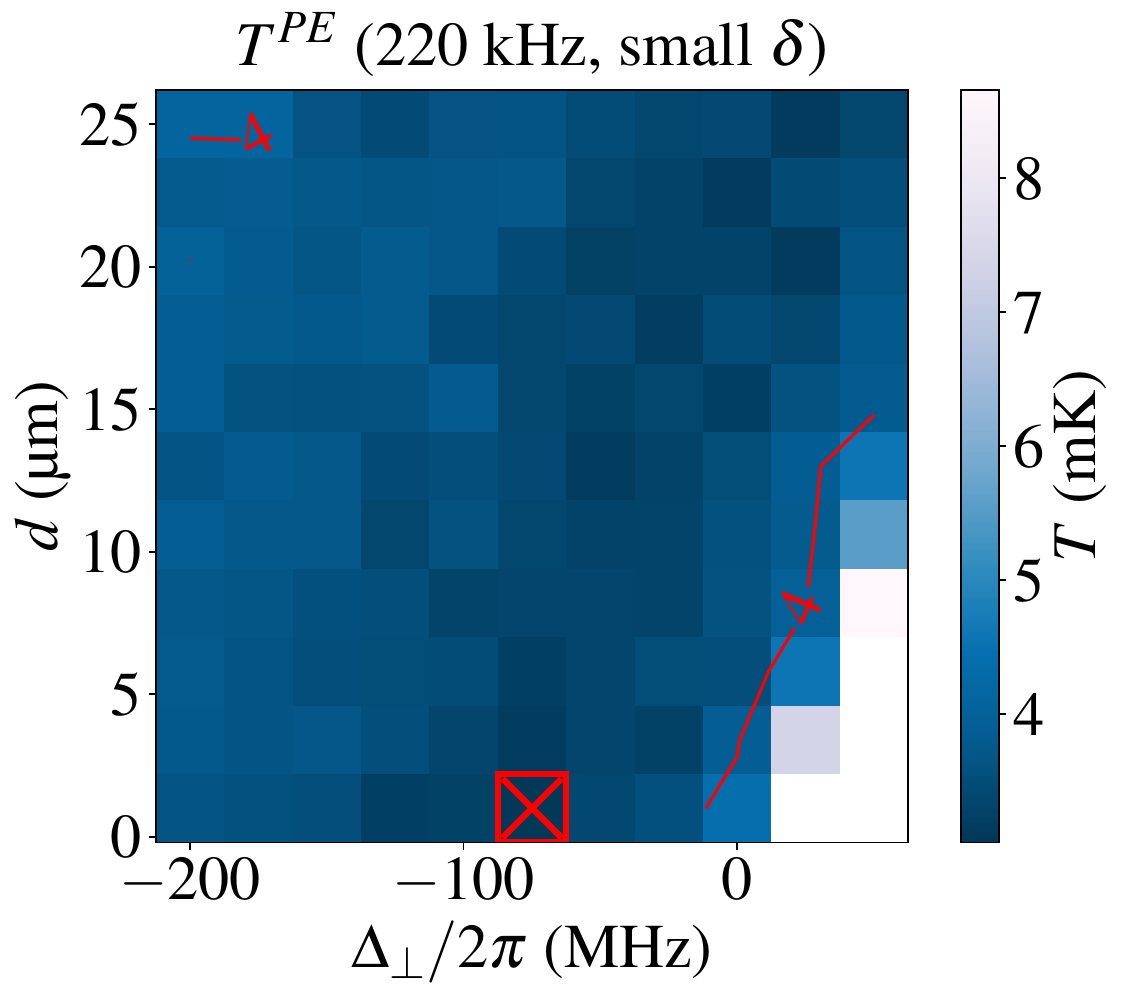}
    \end{subfigure}

    \vspace{0.1em}

    \begin{subfigure}{0.2\textwidth}
        \centering
        \includegraphics[scale=0.23]{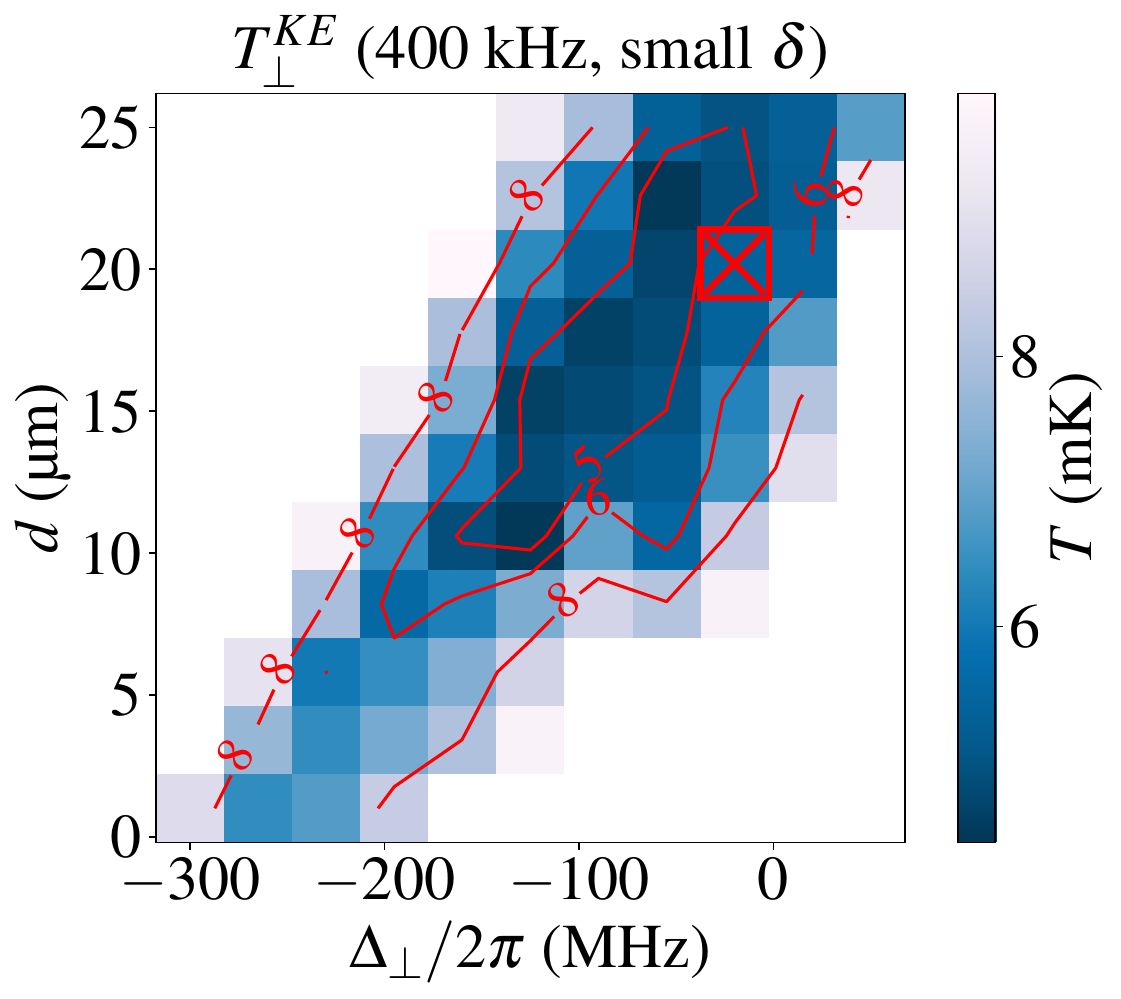}
    \end{subfigure}%
    \hfill
    \begin{subfigure}{0.2\textwidth}
        \centering
        \hspace*{-0.7cm}
        \includegraphics[scale=0.23]{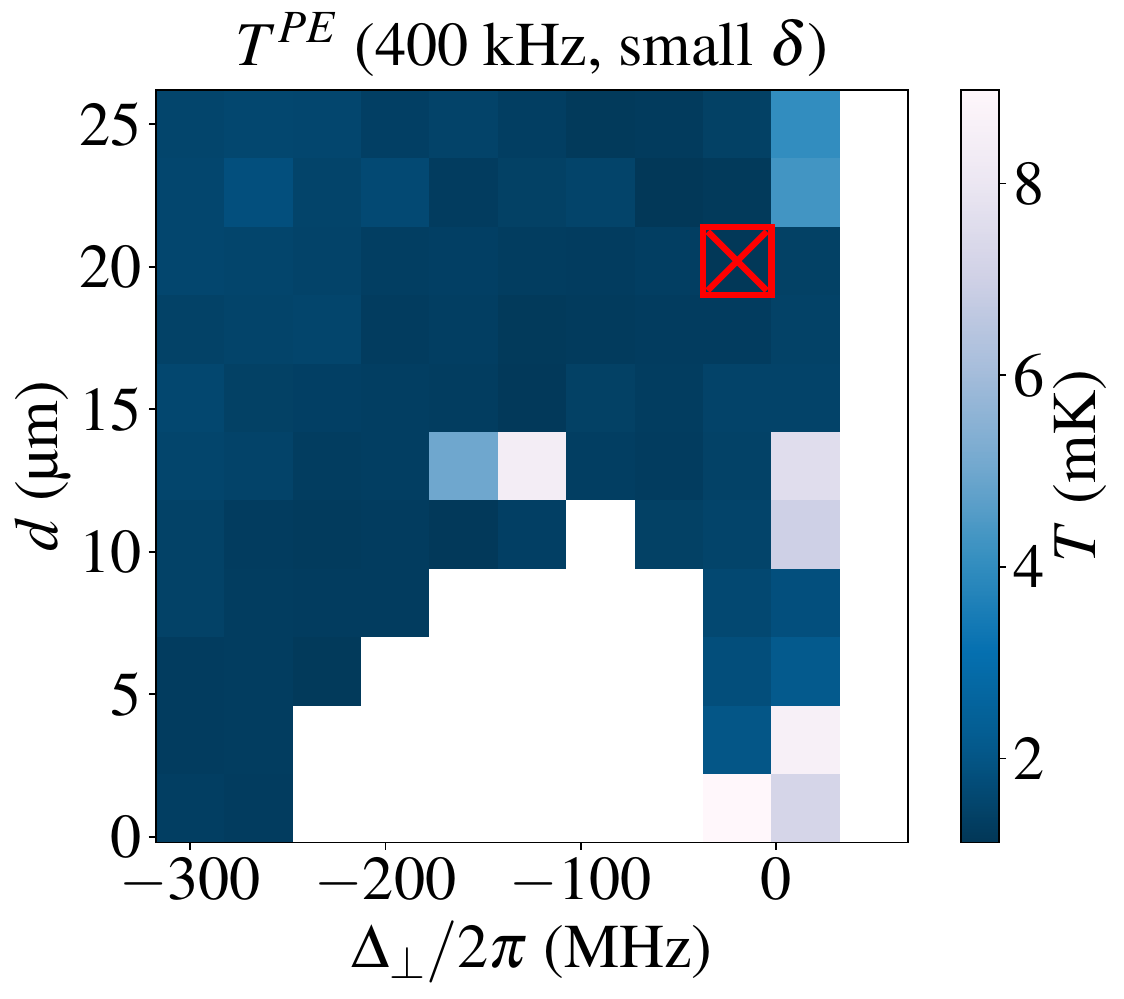}
    \end{subfigure}

    \vspace{0.1em}

    \begin{subfigure}{0.2\textwidth}
        \centering
        \includegraphics[scale=0.23]{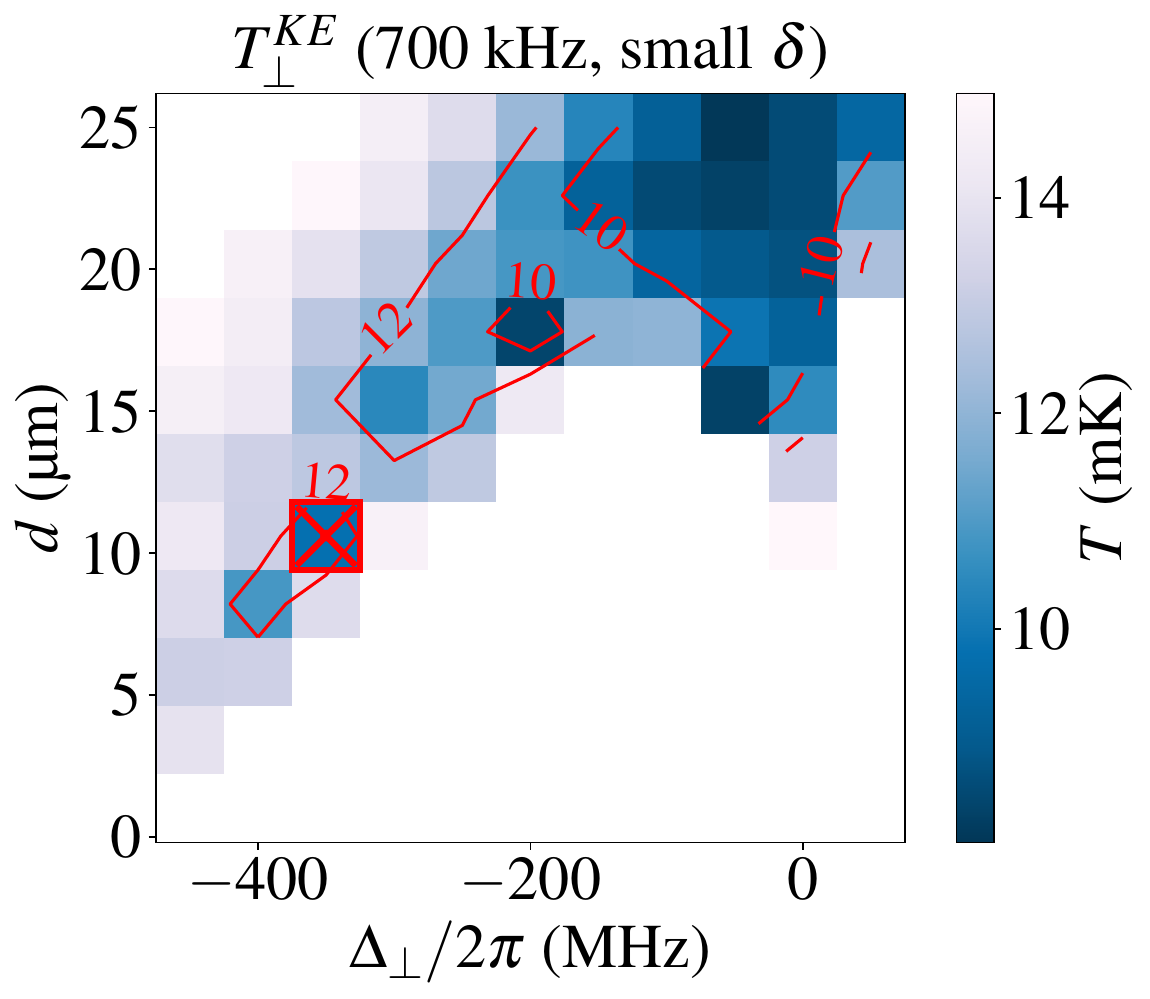}
    \end{subfigure}%
    \hfill
    \begin{subfigure}{0.2\textwidth}
        \centering
        \hspace*{-0.7cm}
        \includegraphics[scale=0.23]{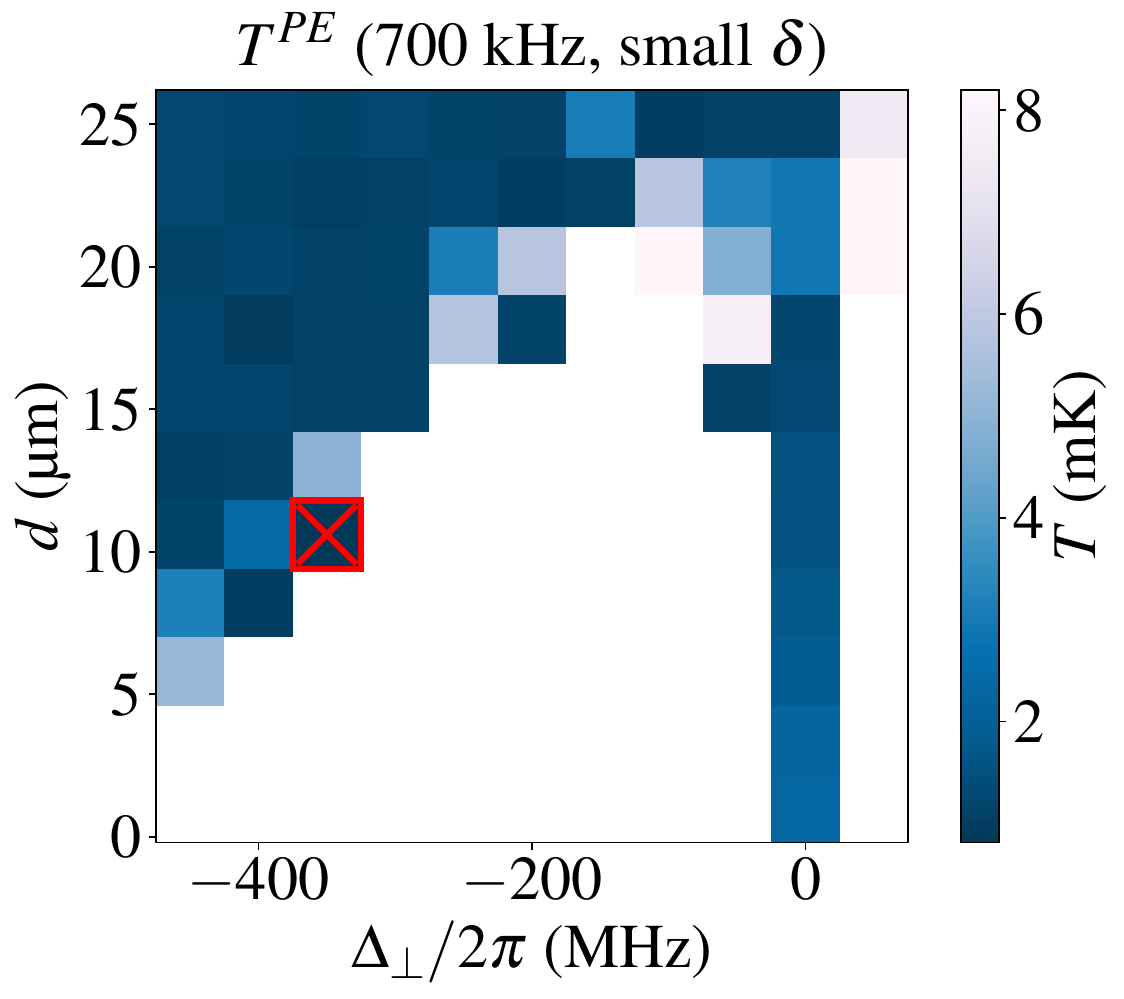}
    \end{subfigure}

    \caption{The final perpendicular kinetic energy and potential energy of the $\delta=0.0104$ ion crystals are plotted as a function of $\Delta_\perp$ and $d$. The squares outlined in red and marked with a red `X' correspond to the simulations used in Figs.~\ref{fig5}, \ref{fig6}, and \ref{fig7}.  White squares correspond to higher temperatures which are omitted from the plot to enhance the color gradient in the regions of interest.}
    \label{fig12}
\end{figure}

\begin{figure}
    \centering

    \begin{subfigure}{0.2\textwidth}
        \centering
        \includegraphics[scale=0.23]{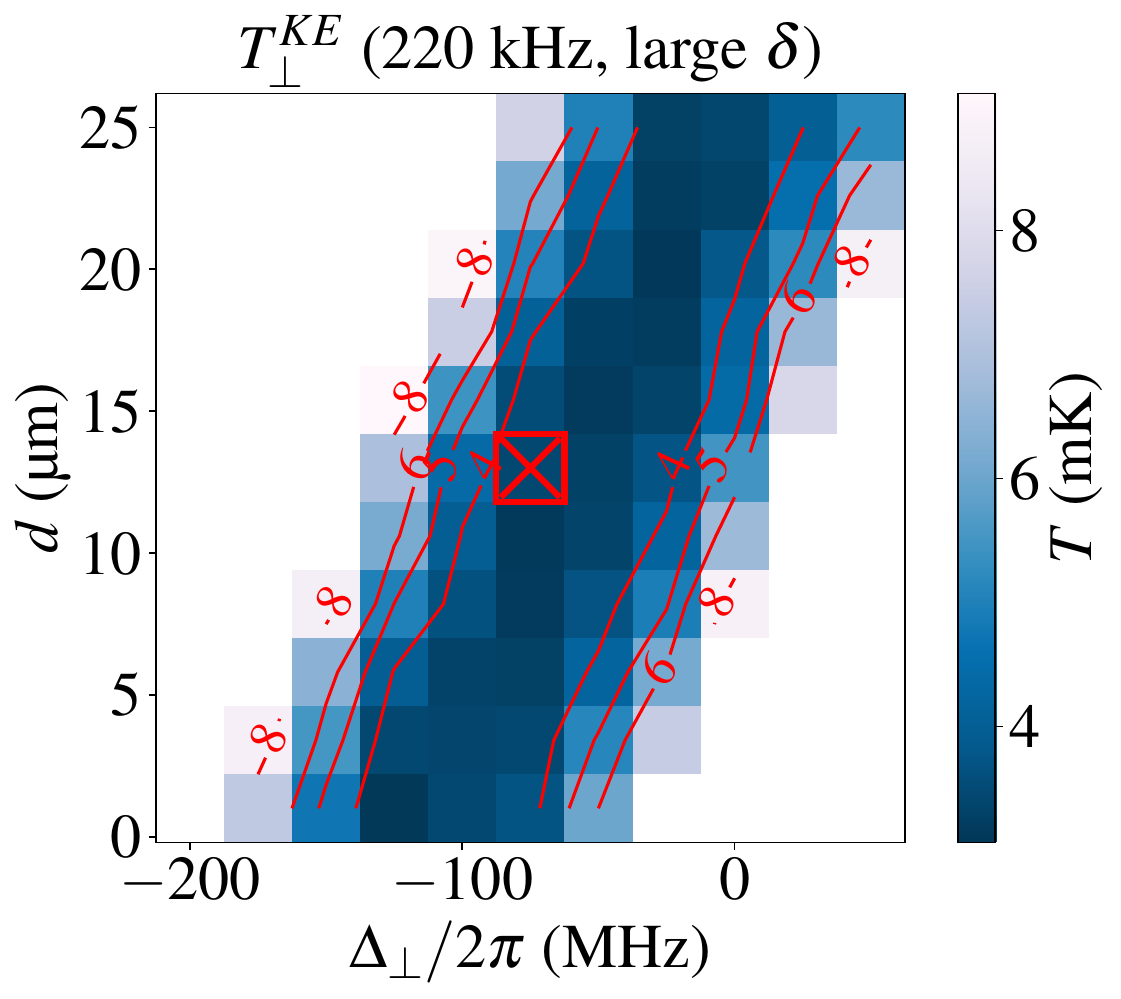}
    \end{subfigure}%
    \hfill
    \begin{subfigure}{0.2\textwidth}
        \centering
        \hspace*{-0.7cm}
        \includegraphics[scale=0.23]{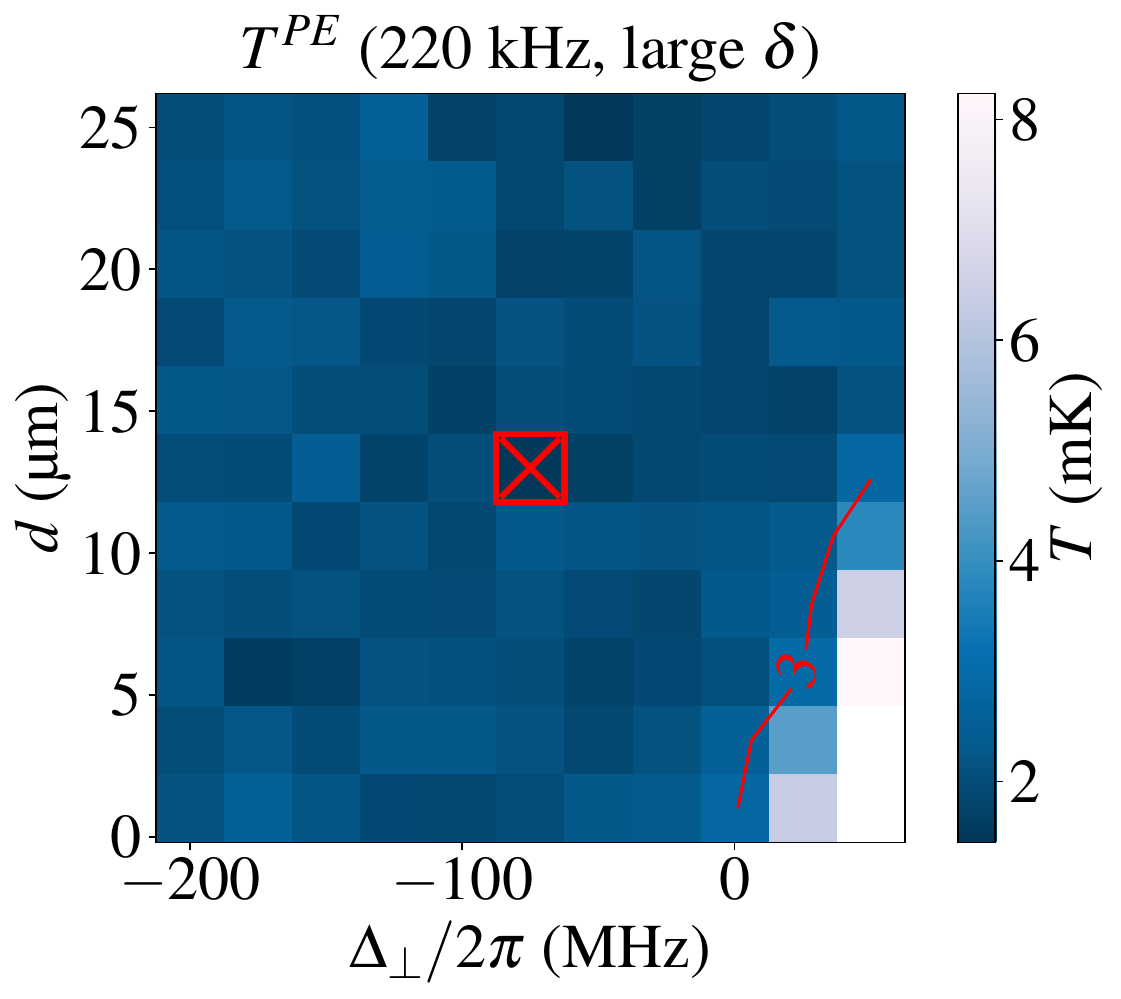}
    \end{subfigure}

    \vspace{0.1em}

    \begin{subfigure}{0.2\textwidth}
        \centering
        \includegraphics[scale=0.23]{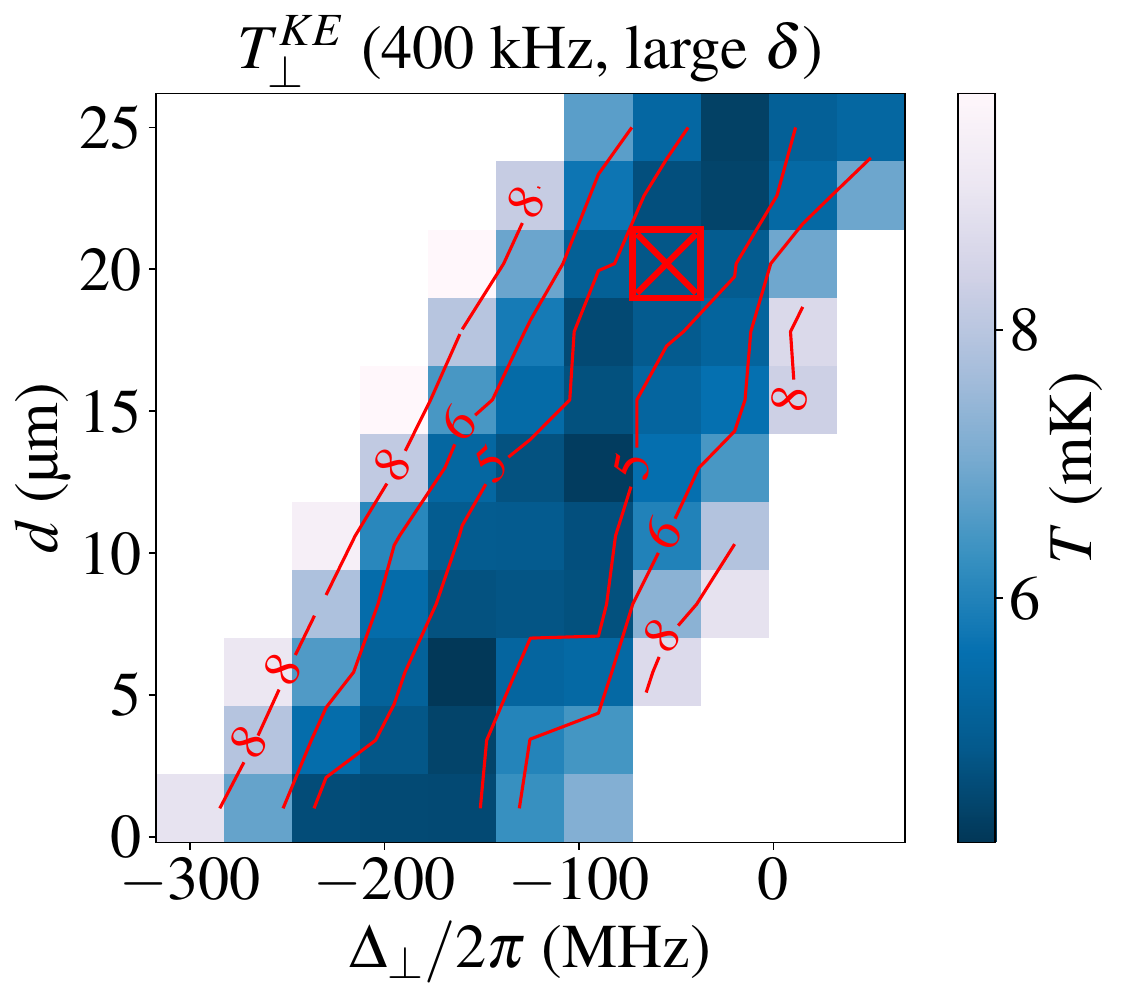}
    \end{subfigure}%
    \hfill
    \begin{subfigure}{0.2\textwidth}
        \centering
        \hspace*{-0.7cm}
        \includegraphics[scale=0.23]{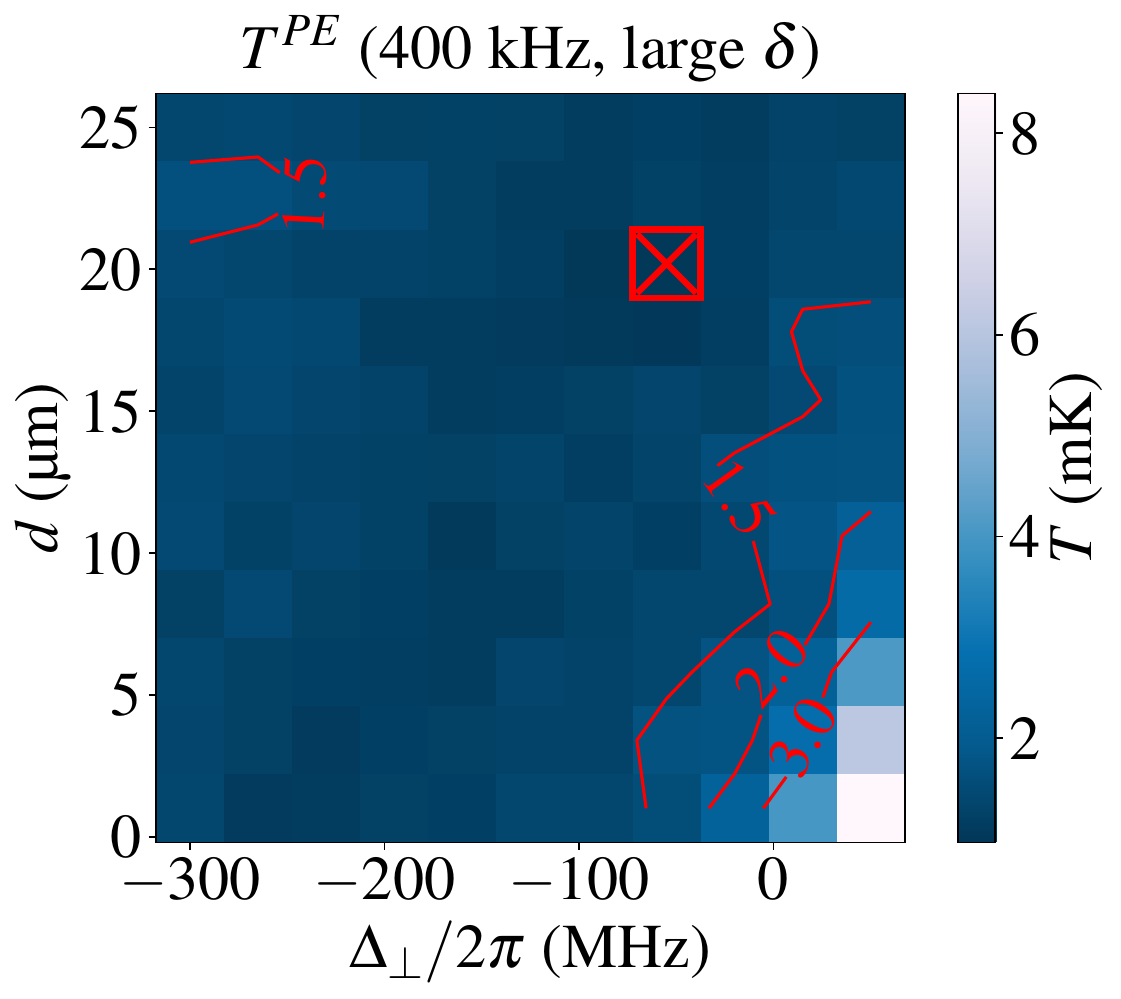}
    \end{subfigure}

    \vspace{0.1em}

    \begin{subfigure}{0.2\textwidth}
        \centering
        \includegraphics[scale=0.23]{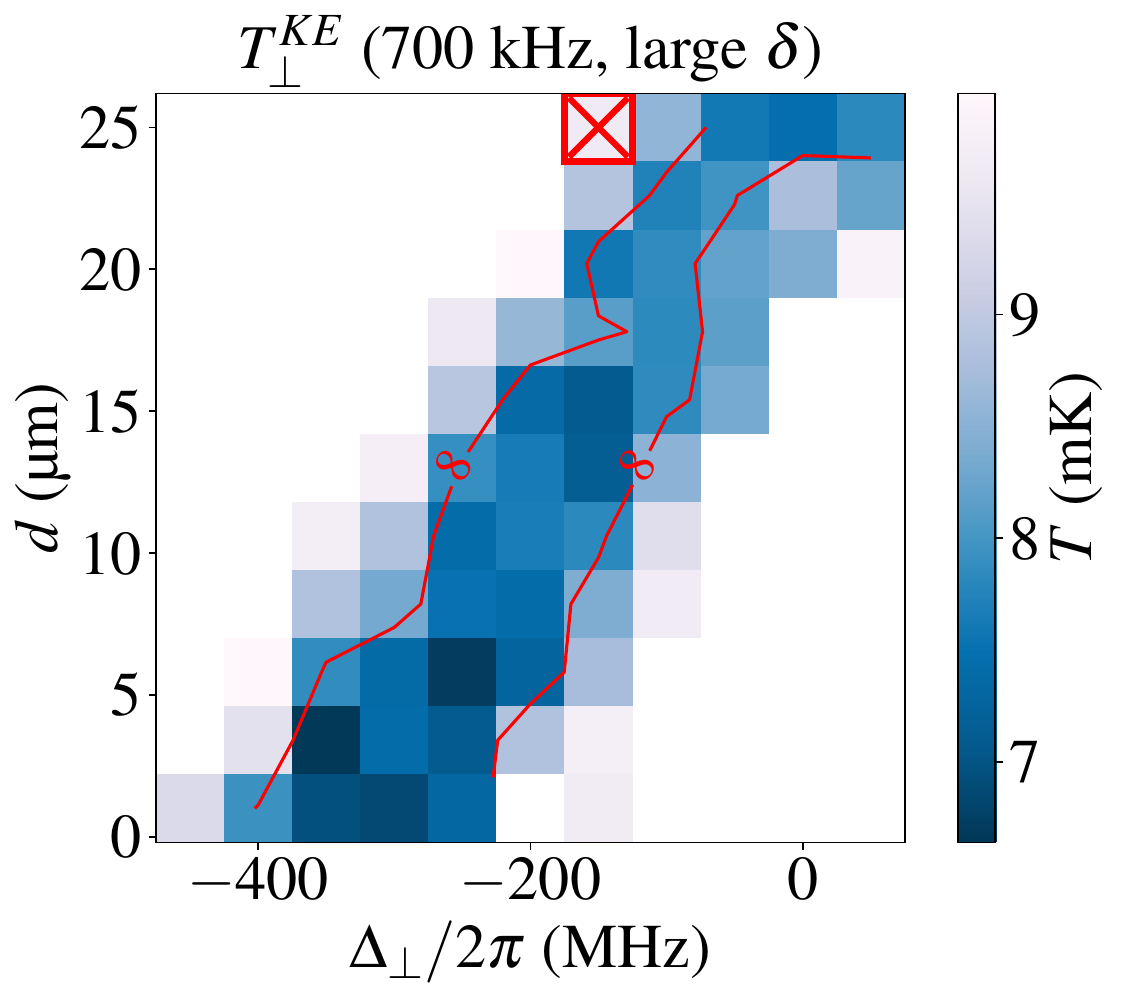}
    \end{subfigure}%
    \hfill
    \begin{subfigure}{0.2\textwidth}
        \centering
        \hspace*{-0.7cm}
        \includegraphics[scale=0.23]{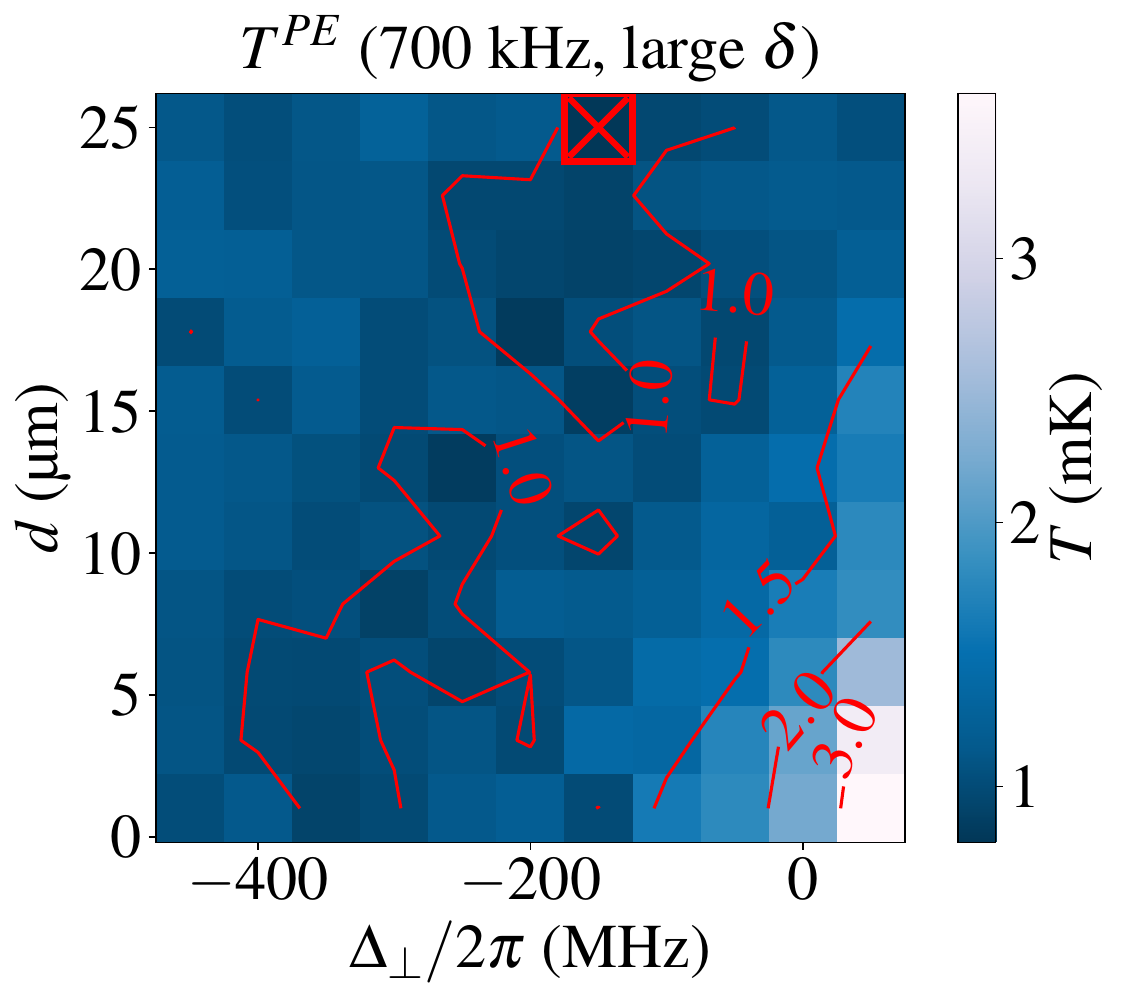}
    \end{subfigure}

    \caption{The final perpendicular kinetic energy and potential energy of the $\delta=0.104$ ion crystals are plotted as a function of $\Delta_\perp$ and $d$.}
    \label{fig13}
\end{figure}

Here, we provide the complete Doppler laser cooling results for the parameter scans which are used to produce Figs.~\ref{fig5}, \ref{fig6}, and \ref{fig7}.  Cooling simulations were performed for a range of perpendicular beam detunings, $\Delta_{\perp}$, and offsets, $d$. We plot the final values of $T^{PE}$ and $T^{KE}_\perp$ for each $(\Delta_{\perp},d)$ parameter pair.  Fig.~\ref{fig12} and Fig.~\ref{fig13} illustrate the final temperatures for the small $\delta$ ($\delta=0.0104$) crystals and large $\delta$ ($\delta=0.104$) crystals, respectively.  In each plot, the square outlined in red and with a red `X' mark denotes the simulation with the lowest final $T^{PE}$.  These are the runs used in Figs.~\ref{fig5}, \ref{fig6}, and \ref{fig7}.

For each crystal, the $(\Delta_{\perp},d)$ parameter space was loosely selected to minimize $T^{KE}_\perp$.  In most of the plots, white squares correspond to simulations in which the final temperature is greater than the initial temperature of 10 mK.  These values are omitted from the plots to enahnce the color gradient in the region of interest.  However, in the case of the small $\delta$ crystals with $\omega_r/2\pi = 178.15$ and $700$ kHz, the square with the best $PE$ cooling corresponds to slight $KE_\perp$ heating, so slightly larger values of $T^{KE}_{\perp}$ are shown.  For the $\omega_r/2\pi = 400$ and $700$ kHz crystals, there exist large regions of parameter space which produce $PE$ (and also $KE_\perp$) heating.  While we do not plot $T^{KE}_\parallel$, we note that the axial kinetic energy cooling is degraded in these regions.  In all other simulation runs, $T^{KE}_\parallel$ is rapidly cooled to $<$ 1 mK.  As seen in Fig.~\ref{fig13}, the regions of $PE$ heating disappear when a stronger rotating wall is applied.

\bibliography{apssamp}

\end{document}